\documentstyle[graphicx,lscape,times]{mn}

\newif\ifAMStwofonts

\ifoldfss
  \newcommand{\rmn}[1] {{\rm #1}}

  \ifCUPmtlplainloaded \else
    \NewTextAlphabet{textbfit} {cmbxti10} {}
    \NewTextAlphabet{textbfss} {cmssbx10} {}
    \NewMathAlphabet{mathbfit} {cmbxti10} {} 
    \NewMathAlphabet{mathbfss} {cmssbx10} {} 
  \fi
  \ifAMStwofonts
    \ifCUPmtlplainloaded \else
      \NewSymbolFont{upmath} {eurm10}
      \NewSymbolFont{AMSa} {msam10}
      \NewMathSymbol{\upi}     {0}{upmath}{19}
      \NewMathSymbol{\umu}     {0}{upmath}{16}
      \NewMathSymbol{\upartial}{0}{upmath}{40}
      \NewMathSymbol{\leqslant}{3}{AMSa}{36}
      \NewMathSymbol{\geqslant}{3}{AMSa}{3E}

       \let\le=\leqslant
       \let\ge=\geqslant
    \fi
  \fi
\fi 

\ifnfssone
  \newmathalphabet{\mathit}
  \addtoversion{normal}{\mathit}{cmr}{m}{it}
  \addtoversion{bold}{\mathit}{cmr}{bx}{it}
  \newcommand{\rmn}[1] {\mathrm{#1}}

  \newmathalphabet{\mathbfit} 
  \addtoversion{normal}{\mathbfit}{cmr}{bx}{it}
  \addtoversion{bold}{\mathbfit}{cmr}{bx}{it}
  \newmathalphabet{\mathbfss} 
  \addtoversion{normal}{\mathbfss}{cmss}{bx}{n}
  \addtoversion{bold}{\mathbfss}{cmss}{bx}{n}
  \ifAMStwofonts
    \ifCUPmtlplainloaded \else
      \UseAMStwoboldmath
      \makeatletter
      \new@mathgroup\upmath@group
      \define@mathgroup\mv@normal\upmath@group{eur}{m}{n}
      \define@mathgroup\mv@bold\upmath@group{eur}{b}{n}
      \edef\UPM{\hexnumber\upmath@group}
      \new@mathgroup\amsa@group
      \define@mathgroup\mv@normal\amsa@group{msa}{m}{n}
      \define@mathgroup\mv@bold\amsa@group{msa}{m}{n}
      \edef\AMSa{\hexnumber\amsa@group}
      \makeatother
      \mathchardef\upi="0\UPM19
      \mathchardef\umu="0\UPM16
      \mathchardef\upartial="0\UPM40
      \mathchardef\leqslant="3\AMSa36
      \mathchardef\geqslant="3\AMSa3E

       \let\le=\leqslant
       \let\ge=\geqslant
    \fi
  \fi
\fi 

\ifnfsstwo
  \newcommand{\rmn}[1] {\mathrm{#1}}

  \DeclareMathAlphabet{\mathbfit}{OT1}{cmr}{bx}{it}
  \SetMathAlphabet\mathbfit{bold}{OT1}{cmr}{bx}{it}
  \DeclareMathAlphabet{\mathbfss}{OT1}{cmss}{bx}{n}
  \SetMathAlphabet\mathbfss{bold}{OT1}{cmss}{bx}{n}
  \ifAMStwofonts
    \ifCUPmtlplainloaded \else
      \DeclareSymbolFont{UPM}{U}{eur}{m}{n}
      \SetSymbolFont{UPM}{bold}{U}{eur}{b}{n}
      \DeclareSymbolFont{AMSa}{U}{msa}{m}{n}
      \DeclareMathSymbol{\upi}{0}{UPM}{"19}
      \DeclareMathSymbol{\umu}{0}{UPM}{"16}
      \DeclareMathSymbol{\upartial}{0}{UPM}{"40}
      \DeclareMathSymbol{\leqslant}{3}{AMSa}{"36}
      \DeclareMathSymbol{\geqslant}{3}{AMSa}{"3E}

       \let\le=\leqslant
       \let\ge=\geqslant
    \fi
  \fi
\fi 

\ifCUPmtlplainloaded \else
  \ifAMStwofonts \else 
    \def\upi{\pi}
    \def\umu{\mu}
    \def\upartial{\partial}
  \fi
\fi

\title[RR Lyrae stars in Fornax clusters]
  {RR Lyrae stars in four globular clusters in the Fornax dwarf galaxy}
\author[A.~D.~Mackey \& G.~F.~Gilmore]
  {A.~D.~Mackey$^1$\thanks{E-mail: dmackey@ast.cam.ac.uk}
  and G.~F.~Gilmore$^1$\\
  $^1$Institute of Astronomy, University of Cambridge, Madingley Road,
  Cambridge CB3 0HA}
\date{Accepted --. Received --}
\pagerange{000--000}
\pubyear{2003}

\def\LaTeX{L\kern-.36em\raise.3ex\hbox{a}\kern-.15em
    T\kern-.1667em\lower.7ex\hbox{E}\kern-.125emX}

\begin{document}

\label{firstpage}

\maketitle

\begin{abstract}
We have surveyed four of the globular clusters in the Fornax dwarf galaxy (clusters 1, 2, 3, and 5)
for RR Lyrae stars, using archival F555W and F814W {\em Hubble Space Telescope} observations.
We identify 197 new RR Lyrae stars in these four clusters, and 13 additional candidate horizontal 
branch variable stars. Although somewhat restricted by our short observational baseline, we derive 
periods and light-curves for all of the stars in the sample, and calculate photometric
parameters such as mean magnitudes and colours. This is the first time that RR Lyrae stars
in the Fornax globular clusters have been quantitatively identified and measured. We find that
the Fornax clusters have exceptionally large specific frequencies of RR Lyrae stars, in comparison
with the galactic globular clusters. It is likely that Fornax 1 has the largest specific 
frequency measured in {\em any} globular cluster. In addition, the Fornax clusters are unusual
in that their RR Lyrae populations possess mean characteristics intermediate between the
two Oosterhoff groups defined by the galactic globular clusters. In this respect the RR Lyrae
populations in the Fornax clusters most closely resemble the {\em field} populations in several
dwarf galaxies. Fornax 5 has an unusually large fraction of RRc stars, and also possesses several
strong RRe (second overtone pulsator) candidates.

With a large sample of horizontal branch variable stars available to us, we revise previous 
measurements of the horizontal branch morphology in each cluster. The Fornax clusters most 
closely resemble the ``young'' galactic halo population defined by Zinn in that their horizontal
branch morphologies are systematically redder than many galactic clusters of similar metallicity.
We also confirm the existence of the second parameter effect among the Fornax clusters, most
markedly between clusters 1 and 3. The edges of the instability strip are well defined in several
of the Fornax clusters, and we are able to make measurements of the intrinsic $V-I$ colours of
these edges. Finally, we determine foreground reddening and distance
estimates for each cluster. We find a mean distance modulus to the Fornax dwarf of 
$(m-M)_0 = 20.66 \pm 0.03$ (random) $\pm 0.15$ (systematic). Our measurements are consistent
with a line of sight depth of $\sim 8-10$ kpc for this galaxy, which is in accordance with its
dimensions as measured in the plane of the sky. This approximately spherical shape for Fornax
is incompatible with tidal model explanations for the observed high internal stellar velocity 
dispersions in many dwarf spheroidal galaxies. Dark matter dominance is suggested.
\end{abstract}


\section{Introduction}
The Fornax dwarf galaxy is one of the most massive of the dwarf spheroidal (dSph) galaxies associated
with the Milky Way (second only to the disrupted Sagittarius dSph), and the only undisturbed local 
galaxy to possess globular clusters (although Kleyna et al. \shortcite{umi} have recently
suggested the presence of a disrupted cluster in the Ursa Minor dSph). In this respect Fornax is 
somewhat unusual, because its population of five means that it has the highest specific frequency 
of globular clusters for any known galaxy. 

As summarized by Strader et al. \shortcite{strader}, although small in number, this globular
cluster system is remarkably complex. Buonanno et al. \shortcite{buonfora,buonforb} have shown
that clusters 1, 2, 3, and 5 appear coeval with both each other and the metal-poor galactic globular
clusters, and that cluster 4 may be up to three gigayears younger.
Strader et al. \shortcite{strader} suggest that Fornax 5 may also be slightly younger.
It has also been found \cite{smith15,buonfora} that the Fornax globular clusters show evidence of
the second parameter effect, whereby another parameter in addition to metallicity appears to
define the morphology of the horizontal branch in a cluster. It is often assumed that age is the
second parameter; however, if the age measurements listed above are correct, then it is possible
that age is not the sole second parameter in the Fornax clusters.
We \cite{fornaxpaper} have studied the surface brightness profiles of the Fornax globular clusters, 
and find a large amount of variation in the cluster structures -- ranging from an extremely 
extended cluster (Fornax 1), to a strong post core-collapse candidate (Fornax 5). Evidently the 
Fornax cluster system is highly complicated; however both this complexity, and the system's isolated 
nature provide unique opportunities for addressing the outstanding difficulties in our understanding of
globular cluster formation and evolution. The Fornax globular clusters are therefore worthy of 
detailed study.

Much can be learned about a stellar population from the variable stars within that population,
and given the relatively close proximity of the Fornax dwarf, it is perhaps surprising that
a detailed investigation of stellar variability in its clusters has not been undertaken.
In fact, we can find no quantitative identification and measurements of any variable stars 
in these objects in the literature -- only several passing mentions of possible RR Lyrae
detections exist \cite{buonold,demers1,buonposter,smith15,smith3,buonfora}, along with one
short article describing a survey in progress, and including cluster 3 \cite{maio}.

In the process of obtaining photometry for colour-magnitude diagrams from archival {\em Hubble
Space Telescope} observations, we noticed the possibility of surveying four of the Fornax
globular clusters for RR Lyrae stars. We present here the results of this survey, which has 
been successful beyond our original expectations. We have discovered 197 RR Lyrae stars in the
four clusters (Fornax 1, 2, 3, and 5), as well as 13 candidate horizontal branch variables.
We describe the data and our survey strategy in Section \ref{obsred}, and present light curves
and mean properties for each star in Section \ref{results}. Finally, we have been able to use
the RR Lyrae discoveries to measure detailed information about each of the four clusters under
consideration. This discussion is presented in Section \ref{discussion}.

\section{Observations and Reductions}
\label{obsred}
\subsection{Data}
\label{data}

\begin{table*}
\begin{minipage}{155mm}
\caption{WFPC2 observations of four of the globular clusters in the Fornax dwarf galaxy (Program ID 5917).}
\begin{tabular}{@{}lllllllll}
\hline \hline
Cluster & Data-group & Reference & Baseline & $N_{im}$ & Exposure & Data-group & $N_{im}$ & Exposure \\
Name & F555W & Image & (days) & & Durations (s) & F814W & & Durations (s) \\
\hline
Fornax 1 & u30m010eb & u30m010et & 0.3595 & 14 & 3$\times$600s, 4$\times$500s, & u30m010ib & 16 & 2$\times$900s, 6$\times$700s, \\
 & & & & & 3$\times$400s, 4$\times$160s \hspace{5mm} & & & 2$\times$500s, 6$\times$120s \\
Fornax 2 & u30m020eb & u30m020et & 0.3602 & 14 & 3$\times$600s, 4$\times$500s, & u30m020ib & 16 & 2$\times$900s, 6$\times$700s, \\
 & & & & & 3$\times$400s, 4$\times$160s \hspace{5mm} & & & 2$\times$500s, 6$\times$120s \\
Fornax 3 & u30m030eb & u30m030et & 0.3602 & 14 & 3$\times$600s, 3$\times$500s, 3$\times$400s, & u30m030ib & 16 & 2$\times$900s, 6$\times$700s, \\
 & & & & & 1$\times$378s, 4$\times$160s \hspace{5mm} & & & 2$\times$500s, 6$\times$120s \\
Fornax 5 & u30m040eb & u30m040et & 0.3553 & 14 & 3$\times$600s, 4$\times$500s, & u30m040ib & 16 & 2$\times$900s, 6$\times$700s, \\
 & & & & & 3$\times$400s, 4$\times$160s \hspace{5mm} & & & 2$\times$500s, 6$\times$120s \\
\hline
\label{observations}
\end{tabular}
\end{minipage}
\end{table*}

Wide Field Planetary Camera 2 (WFPC2) observations of four of the globular clusters in 
the Fornax dwarf galaxy (clusters 1, 2, 3, and 5, according to the notation of Hodge 
\shortcite{hodge}) are available in the {\em HST} archive (Program ID 5917). These observations
were made between 1996 June 4 and 1996 June 6, through the F555W and F814W filters. The observational 
details are listed in Table \ref{observations}. 

Each cluster was imaged 14 times in F555W and 16 times in F814W, with individual exposure
durations ranging from $160-600$s, and $120-900$s, respectively. Long total integration times
are necessary because of the faintness of the Fornax globular clusters, which have distance
moduli $(m-M)_0 \sim 20.68$ \cite{buonforb}. The image sampling, presumably primarily intended 
to facilitate the removal of cosmic rays over the long integrations, renders the data 
suitable for identifying sources with short period variability, such as RR Lyrae stars. The 
exposure durations complement such identification, with horizontal branch (HB) stars being among 
the brightest stars in each image, and neither saturated nor overly faint in any image. 
Furthermore, observations through the two filters are interleaved, so that for any given
F555W observation, it is possible to find an F814W image with a similar exposure duration
and an observation date matching to $\pm 0.02$ days. Such pairing of images allows colour
information over a star's variability cycle to be measured.

Although otherwise well suited for an RR Lyrae star search, the data have very 
short baselines of observation. This means that while variability can be easily detected, period
determination is not straightforward -- especially for RRab type variables, which typically have 
periods in the range $0.4-1.1$ days. Unless such stars are fortuitously measured at a critical 
region of their light curve, the under-sampling renders accurate period determination virtually 
impossible. The shorter cycle RRc type variables, which have periods of $0.2-0.5$ days, are more 
suited to the observation baselines. We will return to the question of period determination in 
Section \ref{templates}.

Finally, we note that there also exist archival WFPC2 observations of Fornax cluster 4 (Program
ID 5637). However, the data consist of only three images per filter, and given the significant 
crowding and field star contamination for this cluster, were not suitable for the present
RR Lyrae work. We do not consider Fornax 4 further.

\subsection{Photometry}
\label{photometry}
All WFPC2 images we retrieved from the {\em HST} archive underwent preliminary reduction 
according to a standard pipeline, using the latest available calibrations. For photometric
measurements from these calibrated images we used Dolphin's HSTphot \cite{hstphot}. This is
a photometric package specifically designed for use on WFPC2 data. In particular, it uses a
library of point spread functions (PSFs) tailored to account for the undersampled WFPC2 PSFs,
allowing very accurate stellar centroiding and photometry.

Before running HSTphot, we completed several preprocessing steps. For each cluster, a reference
image was identified (these are listed in Table \ref{observations}) and all other images aligned
with this image. This was achieved using the {\sc iraf} task {\sc imalign}, treating each
image offset as a simple $x$- and $y$-shift. This is perfectly adequate for the present data --
each image for a given cluster is offset from the reference image due to a simple dithering
pattern of several pixels. There are no significant higher order distortions to account for.

Next, we used the utility software accompanying HSTphot to mask image defects (such as bad
pixels and columns), calculate a background sky image (for use in the photometry calculations),
and attempt to remove cosmic rays and unmasked hot-pixels. We then made photometric measurements
on each image using HSTphot in PSF fitting mode (as opposed to aperture photometry mode). 
We set a minimum threshold for object detection of $3\sigma$ above the background. We also
enabled two additional features of HSTphot. First, we elected to calculate a local adjustment to 
the background image before each photometry measurement, using the pixels just beyond the 
photometry radius \cite{hstphot}. This helped account for regions of rapidly varying background, 
such as near the centres of the very crowded clusters 3 and 5. Second, we chose to run artificial
star tests in conjunction with the photometry measurements. This option generates a large number 
of ``fake'' stars on a CMD, places them one at a time on the image, and solves each just as if it 
were a real star. This allows quantitative estimates of detection completeness as a function of a 
star's magnitude, colour, and position in a cluster.

For each photometric measurement of a star (ideally, 14$\times$F555W measurements and 
16$\times$F814W measurements) HSTphot output nine parameters, including flight magnitudes,
standard magnitudes, and quantities characterizing the goodness-of-fit of the PSF solution
(e.g., classification type, $\chi$, S/N, and sharpness). Magnitudes from HSTphot are calibrated 
according to the recipe of Holtzman et al. \shortcite{holtzman}, and using the latest updates of 
the Dolphin \shortcite{cte} CTE and zero-point calibrations. Each magnitude is corrected to a 
$0\farcs5$ aperture. HSTphot also provided mean positional information for each object, in the 
form of a chip number (where chip $0$ refers to the PC, and chips $1-3$ to WFC2-WFC4 respectively)
and pixel coordinates relative to the frame of the reference image. These coordinates,
in conjunction with the {\sc iraf} task {\sc metric} and the cluster centres measured
in Mackey \& Gilmore \shortcite{fornaxpaper}, allowed us to calculate the radial distance of a 
given object from the centre of its cluster.

We used the goodness-of-fit parameters to select only objects with high quality photometry. 
We retained only measurements for which an object was classified as stellar (HSTphot types
1, 2, and 3), and for which $\chi \le 2.5$, S/N\ $\ge 3.0$, $-0.3 \le$\ sharpness\ $\le 0.3$,
and errors in the flight magnitude $\sigma_{F} \le 0.1$. After the application of this filter,
we kept only stars with five or more F555W-F814W pairs of successful measurements (where
a pair is defined as two observations within $\pm 0.02$ days of each other, as described in
Section \ref{data}). We also passed the artificial star measurements through the quality
filter. This allowed us to assign a detection completeness $\alpha_c$ to each real star, by 
finding the fraction of successful artificial star measurements in a brightness-colour-position 
bin about the real star. Bin widths were $0.25$ mag in brightness, $0.2$ mag in colour, and
$2\arcsec$ in radial distance from the cluster centre. As an example, we consider a typical
Fornax globular cluster RR Lyrae star, of magnitude $V=21.3$, colour $V-I=0.5$, and radial
distance $r=7\arcsec$. The fraction of successful artificial star measurements with 
$21.175 \le V \le 21.425$, $0.4 \le V-I \le 0.6$, and $6\arcsec \le r \le 8\arcsec$ defines 
the detection completeness for this star. We expect the completeness values so derived to be 
accurate to a few per cent, except for very low fractions. Values of $\alpha_c$ less than 
$\sim 0.25$ should be regarded with caution.

\subsection{Identification of Horizontal Branch variable stars}
\label{variability}
The completion of the photometry procedure resulted in a list of stars in each cluster, each with
between 5 and 14 pairs of F555W and F814W measurements. In general, most stars possessed
$\sim 14$ measurement pairs. Using colour-magnitude diagrams (CMDs) we selected stars in
the HB region of each cluster, and determined the Welch \& Stetson \shortcite{varindex} 
variability index $I_{WS}$ for each. This index is calculated as follows. For $N_o$ epochs of 
observations, each resulting in measurements $V_i \pm \sigma_{V,i}$ and $I_i \pm \sigma_{I,i}$, 
the weighted mean magnitudes
\begin{equation}
\overline{V} = \frac{\sum_{i=1}^{N_{o}} \frac{V_i}{\sigma_{V,i}^{2}}}{\sum_{i=1}^{N_{o}} \frac{1}{\sigma_{V,i}^{2}}}
\hspace{15mm}
\overline{I} = \frac{\sum_{i=1}^{N_{o}} \frac{I_i}{\sigma_{I,i}^{2}}}{\sum_{i=1}^{N_{o}} \frac{1}{\sigma_{I,i}^{2}}}
\end{equation}
may be computed. The variability index is then defined as:
\begin{equation}
I_{WS} = \sqrt{\frac{1}{N_{o}(N_{o}-1)}} \sum_{i=1}^{N_{o}} (\delta V_i \delta I_i)
\end{equation}
where the normalized magnitude residuals are
\begin{equation}
\delta V_i = \frac{V_i - \overline{V}}{\sigma_{V,i}}
\hspace{15mm}
\delta I_i = \frac{I_i - \overline{I}}{\sigma_{I,i}}
\end{equation}
The variability index therefore allows one to preferentially search for variations in 
photometric measurements which are correlated in the two colours being measured. For example,
an RR Lyrae star will brighten and fade in both $V$ and $I$, {\em in synchronization} over
its pulsation cycle. Variability such as this results in a large value for $I_{WS}$. In the
case of random errors -- for example where one or more stellar images has been impacted by 
a cosmic ray, causing spurious variability in one colour band, or in the case where crowding
introduces large random errors into each measurement -- then $\delta V_i$ and $\delta I_i$ 
are uncorrelated, and the expectation value of $I_{WS}$ is zero. 

\begin{figure}
\includegraphics[width=0.5\textwidth]{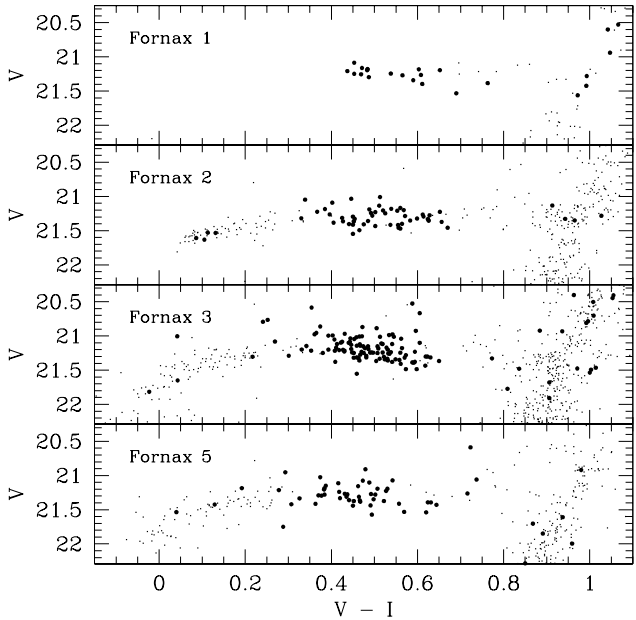}
\caption{Colour-magnitude diagrams showing the horizontal branches of each of the four clusters. Stars with $I_{WS} \ge 3.0$ are marked as bold points -- these are the candidate variables. The RR Lyrae region on each horizontal branch is well defined. We identified a total of 283 candidate variables -- 24, 56, 137, and 66, in the four clusters respectively.}
\label{candvar}
\end{figure}
We therefore selected stars for which $I_{WS}$ was greater than some positive constant as
candidate variables. A small amount of experimentation showed $I_{WS} \ge 3.0$ to
be a suitable value. The resultant candidate variable stars are shown in Fig. \ref{candvar}.
We identified 283 candidates in total (24, 56, 137, and 66, in clusters 1, 2, 3, and 5,
respectively). In each of the HB regions on the CMDs, the RR Lyrae strip is evident, and in 
most cases clearly defined on both edges. For each cluster, there also in general exist some 
candidate variable stars in the blue HB (BHB) region (except Fornax 1, for which no BHB extension 
is observed), and some along the red giant branch (RGB). Because we did not want to assume any 
information, a priori, about the exact limits of the instability strip on the HB, we tested all 
flagged variable stars equally as RR Lyrae candidates.

\subsection{Light-curve fitting}
\label{templates}
The best way to identify a candidate star as an RR Lyrae, and in addition obtain quantitative
information about its variability parameters, is to attempt to fit a light curve to its
photometry. For very under-sampled data, such as the present measurements, this can be tricky.
Eschewing elegance, the most reliable method is the simple application of brute force. 
We used a routine written by Andrew Layden (described in Layden \& Sarajedini \shortcite{laysar};
Layden et al. \shortcite{layden2}; Layden \shortcite{layden1}; and the references therein)
for our period measurements. 

This program works by taking a set of 10 variable star templates and attempting to fit them
to the $V$-band photometric measurements for a number of different periods. The periods are 
selected by an incremental increase $\Delta P$ over a range $P_1$ to $P_2$. For each candidate
period, the measurements are folded, and a 3-parameter fit (magnitude zero-point, phase 
zero-point, and light-curve amplitude) made to each template. Six of the ten templates 
represent RRab type variable stars, while two represent RRc type pulsators, and two
represent variable binary stars (a W Ursae Majoris contact binary, and an Algol eclipsing 
binary). The RRab curves (templates $1-6$), and one of the RRc curves (template $7$) are
derived from high quality measured light curves, as described in Layden \& Sarajedini 
\shortcite{laysar}, and Layden \shortcite{layden1}. The second RRc template (number 8) is
a simple cosine curve. From each template fit, $\chi^2$ is calculated. Period-template
combinations with small $\chi^2$ are likely to be the best representations of the measured data.

Such a procedure works well when the sampling is spread over a much longer baseline than the period
of the variable star under consideration. In the present case however, our $\sim 0.36$ day 
baseline is in general {\em less} than one RR Lyrae pulsation cycle. For the short period (RRc) 
variables, which typically have $P < 0.45$ days, this does not pose too much of a problem. It 
is possible to derive a reliable period estimate for these stars, although naturally, accuracy 
would be greatly increased by having a baseline of many periods. 

The situation is not so good for the RRab variables, which might have periods typically about 
twice that of our baseline. It is possible to determine a reliable period estimate for such stars, 
but only under certain circumstances. Specifically, it is necessary to observe the full amplitude 
of a star's variability (this allows the three parameters in a template fit to be well determined)
and some portion of its decline from maximum (which allows discrimination between different
potential periods). For a mean RRab period of $0.6$ days (in which case our baseline is $0.6P$),
examination of Figure 1 in Layden \shortcite{layden1} shows that we would observe the
appropriate portion of the light curve in approximately $\sim 40$ per cent of cases. For shorter
period RRab stars this fraction is greater, and for longer period stars it is smaller.
Another $\sim 10-20$ per cent of the time, we might observe {\em most} of the portion of the
cycle which we require, in which case the derived period will be a good estimate. In the 
remainder of cases, without assuming some prior knowledge of either the star's period or its
amplitude, we cannot make an accurate period determination. In these incidences, we must make do 
with setting a lower limit to the period, which we achieve by assuming that the observed 
amplitude is the real amplitude of the star. The presence of these lower limits in our
period measurements must be borne in mind if the data is put to any interpretive uses (see e.g.,
Section \ref{discussion}). Measurements at future epochs are required to pin down accurately
the periods of such stars.

Since we were searching for RR Lyrae stars, we chose a period range of $0.2 - 1.0$ days, 
with $\Delta P = 0.01$ per cent. For each candidate, we examined the few fits to the $V$-band 
photometry with the smallest $\chi^2$. We used the $I$-band data as a consistency check, to 
verify that a fit which appeared suitable for the $V$-band photometry also resulted in a good 
$I$-band light-curve. In general the $I$-band data was of significantly poorer quality than the 
$V$-band data, so fitting templates directly to these measurements did not add useful information 
to the procedure. In the vast majority of cases, the $V$-band solution with the very smallest 
$\chi^2$ provided evidently the best fit. We conservatively estimate the typical random uncertainty 
in any of our fitted periods to be $\sim 0.005$ days. While we identified plenty of RR Lyrae 
stars, we did not find any stars with data clearly best fit by either of the variable binary star 
templates. However, given our restriction to a small portion of the CMD, and our very short 
observation baseline, this is not too surprising. Most of the candidate variables flagged on the 
BHB or RGB seemed to have either poor measurements or some form of irregular variability. Certainly 
none had values of $I_{WS}$ as large as those for typical RR Lyrae stars.

\section{Results}
\label{results}

\begin{figure*}
\begin{minipage}{175mm}
\begin{center}
\includegraphics[width=150mm,height=127mm]{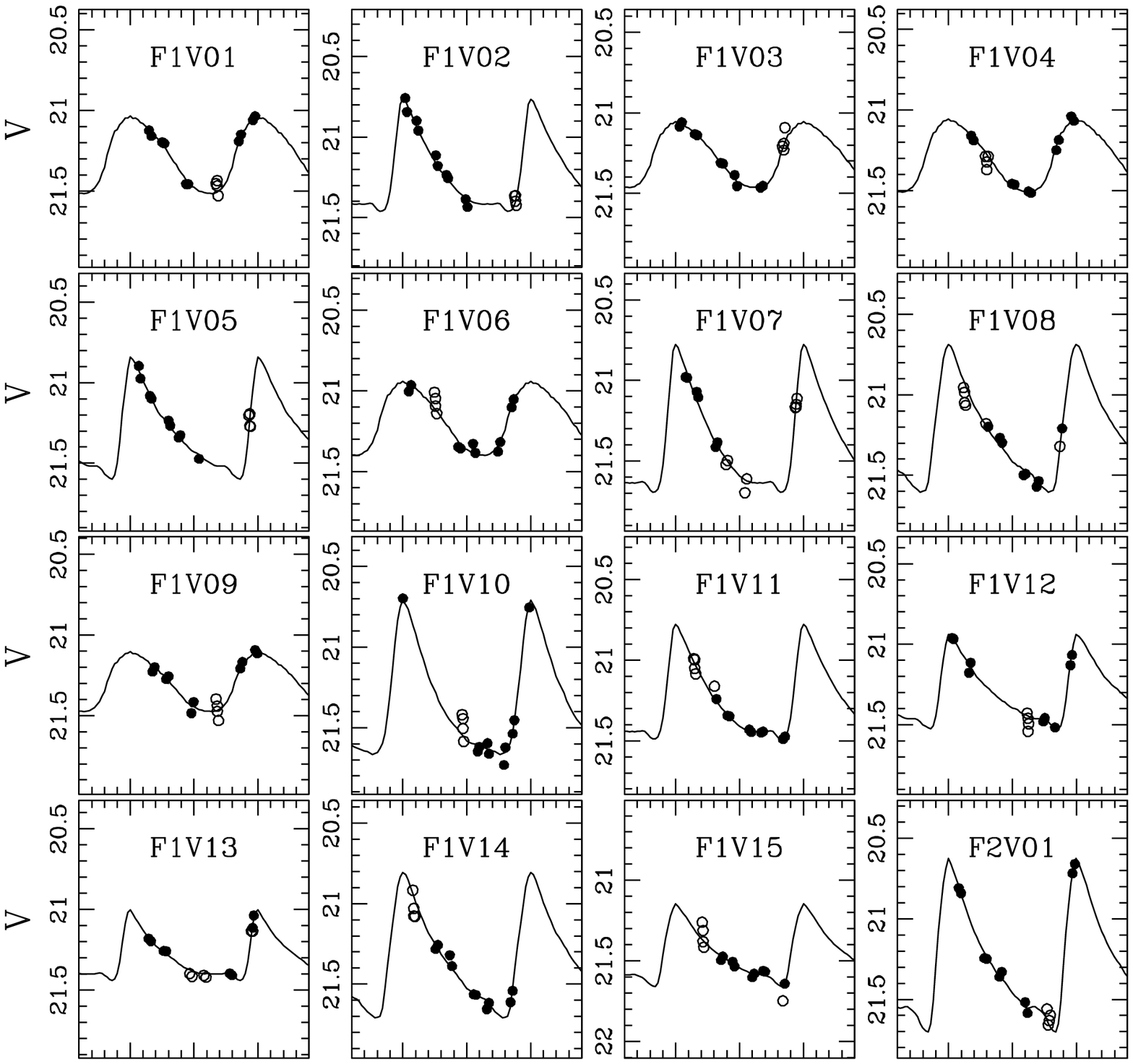} \\
\vspace{-16mm}
\includegraphics[width=150mm,height=127mm]{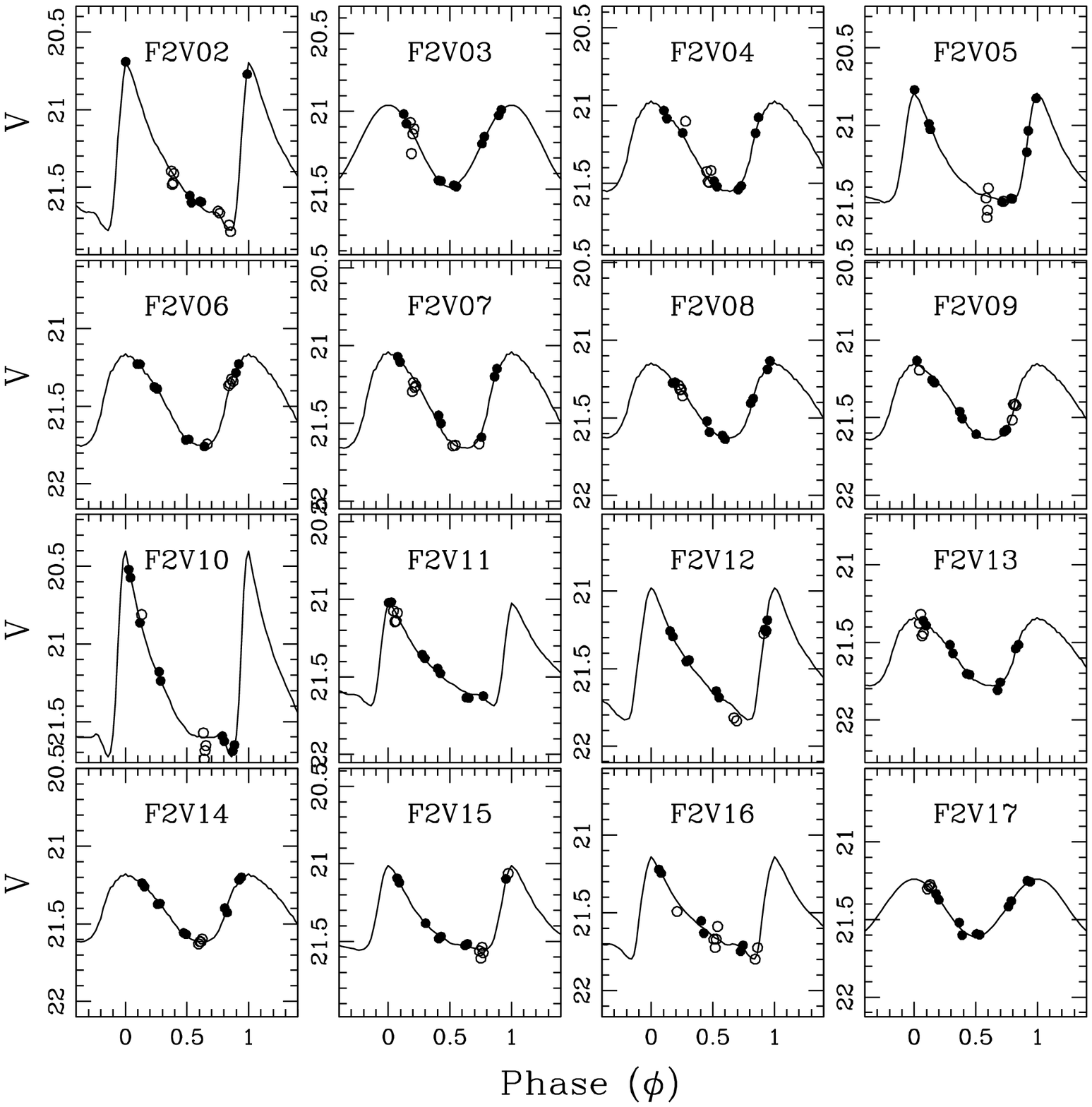}
\end{center}
\end{minipage}
\end{figure*}

\begin{figure*}
\begin{minipage}{175mm}
\begin{center}
\includegraphics[width=150mm,height=127mm]{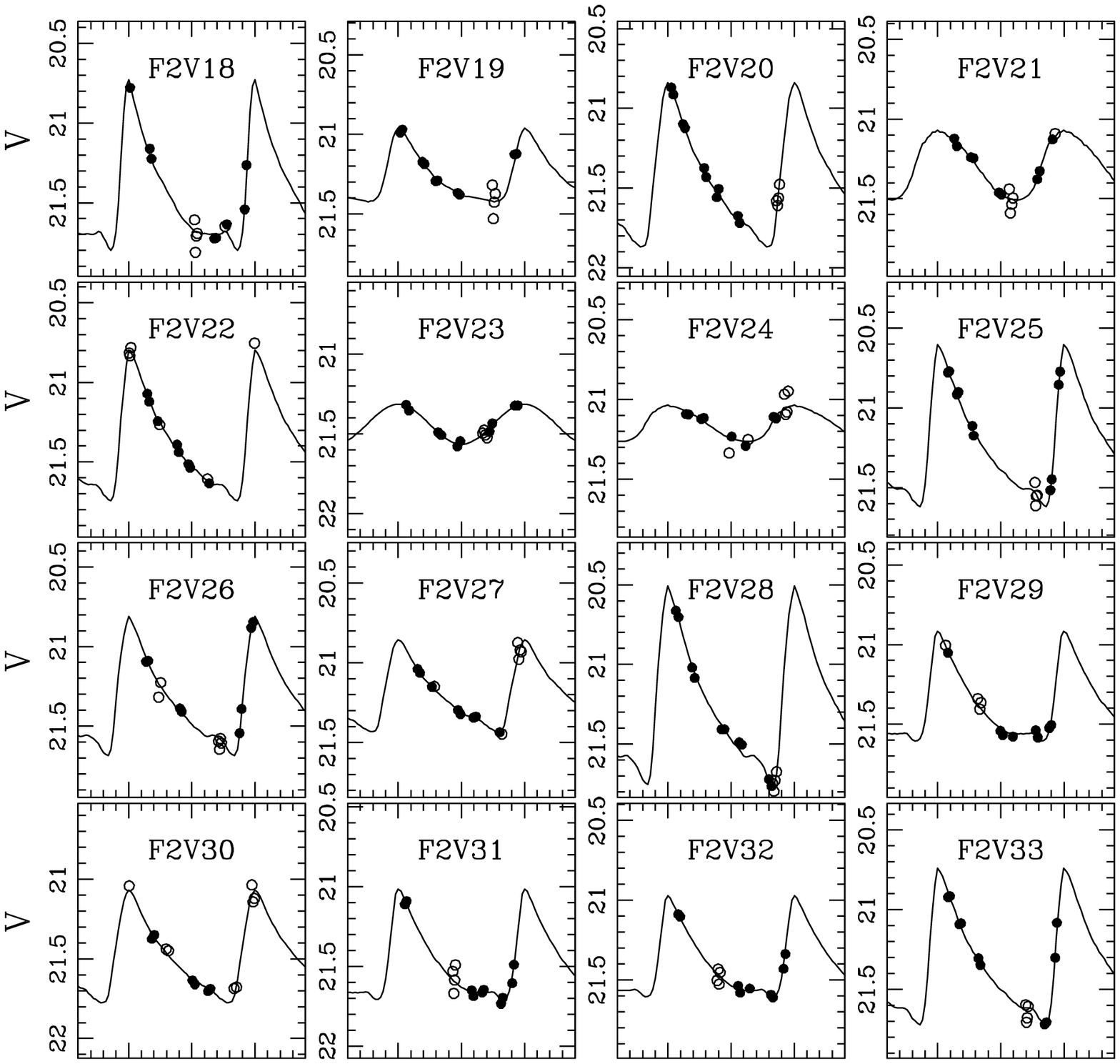} \\
\vspace{-16mm}
\includegraphics[width=150mm,height=127mm]{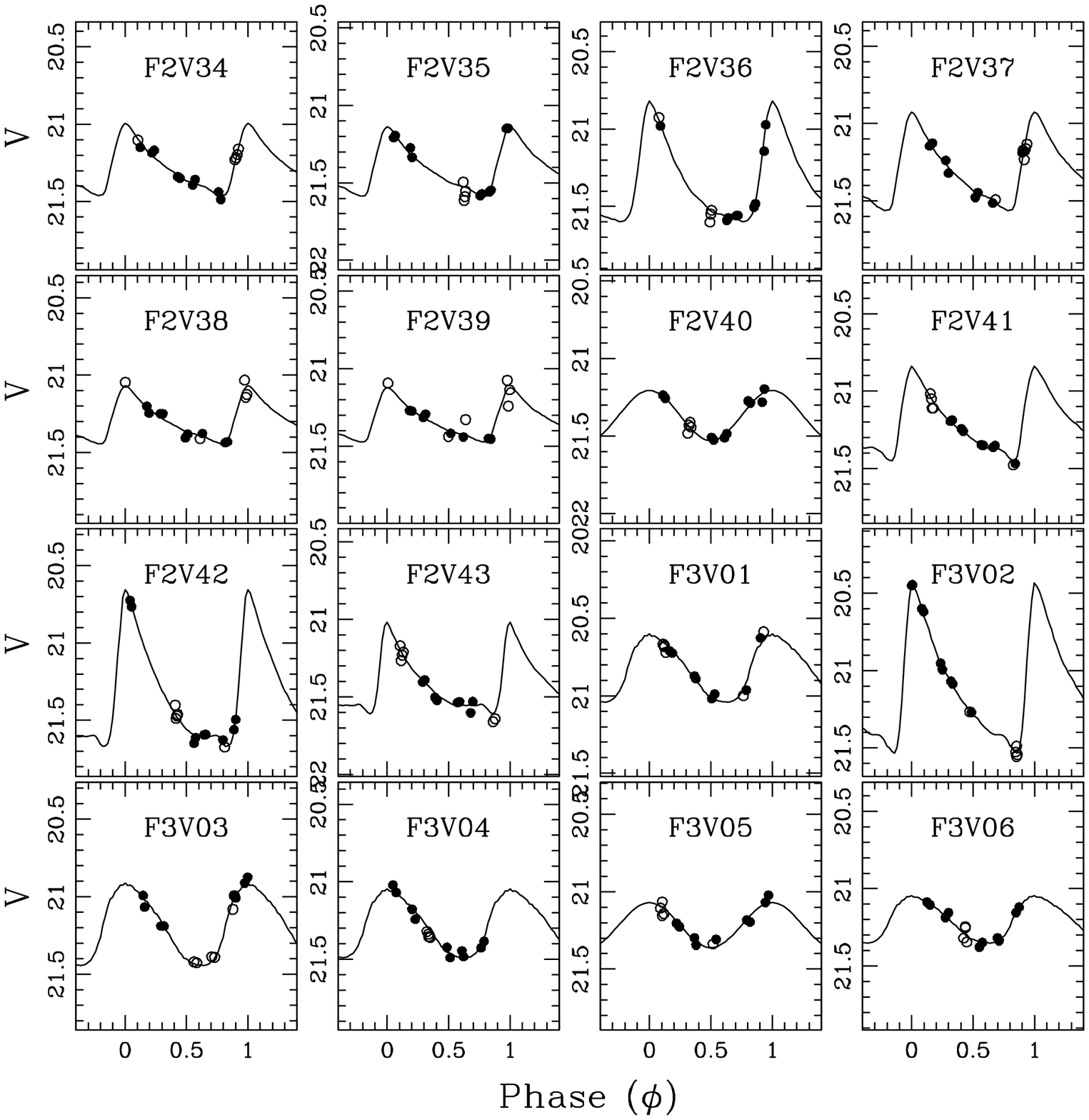}
\end{center}
\end{minipage}
\end{figure*}

\begin{figure*}
\begin{minipage}{175mm}
\begin{center}
\includegraphics[width=150mm,height=127mm]{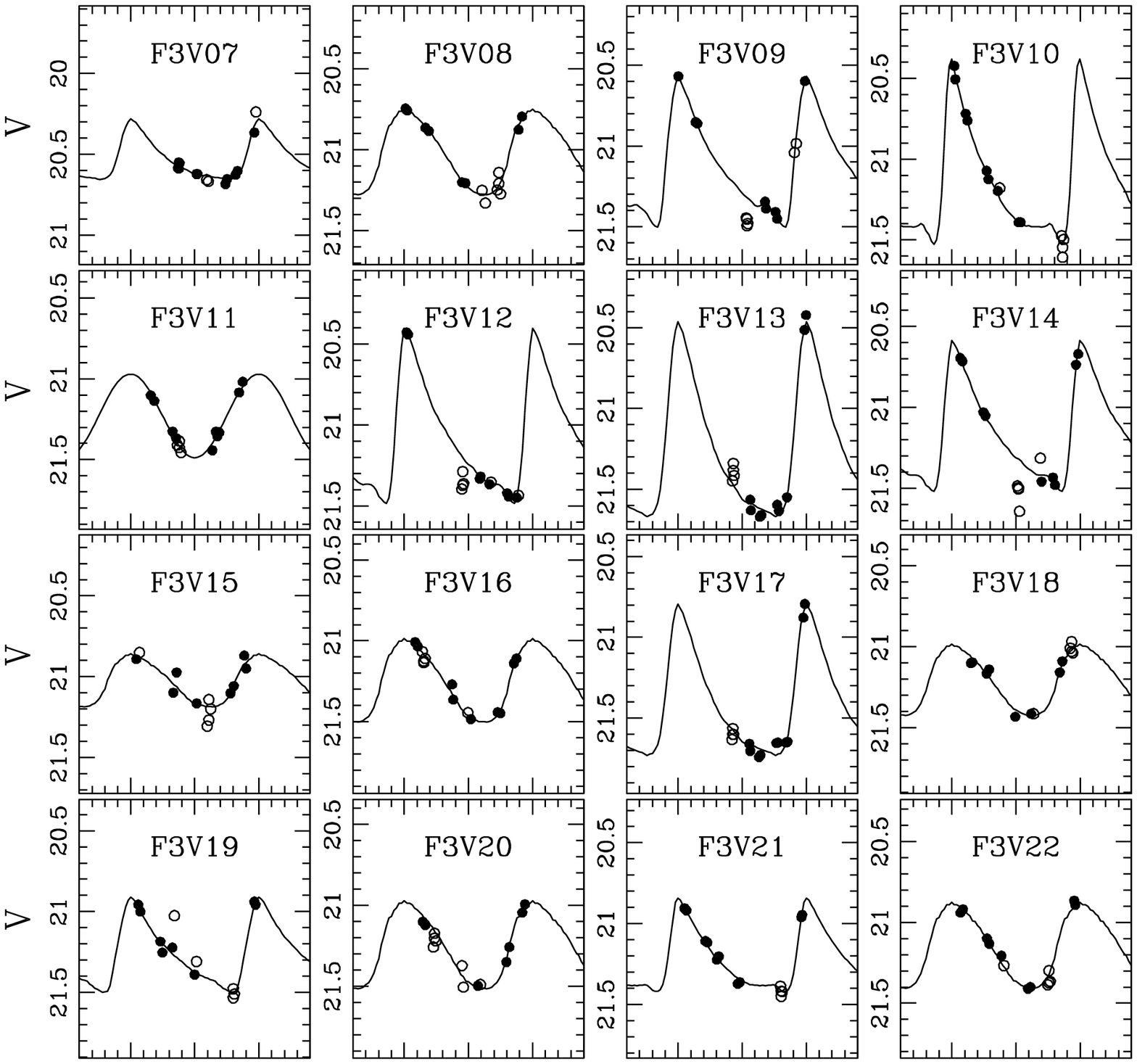} \\
\vspace{-16mm}
\includegraphics[width=150mm,height=127mm]{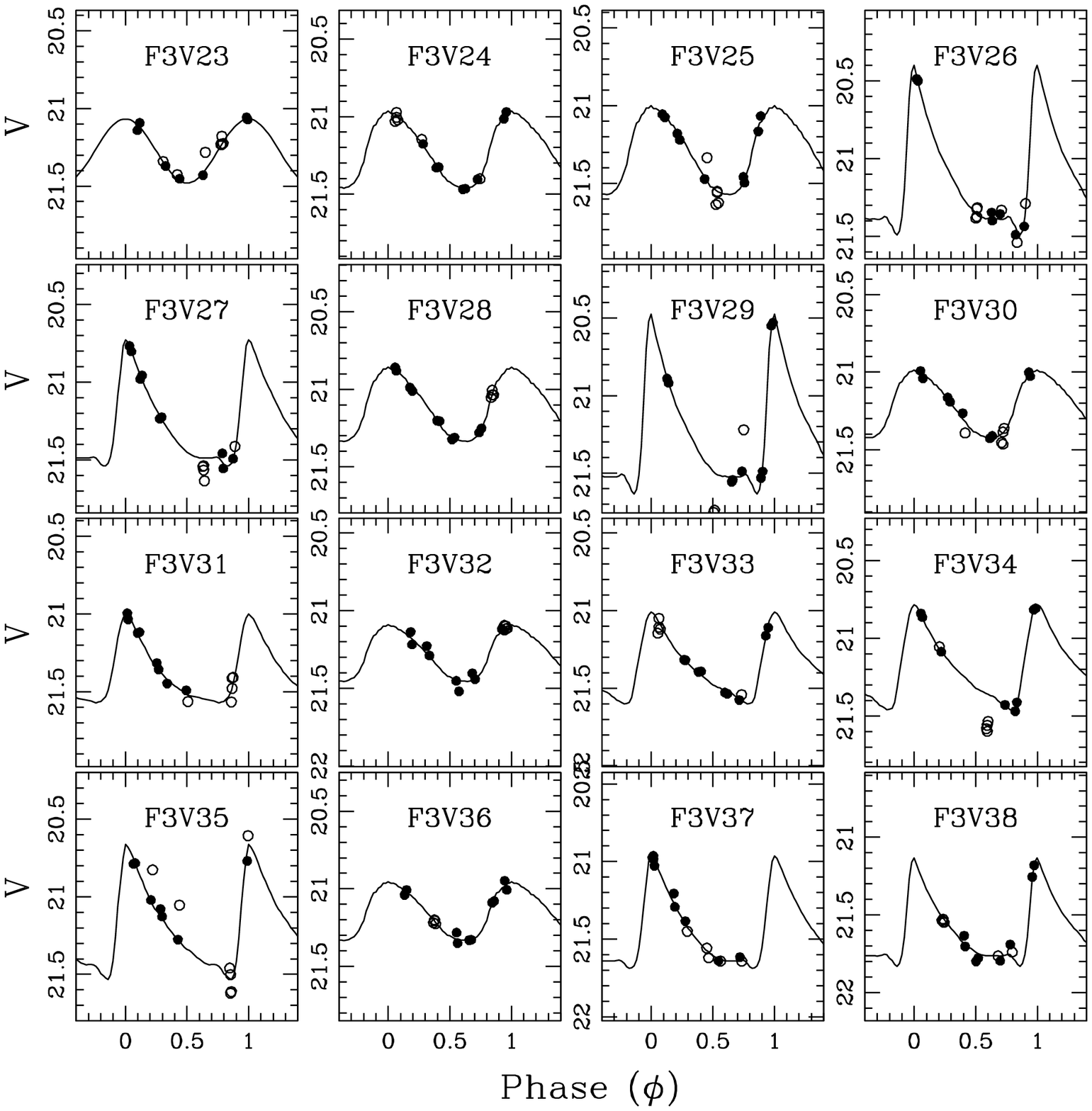}
\end{center}
\end{minipage}
\end{figure*}

\begin{figure*}
\begin{minipage}{175mm}
\begin{center}
\includegraphics[width=150mm,height=127mm]{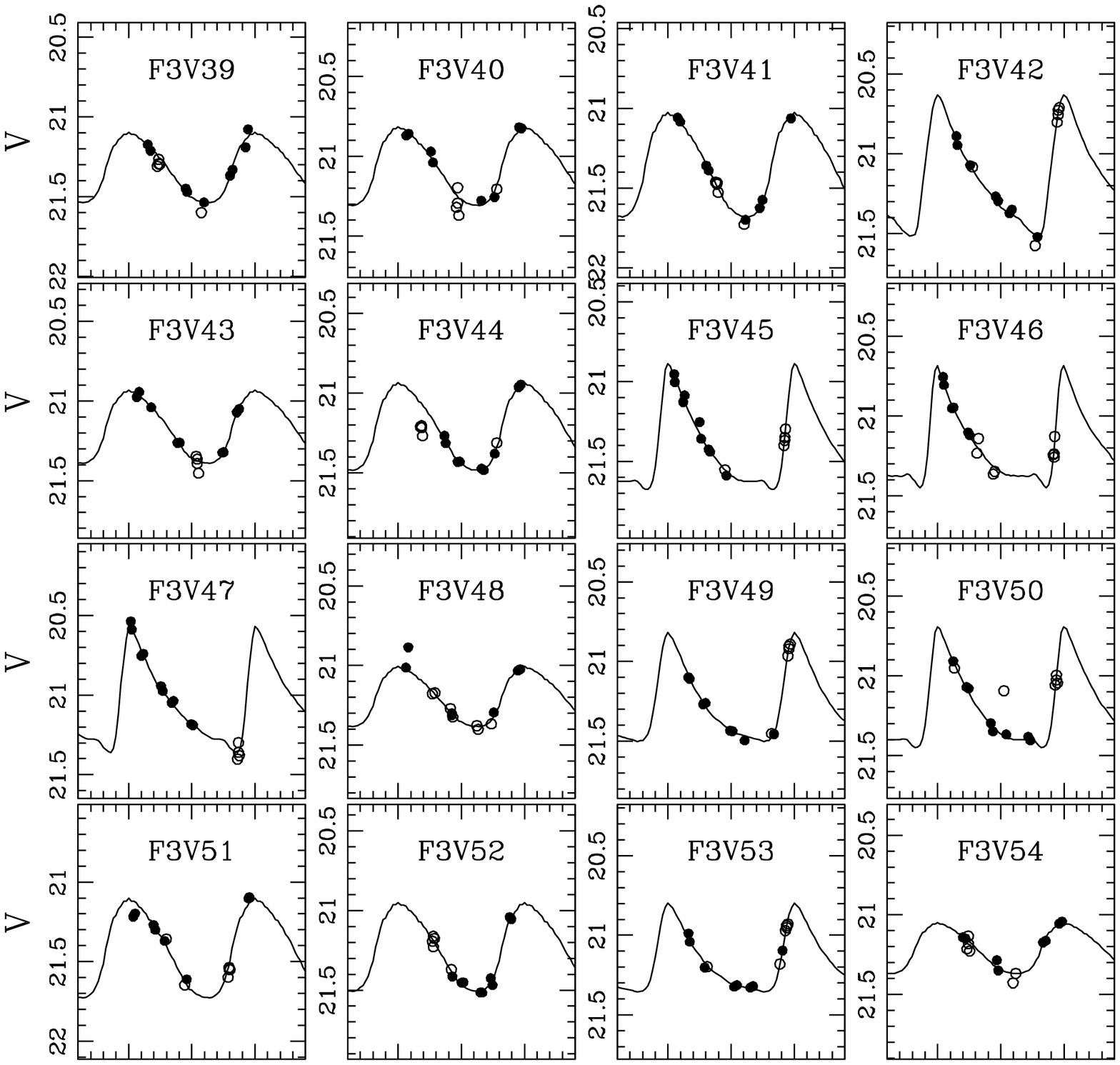} \\
\vspace{-16mm}
\includegraphics[width=150mm,height=127mm]{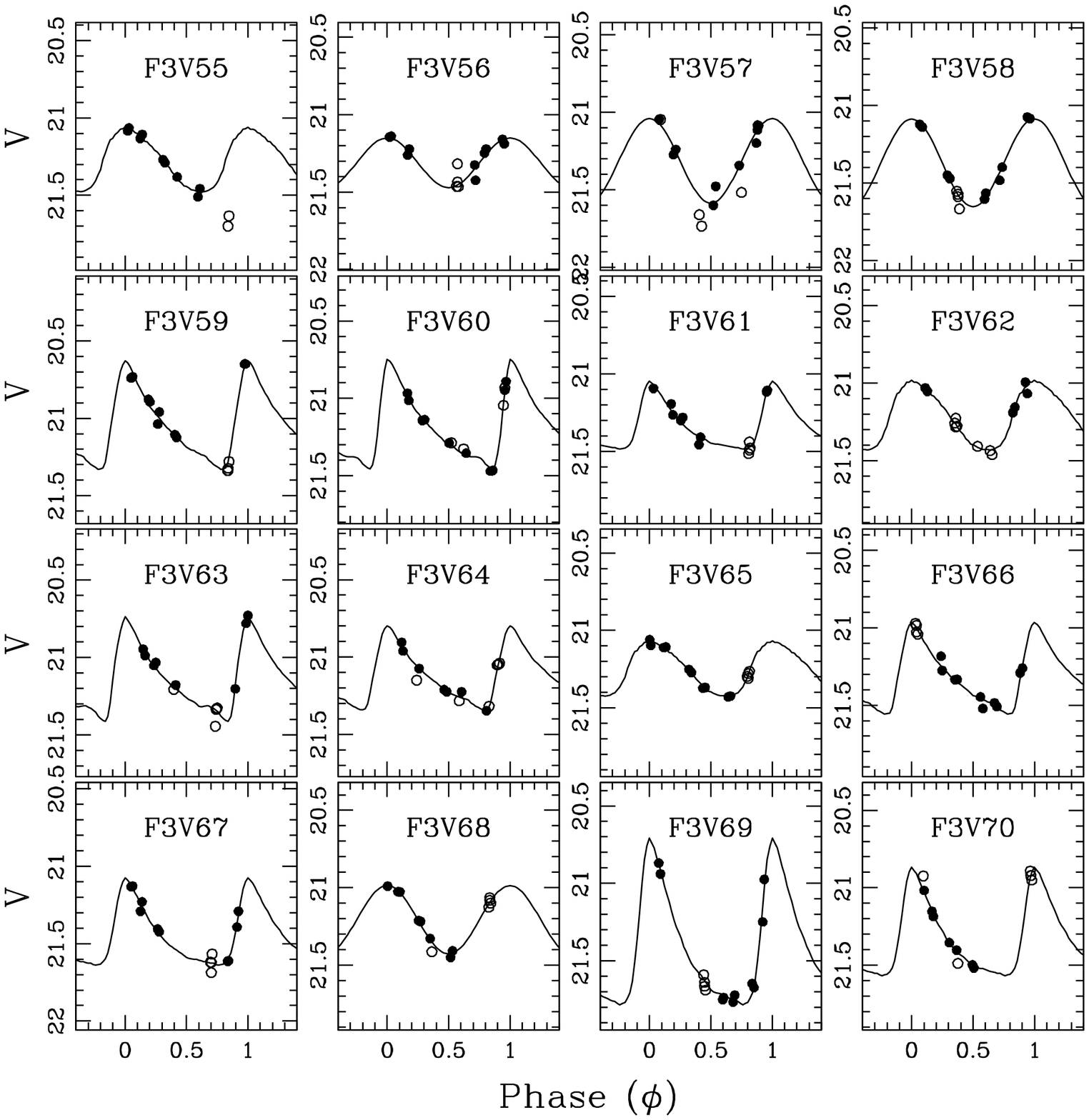}
\end{center}
\end{minipage}
\end{figure*}

\begin{figure*}
\begin{minipage}{175mm}
\begin{center}
\includegraphics[width=150mm,height=127mm]{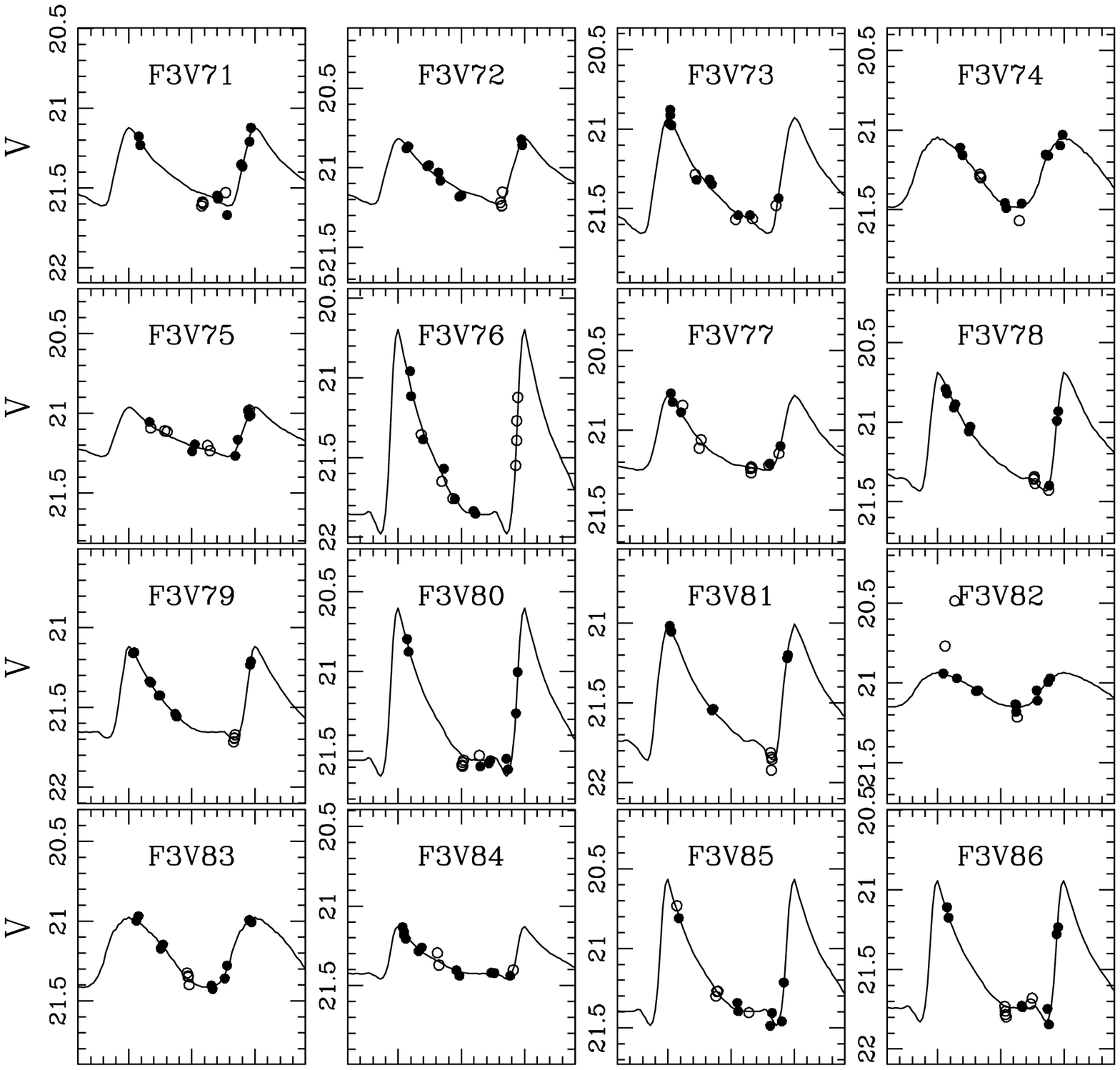} \\
\vspace{-16mm}
\includegraphics[width=150mm,height=127mm]{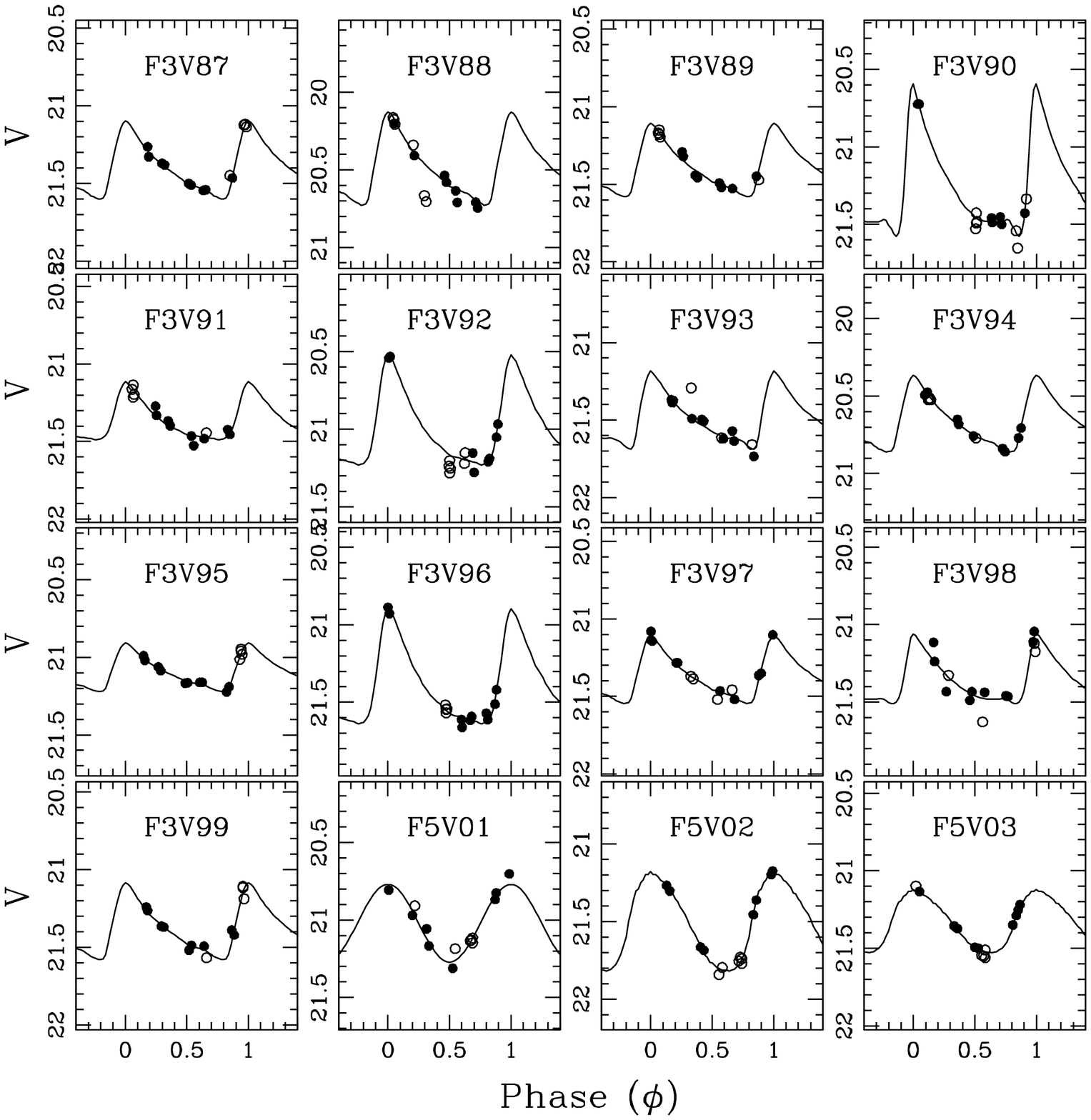}
\end{center}
\end{minipage}
\end{figure*}

\begin{figure*}
\begin{minipage}{175mm}
\begin{center}
\includegraphics[width=150mm,height=127mm]{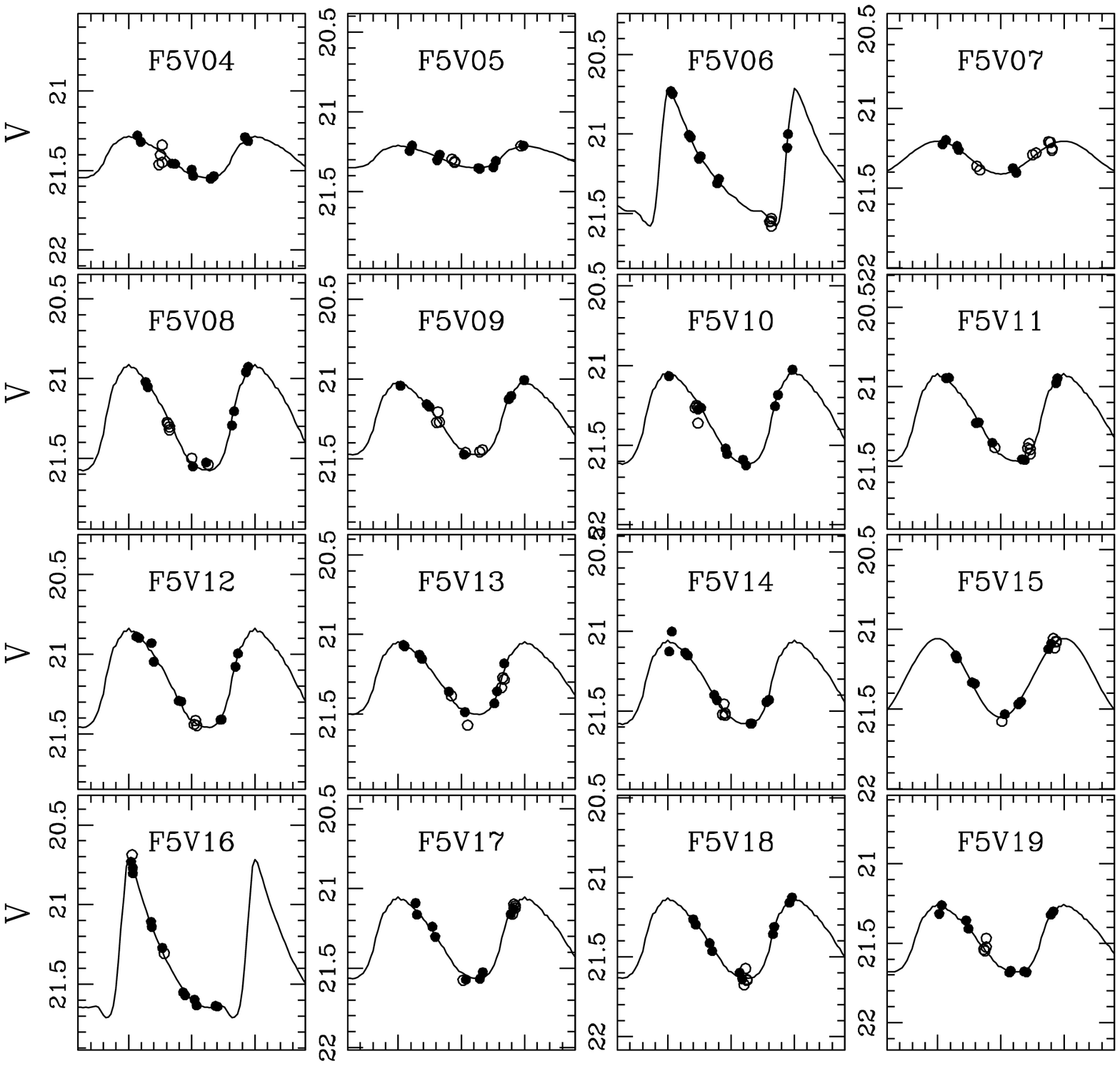} \\
\vspace{-16mm}
\includegraphics[width=150mm,height=127mm]{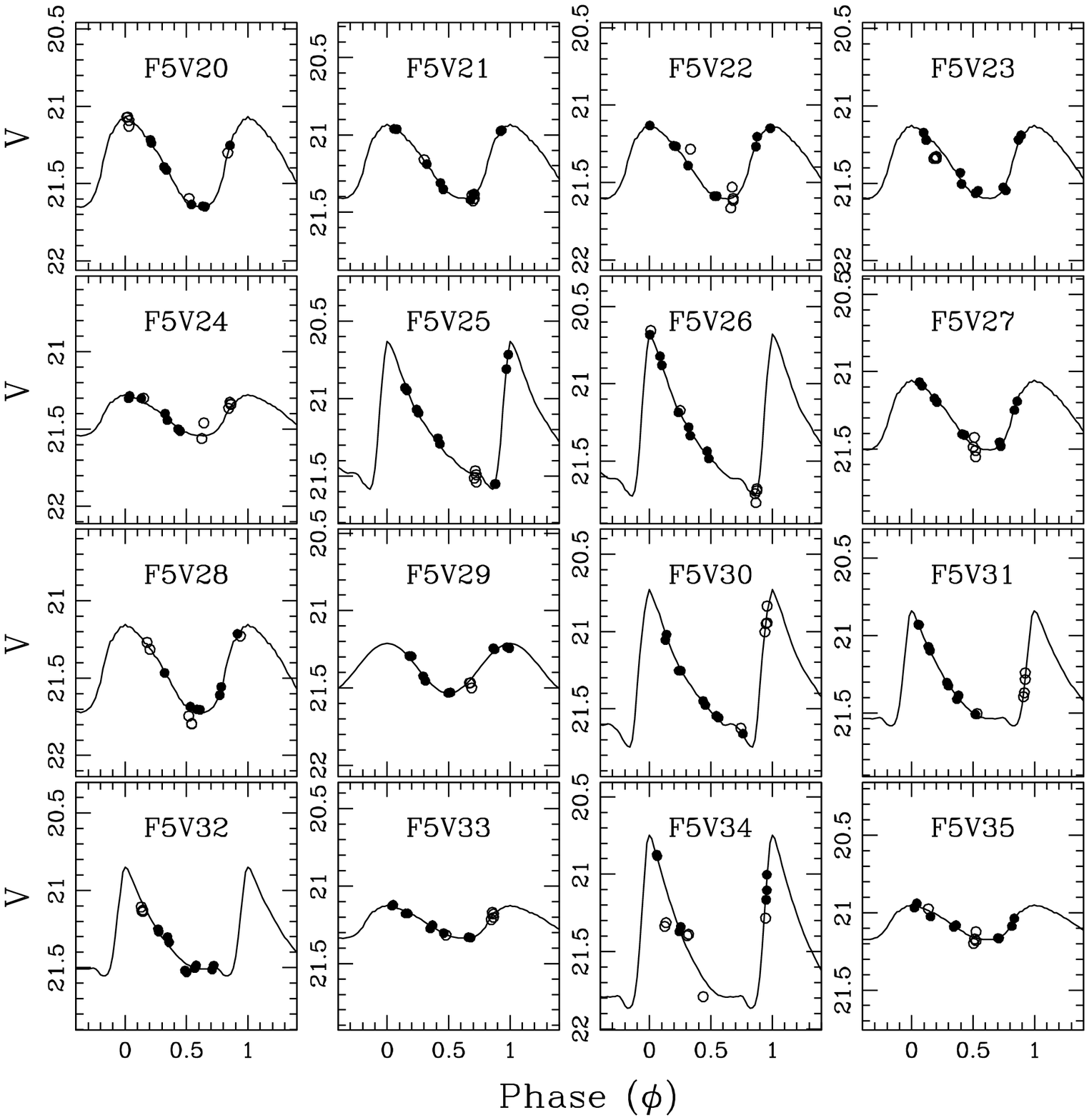}
\end{center}
\end{minipage}
\end{figure*}

\begin{figure*}
\begin{minipage}{175mm}
\begin{center}
\includegraphics[width=150mm,height=127mm]{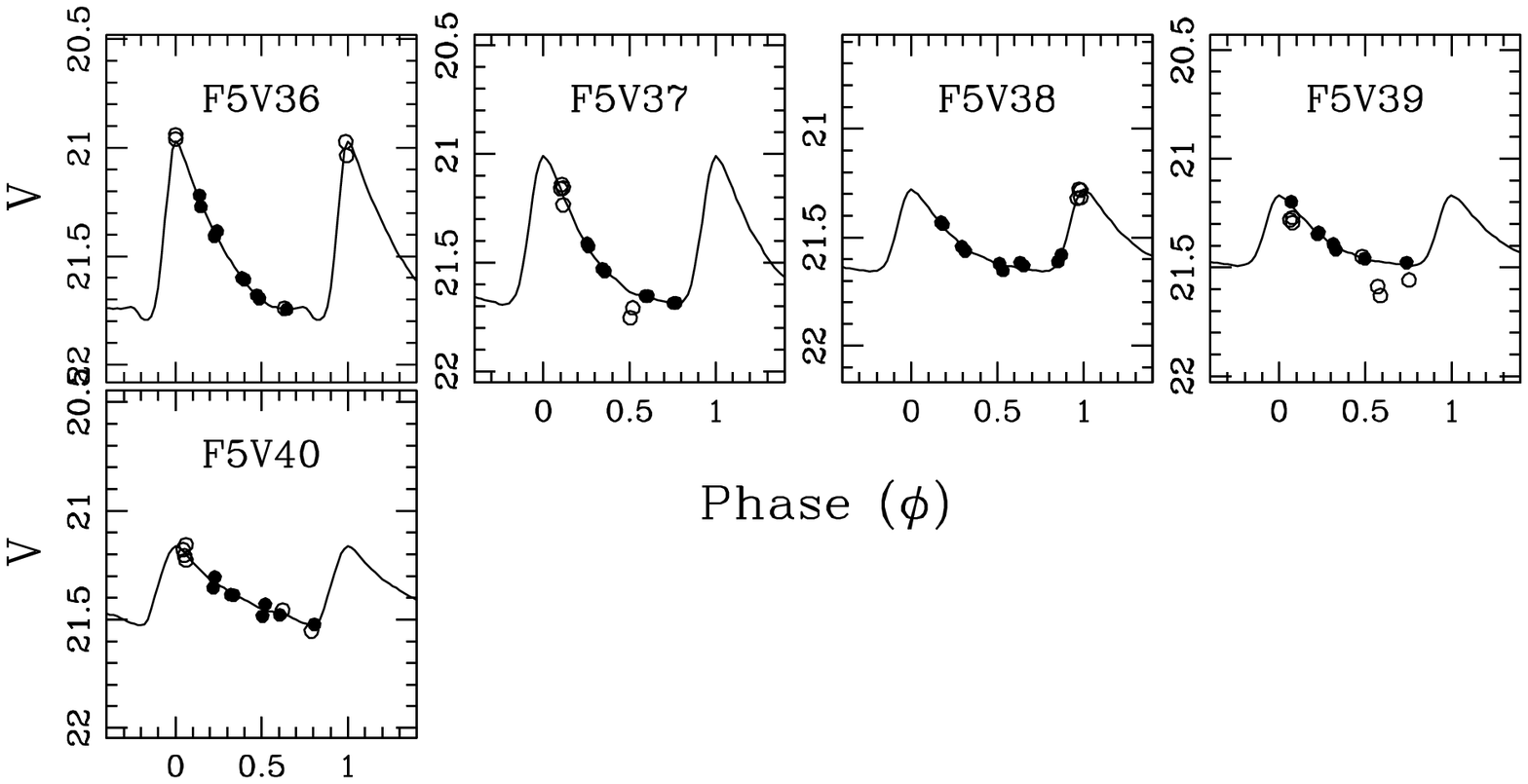} \\
\vspace{-66mm}
\includegraphics[width=150mm,height=127mm]{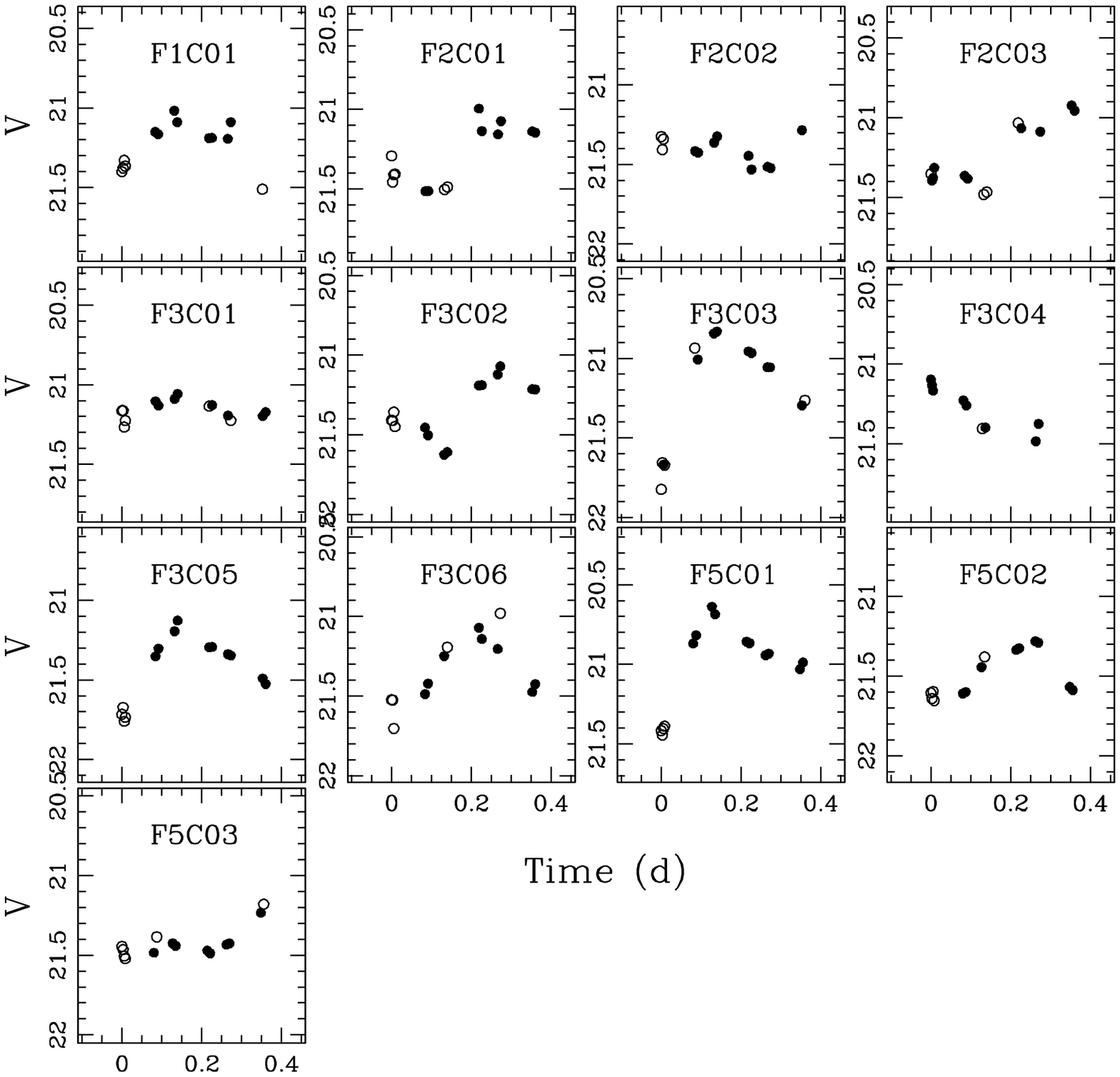}
\caption{$V$-band photometric measurements and best fitting light-curves for the 197 RR Lyrae stars. In addition the $V$-band photometric measurements for the 13 candidate variable stars are plotted. In each plot, the hollow circles represent points which have measurement errors more than $1.5$ times larger than the mean photometric error for observations of the given star. These usually represent the shortest duration exposures in each set. Typical measurement errors are just within the plotted points. This photometry is available on-line at {\tt http://www.ast.cam.ac.uk/STELLARPOPS/Fornax\_RRLyr/}.}
\label{lightcurves}
\end{center}
\end{minipage}
\end{figure*}

\subsection{RR Lyrae stars}
\label{rrlyraestars}

\begin{figure*}
\begin{minipage}{175mm}
\begin{center}
\includegraphics[width=41mm]{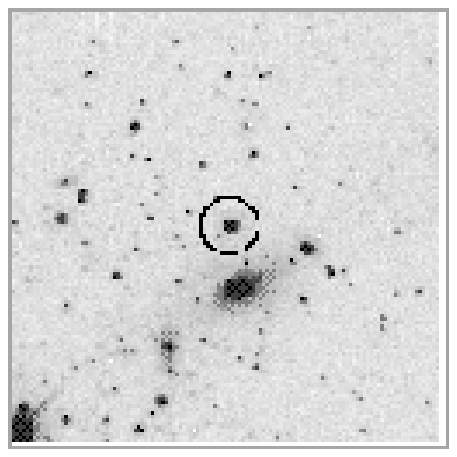}
\hspace{0mm}
\includegraphics[width=41mm]{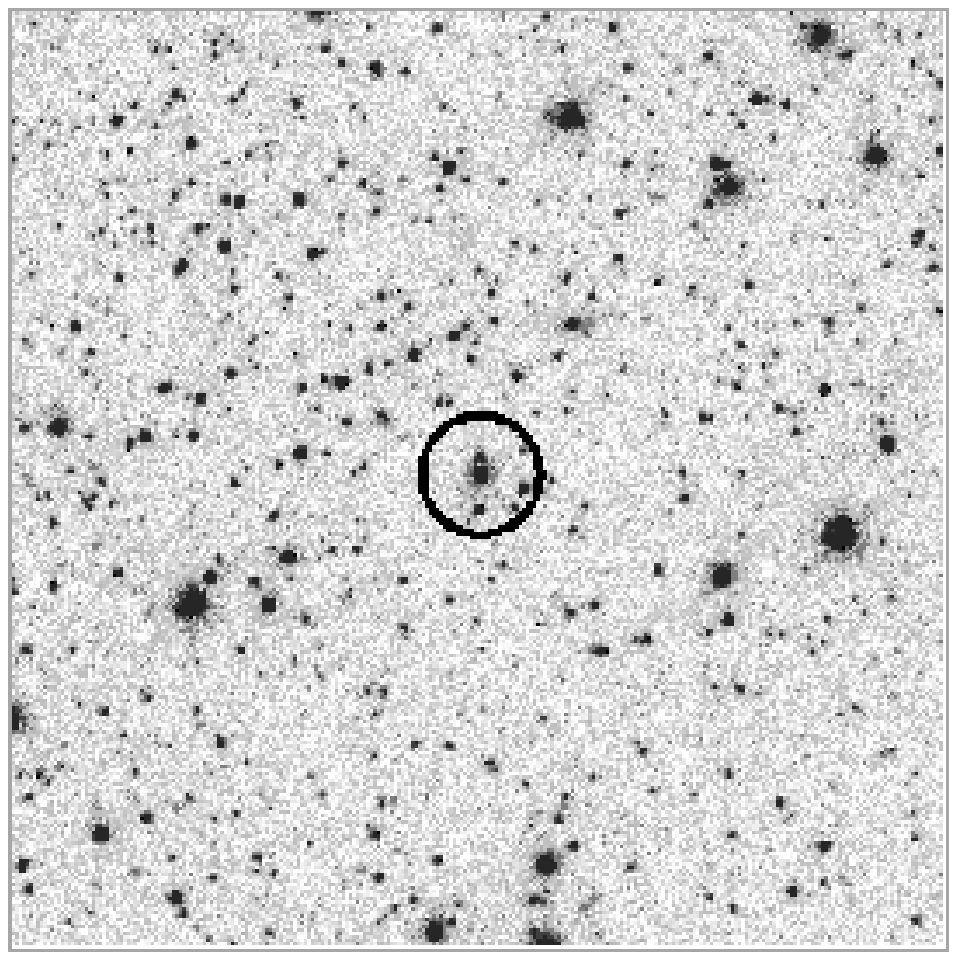}
\hspace{0mm}
\includegraphics[width=41mm]{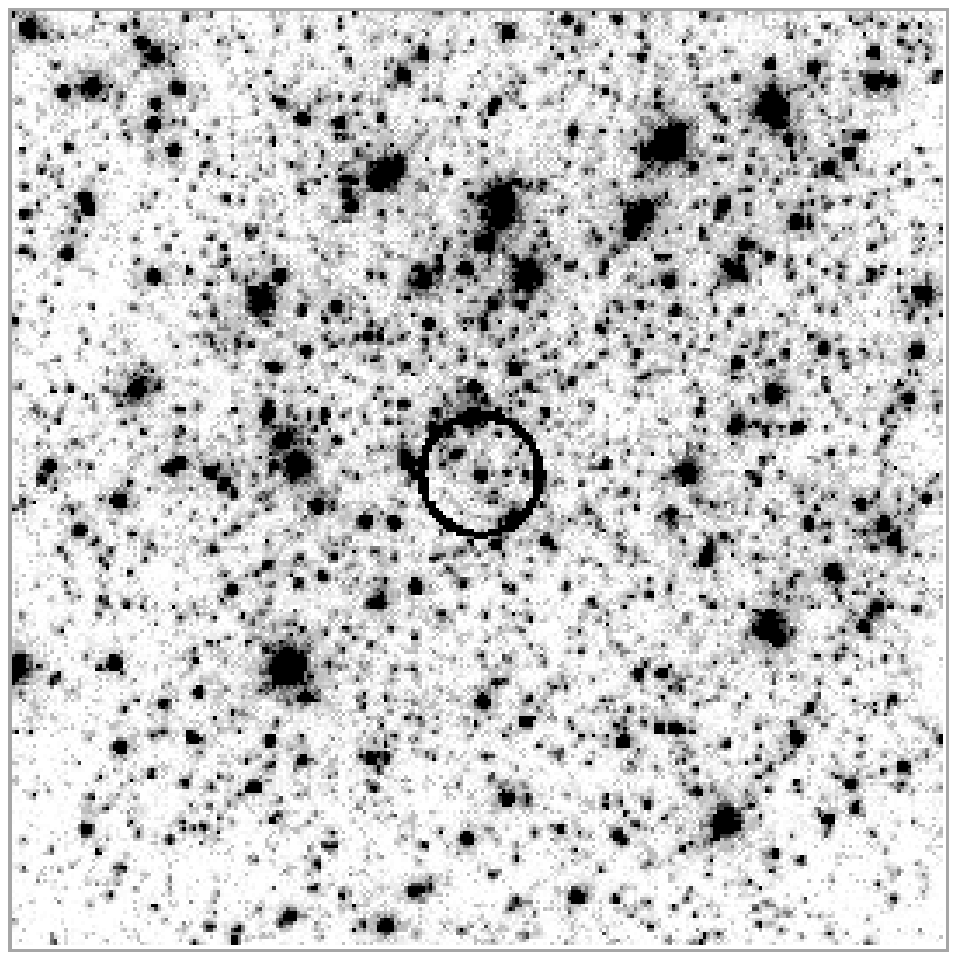}
\hspace{0mm}
\includegraphics[width=41mm]{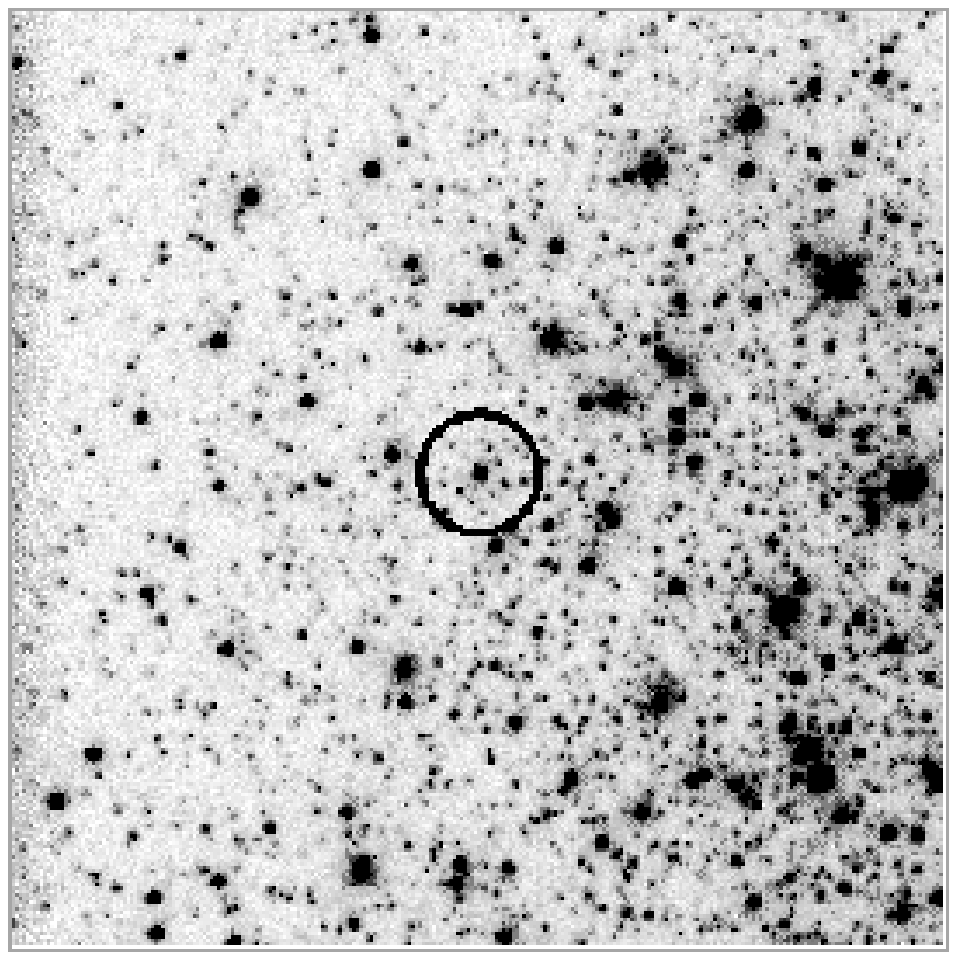} \\
\vspace{1mm}
\includegraphics[width=41mm]{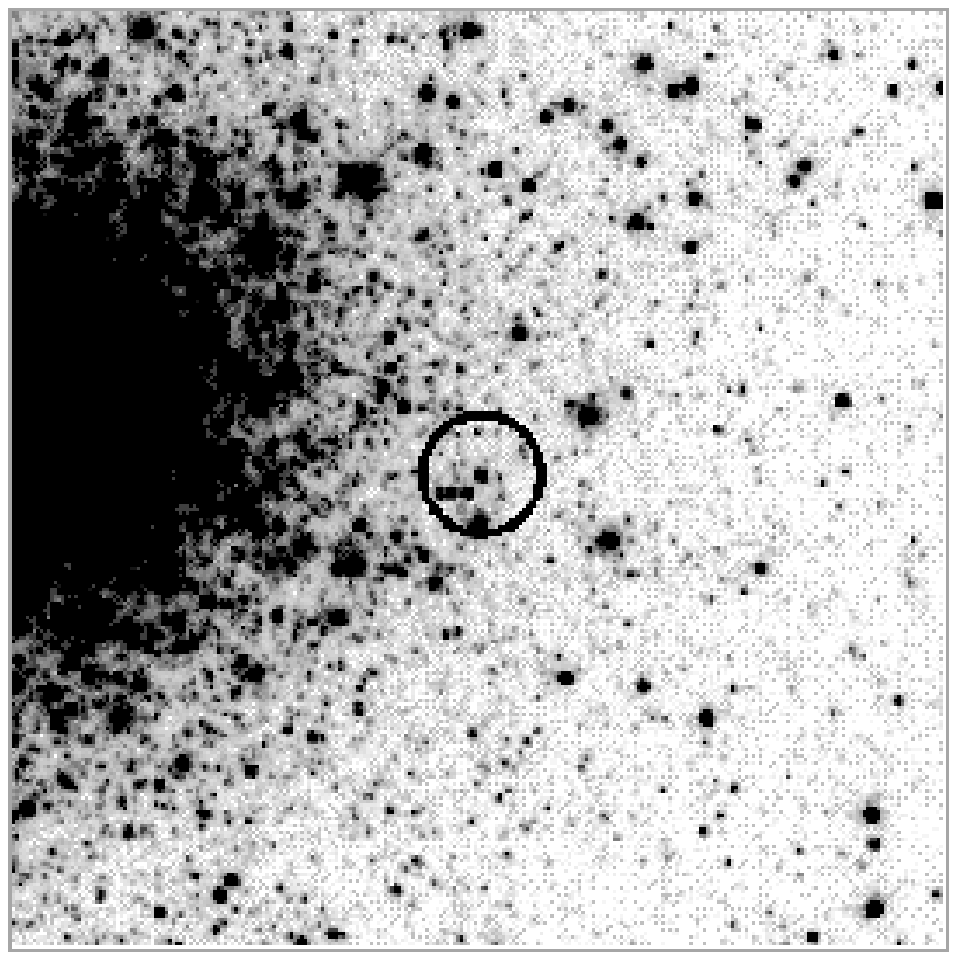}
\hspace{0mm}
\includegraphics[width=41mm]{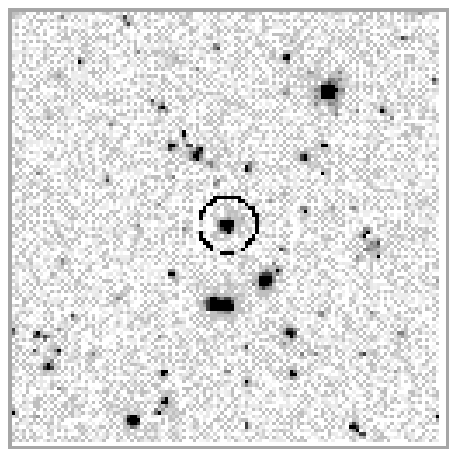}
\hspace{0mm}
\includegraphics[width=41mm]{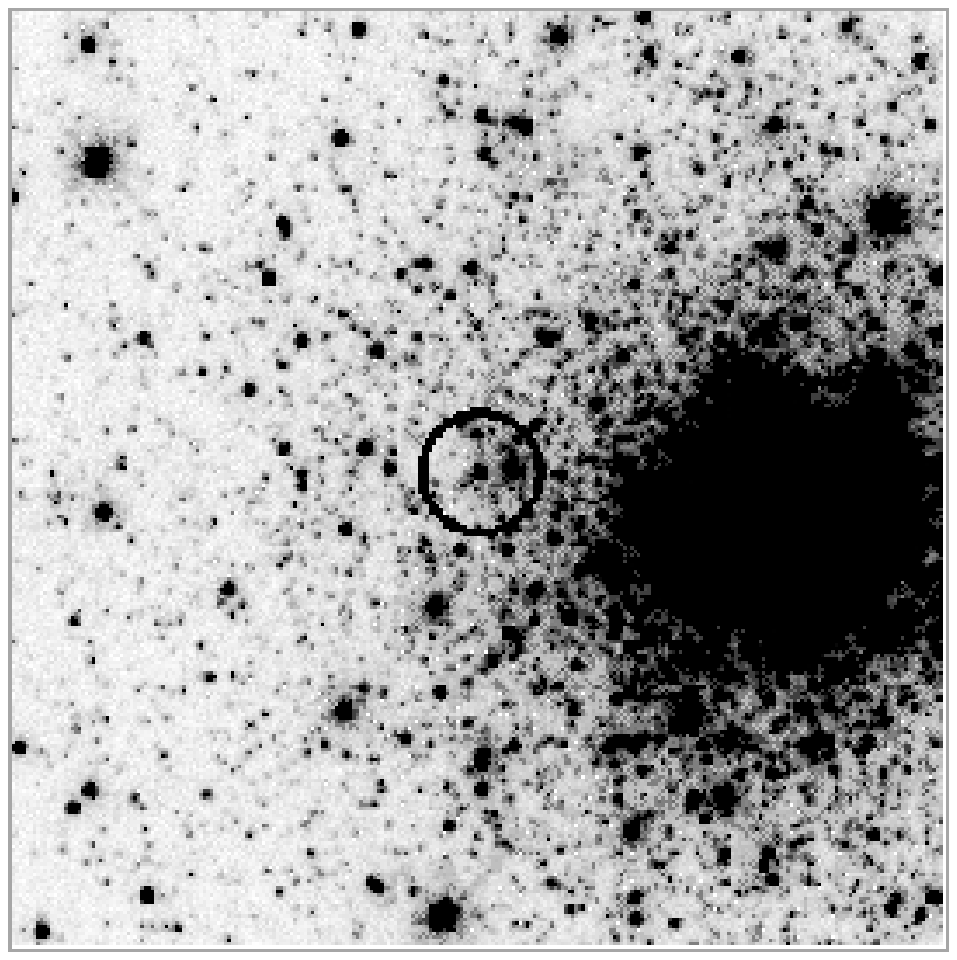}
\hspace{0mm}
\includegraphics[width=41mm]{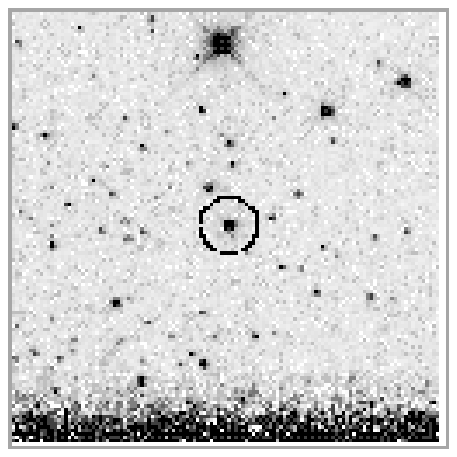}
\caption{Example finder charts for eight of the RR Lyrae stars -- two per cluster. From left to right: (top) F1V02, F1V14, F2V21, F2V28; (bottom) F3V42, F4V49, F5V06, F5V18. Each chart is $15\arcsec$ on a side. Three of the stars (F1V02, F3V49, and F5V18) are on WFC chips; the remainder are on the PC.}
\label{finders}
\end{center}
\end{minipage}
\end{figure*}

Of the 283 candidate objects, we identified 197 as RR Lyrae stars (15, 43, 99, and 40, in 
clusters 1, 2, 3, and 5, respectively). For each of these stars, we obtained
template fits and period measurements. The best fitting $V$-band light curves are presented
in Fig. \ref{lightcurves}, and the positions, periods, and fit parameters (template, rms scatter)
listed in Table \ref{rrparams}. In this table, the coordinates (chip, $X_c$, $Y_c$) are relative 
to the reference images listed in Table \ref{observations}, and are in units of pixels. 
``Chip'' refers to the camera on which the star was imaged -- $0$ is the PC, and $1-3$ are 
WFC2--WFC4. The radii ($r$) and completeness fractions ($\alpha_c$) are calculated as described in 
Section \ref{photometry}. $N_t$ refers to the label of the best fitting template -- numbers
$1-6$ are RRab templates, and $7-8$ are RRc templates. We consider periods marked with a star
to be lower limit measurements. Note that on average, $\sim 53$ per cent of RRab stars have 
periods which are not lower limits -- in accordance with our estimate in Section \ref{templates}.
In addition, we have marked several stars with question marks next to their type. In these
cases, the light curves were best fit by a template of the RR Lyrae type listed (i.e., RRab or RRc),
but could be fit by a template of the other RR Lyrae type almost equally as well. Further 
measurements are required to confirm the listed classification for these stars.

It is clear from the $\alpha_c$ values that the only regions significantly affected by detection
incompleteness are the very centres of the very crowded clusters 3 and 5. For cluster 3, 
completeness is significantly degraded within $\sim 4\arcsec$, and there are only two detections 
within $2\arcsec$. Cluster 5 appears even more compact (cf. Mackey \& Gilmore 
\shortcite{fornaxpaper}), with completeness only degraded within $\sim 2\arcsec$; however, we 
note that there is only one detection within $1.5\arcsec$. 

From the best fitting light curves, we computed an intensity mean $V$ magnitude, denoted 
$\langle V \rangle$, and the maximum, minimum, and amplitude of the variability ($V_{+}$, $V_{-}$,
and $\Delta V$, respectively). We note that these quantities are in general not significantly
different from the same parameters derived directly from the $V$-band measurements. 
From the $I$-band measurements, we calculated the intensity mean $\langle I \rangle$. $N_o$
is the number of pairs of observations for a given star. The parameter $\degr E_m$ is the epoch 
of maximum, in days since MJD $50\,000$. The reader should bear in mind that this parameter is 
derived using a star's measured period. These quantities are all listed in Table \ref{rrparams}.

We also characterized the colour of each RR Lyrae star. In accordance with Sandage 
\shortcite{sandage1,sandage2}, we have calculated the magnitude mean of the $N_o$ 
$(V-I)$ measurements per RR Lyrae to best represent the colour of an ``equivalent static star''. 
This quantity, $\overline{(V-I)}$, is in preference to the two possible intensity mean colours
$\langle V \rangle - \langle I \rangle$ and $\langle V-I \rangle$. Finally, the magnitude 
mean colour of an RR Lyrae star at its minimum light (defined to be phases $0.5 < \phi < 0.8$)
can be of use (see Section \ref{reddening}). For stars with measurements falling in this 
phase range, we also calculated this colour (denoted $(V-I)_m$). Both colours
are listed in Table \ref{rrparams}, together with the calculated error $\sigma_m$ in 
$(V-I)_m$, and the number of observations at minimum light $N_m$ used in its 
calculation.

Finally, using the coordinates listed in Table \ref{rrparams}, we constructed a finder chart
for each RR Lyrae star. For reasons of practicality, we have not included the full set of 
charts in this paper; however, some examples are displayed in Fig. \ref{finders}. The remainder
are available from the author on request. Alternatively, they may be downloaded from
{\it http://www.ast.cam.ac.uk/STELLARPOPS/Fornax\_RRLyr/} along with the light-curves, and data 
from Table \ref{rrparams}.

\begin{table*}
\centerline{\resizebox{!}{!}{\rotatebox{90}{
\begin{minipage}{238mm}
\caption{Measured and calculated properties of the RR Lyrae stars.}
\label{rrparams}
\begin{tabular}{@{}lcccccclccccccccccccc}
\hline \hline
Name & Chip & $X_c$ & $Y_c$ & $r$ & $\alpha_c$ & Period & Type & $N_t$ & rms & $\langle V \rangle$ & $\langle I \rangle$ & $\overline{(V-I)}$ & $N_{o}$ & $V_{+}$ & $V_{-}$ & $\Delta V$ & $\degr E_{m}$ & $(V-I)_{m}$ & $\sigma_m$ & $N_{m}$ \\
 & & (pix) & (pix) & $(\arcsec)$ & & (days) & & & & & & & & & & & & & & \\
\hline
F1V01 & 1 & 206.81 & 310.88 & 43.50 & 1.00 & 0.459 & RRc & 7 & 0.016 & 21.275 & 20.822 & 0.452 & 14 & 21.034 & 21.516 & 0.482 & 237.4963 & 0.489 & 0.034 & 4 \\
F1V02 & 1 & 383.22 & 130.61 & 29.80 & 1.00 & 0.570 & RRab & 3 & 0.030 & 21.206 & 20.687 & 0.492 & 14 & 20.765 & 21.460 & 0.695 & 237.8765 & 0.652 & 0.025 & 1 \\
F1V03 & 3 & 118.48 & 282.34 & 26.11 & 1.00 & 0.425 & RRc & 7 & 0.015 & 21.261 & 20.798 & 0.450 & 14 & 21.055 & 21.465 & 0.410 & 237.4490 & 0.512 & 0.021 & 2 \\
F1V04 & 3 & 209.32 & 86.92 & 35.78 & 1.00 & 0.396 & RRc & 7 & 0.017 & 21.281 & 20.794 & 0.495 & 14 & 21.054 & 21.508 & 0.454 & 237.6898 & 0.582 & 0.016 & 3 \\
F1V05 & 1 & 410.36 & 122.48 & 30.62 & 1.00 & 0.589 & RRab & 2 & 0.026 & 21.285 & 20.697 & 0.503 & 13 & 20.842 & 21.599 & 0.757 & 237.2573 & 0.652 & 0.027 & 1 \\
F1V06 & 0 & 464.51 & 285.29 & 8.00 & 1.00 & 0.440 & RRc & 7 & 0.026 & 21.169 & 20.691 & 0.483 & 14 & 20.939 & 21.399 & 0.460 & 237.6960 & 0.500 & 0.015 & 4 \\
F1V07 & 0 & 586.78 & 766.38 & 17.53 & 1.00 & 0.580 & RRab & 3 & 0.039 & 21.348 & 20.690 & 0.576 & 14 & 20.775 & 21.693 & 0.918 & 237.2618 & 0.738 & 0.032 & 2 \\
F1V08 & 0 & 709.54 & 616.83 & 16.99 & 1.00 & 0.465$^{*}$ & RRab & 6 & 0.034 & 21.195 & 20.756 & 0.496 & 14 & 20.687 & 21.604 & 0.917 & 237.7498 & 0.598 & 0.017 & 4 \\
F1V09 & 0 & 402.46 & 183.91 & 11.64 & 1.00 & 0.436 & RRc & 7 & 0.023 & 21.293 & 20.748 & 0.555 & 14 & 21.104 & 21.477 & 0.373 & 237.5112 & 0.569 & 0.040 & 4 \\
F1V10 & 0 & 362.44 & 160.38 & 12.68 & 1.00 & 0.670 & RRab & 5 & 0.032 & 21.299 & 20.786 & 0.641 & 14 & 20.710 & 21.667 & 0.957 & 237.4937 & 0.715 & 0.018 & 5 \\
F1V11 & 1 & 242.83 & 101.56 & 22.72 & 1.00 & 0.503$^{*}$ & RRab & 3 & 0.011 & 21.228 & 20.716 & 0.562 & 14 & 20.777 & 21.488 & 0.711 & 237.7337 & 0.654 & 0.014 & 4 \\
F1V12 & 0 & 587.09 & 652.50 & 13.52 & 1.00 & 0.644 & RRab & 2 & 0.019 & 21.287 & 20.619 & 0.664 & 13 & 20.940 & 21.525 & 0.585 & 237.4082 & 0.733 & 0.020 & 6 \\
F1V13 & 3 & 132.83 & 53.26 & 30.19 & 1.00 & 0.424$^{*}$ & RRab? & 1 & 0.012 & 21.290 & 20.673 & 0.592 & 14 & 21.002 & 21.442 & 0.440 & 237.4022 & 0.638 & 0.021 & 4 \\
F1V14 & 0 & 379.05 & 321.76 & 5.34 & 1.00 & 0.460$^{*}$ & RRab & 6 & 0.032 & 21.307 & 20.735 & 0.606 & 14 & 20.806 & 21.710 & 0.904 & 237.7688 & 0.704 & 0.016 & 4 \\
F1V15 & 0 & 736.93 & 535.03 & 16.79 & 0.98 & 0.556$^{*}$ & RRab & 4 & 0.022 & 21.444 & 20.802 & 0.700 & 14 & 21.147 & 21.659 & 0.512 & 237.6896 & 0.706 & 0.016 & 4 \\
F2V01 & 0 & 386.68 & 480.06 & 2.25 & 0.96 & 0.429$^{*}$ & RRab & 4 & 0.019 & 21.223 & 20.786 & 0.469 & 14 & 20.626 & 21.701 & 1.075 & 239.5524 & 0.620 & 0.021 & 6 \\
F2V02 & 0 & 538.01 & 546.13 & 9.66 & 0.98 & 0.578 & RRab & 2 & 0.024 & 21.315 & 20.890 & 0.560 & 14 & 20.698 & 21.778 & 1.08 & 239.6668 & 0.662 & 0.015 & 6 \\
F2V03 & 0 & 282.08 & 409.83 & 3.52 & 0.96 & 0.375 & RRc & 8 & 0.021 & 21.209 & 20.786 & 0.418 & 14 & 20.962 & 21.488 & 0.526 & 239.8160 & 0.411 & 0.016 & 4 \\
F2V04 & 0 & 368.29 & 260.99 & 8.50 & 1.00 & 0.335 & RRc & 7 & 0.021 & 21.258 & 20.926 & 0.389 & 14 & 20.971 & 21.552 & 0.581 & 239.7316 & 0.479 & 0.017 & 4 \\
F2V05 & 0 & 617.42 & 580.55 & 13.57 & 0.97 & 0.664 & RRab & 5 & 0.024 & 21.237 & 20.715 & 0.542 & 14 & 20.795 & 21.498 & 0.703 & 239.4942 & 0.610 & 0.017 & 8 \\
F2V06 & 3 & 141.64 & 576.57 & 47.35 & 1.00 & 0.336 & RRc & 7 & 0.009 & 21.455 & 20.931 & 0.494 & 14 & 21.162 & 21.757 & 0.595 & 239.6009 & 0.693 & 0.021 & 3 \\
F2V07 & 1 & 59.96 & 77.64 & 25.54 & 1.00 & 0.403 & RRc & 7 & 0.032 & 21.345 & 20.930 & 0.421 & 14 & 21.040 & 21.661 & 0.621 & 239.8042 & 0.516 & 0.018 & 4 \\
F2V08 & 3 & 325.76 & 346.47 & 48.41 & 1.00 & 0.374 & RRc & 7 & 0.018 & 21.387 & 20.932 & 0.440 & 14 & 21.147 & 21.629 & 0.482 & 239.7992 & 0.529 & 0.022 & 2 \\
F2V09 & 1 & 171.42 & 249.32 & 36.51 & 1.00 & 0.381 & RRc & 7 & 0.019 & 21.396 & 20.956 & 0.453 & 13 & 21.151 & 21.642 & 0.491 & 239.9604 & 0.500 & 0.017 & 4 \\
F2V10 & 0 & 283.13 & 522.39 & 4.58 & 1.00 & 0.556 & RRab & 1 & 0.020 & 21.219 & 20.731 & 0.543 & 14 & 20.404 & 21.726 & 1.322 & 239.5312 & 0.612 & 0.029 & 5 \\
F2V11 & 0 & 253.69 & 501.65 & 5.03 & 1.00 & 0.370$^{*}$ & RRab & 2 & 0.023 & 21.416 & 20.796 & 0.506 & 13 & 21.026 & 21.686 & 0.660 & 239.8671 & 0.626 & 0.026 & 3 \\
F2V12 & 0 & 656.24 & 746.22 & 19.36 & 0.97 & 0.354$^{*}$ & RRab? & 6 & 0.026 & 21.455 & 20.950 & 0.477 & 13 & 20.981 & 21.831 & 0.850 & 239.5608 & 0.580 & 0.026 & 4 \\
F2V13 & 3 & 183.13 & 764.91 & 65.34 & 1.00 & 0.345 & RRc & 7 & 0.020 & 21.561 & 21.061 & 0.467 & 14 & 21.341 & 21.779 & 0.438 & 239.8683 & 0.535 & 0.024 & 2 \\
F2V14 & 1 & 219.03 & 428.71 & 53.86 & 1.00 & 0.404 & RRc & 7 & 0.020 & 21.399 & 20.964 & 0.463 & 14 & 21.179 & 21.617 & 0.438 & 239.6427 & 0.501 & 0.038 & 4 \\
F2V15 & 0 & 442.63 & 313.64 & 7.39 & 0.97 & 0.403$^{*}$ & RRab & 5 & 0.012 & 21.358 & 20.920 & 0.461 & 13 & 21.011 & 21.557 & 0.546 & 239.5840 & 0.501 & 0.021 & 6 \\
F2V16 & 0 & 178.65 & 79.80 & 18.36 & 0.97 & 0.396$^{*}$ & RRab? & 4 & 0.025 & 21.516 & 21.028 & 0.565 & 13 & 21.139 & 21.798 & 0.659 & 239.6814 & 0.594 & 0.033 & 6 \\
F2V17 & 1 & 406.27 & 276.45 & 42.38 & 1.00 & 0.333 & RRc & 8 & 0.017 & 21.418 & 20.925 & 0.464 & 14 & 21.240 & 21.611 & 0.371 & 239.8469 & 0.487 & 0.014 & 4 \\
F2V18 & 0 & 291.35 & 145.58 & 13.95 & 0.98 & 0.551 & RRab & 1 & 0.035 & 21.399 & 20.989 & 0.498 & 13 & 20.727 & 21.798 & 1.071 & 239.5952 & 0.605 & 0.021 & 8 \\
F2V19 & 0 & 389.24 & 438.00 & 1.83 & 1.00 & 0.490$^{*}$ & RRab & 5 & 0.022 & 21.256 & 20.736 & 0.511 & 14 & 20.960 & 21.423 & 0.463 & 239.5182 & 0.569 & 0.041 & 4 \\
F2V20 & 0 & 366.18 & 412.15 & 1.75 & 1.00 & 0.514 & RRab & 6 & 0.018 & 21.405 & 20.765 & 0.617 & 14 & 20.840 & 21.870 & 1.030 & 239.4404 & 0.717 & 0.025 & 2 \\
F2V21 & 0 & 391.94 & 407.81 & 2.61 & 0.97 & 0.387 & RRc & 7 & 0.014 & 21.289 & 20.811 & 0.520 & 14 & 21.069 & 21.507 & 0.438 & 239.6635 & 0.557 & 0.025 & 6 \\
F2V22 & 0 & 168.22 & 625.25 & 11.54 & 1.00 & 0.567 & RRab & 2 & 0.014 & 21.344 & 20.655 & 0.544 & 14 & 20.797 & 21.744 & 0.947 & 239.8844 & 0.706 & 0.025 & 2 \\
F2V23 & 3 & 54.41 & 232.32 & 19.43 & 0.97 & 0.337 & RRc & 8 & 0.014 & 21.436 & 20.942 & 0.510 & 14 & 21.315 & 21.565 & 0.250 & 239.6575 & 0.485 & 0.018 & 6 \\
F2V24 & 0 & 538.69 & 384.36 & 9.05 & 0.98 & 0.389 & RRc? & 7 & 0.020 & 21.152 & 20.607 & 0.521 & 14 & 21.035 & 21.262 & 0.227 & 239.5243 & 0.541 & 0.023 & 3 \\
F2V25 & 0 & 522.55 & 521.33 & 8.54 & 0.98 & 0.706 & RRab & 2 & 0.021 & 21.186 & 20.613 & 0.532 & 14 & 20.602 & 21.619 & 1.017 & 240.0444 & 0.632 & 0.044 & 4 \\
\hline
\end{tabular}
\end{minipage}}}}
\end{table*}

\begin{table*}
\centerline{\resizebox{!}{!}{\rotatebox{90}{
\begin{minipage}{238mm}
\begin{tabular}{@{}lcccccclccccccccccccc}
\hline \hline
Name & Chip & $X_c$ & $Y_c$ & $r$ & $\alpha_c$ & Period & Type & $N_t$ & rms & $\langle V \rangle$ & $\langle I \rangle$ & $\overline{(V-I)}$ & $N_{o}$ & $V_{+}$ & $V_{-}$ & $\Delta V$ & $\degr E_{m}$ & $(V-I)_{m}$ & $\sigma_m$ & $N_{m}$ \\
 & & (pix) & (pix) & $(\arcsec)$ & & (days) & & & & & & & & & & & & & & \\
\hline
F2V26 & 0 & 614.26 & 763.72 & 18.65 & 1.00 & 0.512 & RRab & 4 & 0.018 & 21.303 & 20.750 & 0.561 & 14 & 20.810 & 21.685 & 0.875 & 239.5189 & 0.785 & 0.049 & 4 \\
F2V27 & 0 & 630.92 & 316.23 & 14.09 & 0.98 & 0.415$^{*}$ & RRab & 6 & 0.010 & 21.184 & 20.591 & 0.566 & 14 & 20.854 & 21.433 & 0.579 & 239.4905 & 0.640 & 0.026 & 2 \\
F2V28 & 0 & 206.64 & 518.13 & 7.27 & 0.97 & 0.366$^{*}$ & RRab & 4 & 0.020 & 21.190 & 20.772 & 0.609 & 14 & 20.507 & 21.753 & 1.246 & 239.5796 & 0.608 & 0.024 & 3 \\
F2V29 & 0 & 594.92 & 427.57 & 11.17 & 1.00 & 0.476$^{*}$ & RRab & 3 & 0.025 & 21.353 & 20.841 & 0.571 & 12 & 20.916 & 21.604 & 0.688 & 239.7314 & 0.675 & 0.017 & 4 \\
F2V30 & 0 & 736.48 & 463.22 & 17.55 & 1.00 & 0.414$^{*}$ & RRab & 6 & 0.028 & 21.468 & 20.875 & 0.544 & 14 & 21.069 & 21.776 & 0.707 & 239.4791 & 0.663 & 0.018 & 4 \\
F2V31 & 0 & 117.45 & 455.80 & 10.56 & 1.00 & 0.574 & RRab & 3 & 0.023 & 21.455 & 20.911 & 0.628 & 14 & 21.015 & 21.708 & 0.693 & 239.6350 & 0.625 & 0.018 & 4 \\
F2V32 & 1 & 234.56 & 75.95 & 18.83 & 0.97 & 0.514 & RRab & 3 & 0.013 & 21.377 & 20.839 & 0.598 & 13 & 20.972 & 21.606 & 0.634 & 239.6824 & 0.653 & 0.017 & 3 \\
F2V33 & 0 & 212.75 & 492.55 & 6.57 & 0.97 & 0.566 & RRab & 2 & 0.023 & 21.306 & 20.711 & 0.632 & 14 & 20.740 & 21.722 & 0.982 & 240.0546 & 0.755 & 0.050 & 4 \\
F2V34 & 0 & 324.52 & 251.83 & 8.95 & 0.98 & 0.410$^{*}$ & RRab & 6 & 0.025 & 21.263 & 20.695 & 0.572 & 14 & 20.996 & 21.461 & 0.465 & 239.9261 & 0.638 & 0.015 & 4 \\
F2V35 & 3 & 82.59 & 311.95 & 24.90 & 1.00 & 0.624 & RRab & 6 & 0.020 & 21.396 & 20.808 & 0.602 & 14 & 21.139 & 21.586 & 0.447 & 239.4964 & 0.630 & 0.020 & 6 \\
F2V36 & 0 & 325.10 & 654.98 & 9.51 & 0.98 & 0.602 & RRab & 5 & 0.027 & 21.307 & 20.788 & 0.576 & 13 & 20.819 & 21.600 & 0.781 & 239.5901 & 0.623 & 0.018 & 5 \\
F2V37 & 0 & 284.47 & 544.16 & 5.33 & 1.00 & 0.357$^{*}$ & RRab? & 6 & 0.034 & 21.287 & 20.645 & 0.625 & 14 & 20.927 & 21.563 & 0.636 & 239.5592 & 0.719 & 0.032 & 4 \\
F2V38 & 0 & 375.04 & 424.16 & 1.54 & 1.00 & 0.423$^{*}$ & RRab & 6 & 0.022 & 21.285 & 20.634 & 0.626 & 14 & 21.069 & 21.443 & 0.374 & 239.4710 & 0.678 & 0.018 & 3 \\
F2V39 & 0 & 473.83 & 579.99 & 8.26 & 0.98 & 0.420$^{*}$ & RRab & 6 & 0.019 & 21.326 & 20.666 & 0.633 & 14 & 21.122 & 21.474 & 0.352 & 239.4730 & 0.654 & 0.021 & 3 \\
F2V40 & 2 & 376.04 & 267.21 & 62.42 & 1.00 & 0.445 & RRc? & 8 & 0.024 & 21.362 & 20.743 & 0.636 & 14 & 21.206 & 21.529 & 0.323 & 239.7434 & 0.715 & 0.016 & 4 \\
F2V41 & 3 & 165.13 & 97.46 & 31.25 & 1.00 & 0.526 & RRab & 4 & 0.019 & 21.191 & 20.618 & 0.630 & 14 & 20.839 & 21.451 & 0.612 & 239.8026 & 0.665 & 0.014 & 4 \\
F2V42 & 0 & 485.25 & 489.17 & 6.45 & 0.99 & 0.562 & RRab & 3 & 0.020 & 21.283 & 20.782 & 0.639 & 14 & 20.657 & 21.668 & 1.011 & 239.6534 & 0.661 & 0.015 & 5 \\
F2V43 & 0 & 518.56 & 687.93 & 13.34 & 1.00 & 0.470$^{*}$ & RRab & 1 & 0.028 & 21.403 & 20.767 & 0.671 & 14 & 21.018 & 21.611 & 0.593 & 239.8331 & 0.736 & 0.018 & 4 \\
F3V01 & 0 & 336.08 & 534.52 & 5.13 & 1.00 & 0.334 & RRc & 7 & 0.028 & 20.822 & 20.553 & 0.235 & 14 & 20.602 & 21.041 & 0.439 & 238.8422 & 0.285 & 0.016 & 4 \\
F3V02 & 0 & 353.43 & 332.31 & 4.80 & 1.00 & 0.562 & RRab & 2 & 0.010 & 21.062 & 20.585 & 0.457 & 14 & 20.435 & 21.533 & 1.098 & 238.4032 & $-$ & $-$ & 0 \\
F3V03 & 0 & 410.86 & 470.41 & 2.07 & 0.90 & 0.321 & RRc & 7 & 0.027 & 21.176 & 20.822 & 0.308 & 14 & 20.911 & 21.445 & 0.534 & 238.5972 & 0.416 & 0.046 & 4 \\
F3V04 & 1 & 113.33 & 98.49 & 25.28 & 1.00 & 0.303 & RRc & 7 & 0.027 & 21.270 & 20.974 & 0.348 & 14 & 21.047 & 21.491 & 0.444 & 238.7809 & 0.430 & 0.014 & 5 \\
F3V05 & 0 & 432.93 & 526.50 & 4.78 & 0.86 & 0.311 & RRc & 8 & 0.030 & 21.211 & 20.828 & 0.348 & 14 & 21.071 & 21.361 & 0.290 & 238.8506 & 0.335 & 0.021 & 3 \\
F3V06 & 0 & 535.69 & 539.75 & 8.39 & 1.00 & 0.312 & RRc & 7 & 0.017 & 21.201 & 20.846 & 0.382 & 14 & 21.045 & 21.352 & 0.307 & 238.7466 & 0.387 & 0.019 & 4 \\
F3V07 & 0 & 341.33 & 445.33 & 2.07 & 0.75 & 0.589 & RRab & 5 & 0.026 & 20.522 & 20.240 & 0.329 & 14 & 20.281 & 20.655 & 0.374 & 238.6618 & 0.392 & 0.020 & 6 \\
F3V08 & 0 & 394.59 & 462.39 & 1.41 & 0.20 & 0.301 & RRc & 7 & 0.024 & 21.011 & 20.659 & 0.383 & 14 & 20.748 & 21.277 & 0.529 & 238.6608 & 0.525 & 0.042 & 6 \\
F3V09 & 0 & 373.71 & 572.11 & 6.35 & 1.00 & 0.580 & RRab & 4 & 0.013 & 21.093 & 20.618 & 0.534 & 14 & 20.568 & 21.502 & 0.934 & 238.5710 & 0.629 & 0.018 & 8 \\
F3V10 & 0 & 421.96 & 348.88 & 4.18 & 1.00 & 0.528 & RRab & 1 & 0.018 & 21.095 & 20.591 & 0.517 & 14 & 20.380 & 21.526 & 1.146 & 238.4261 & 0.632 & 0.027 & 2 \\
F3V11 & 0 & 466.71 & 440.11 & 3.73 & 0.90 & 0.277 & RRc & 8 & 0.022 & 21.215 & 20.931 & 0.359 & 14 & 20.972 & 21.488 & 0.516 & 238.7777 & 0.351 & 0.022 & 4 \\
F3V12 & 0 & 404.05 & 518.93 & 4.00 & 1.00 & 0.620 & RRab & 2 & 0.015 & 21.021 & 20.652 & 0.578 & 14 & 20.403 & 21.484 & 1.081 & 238.6008 & 0.619 & 0.022 & 4 \\
F3V13 & 0 & 206.01 & 302.14 & 10.08 & 1.00 & 0.627 & RRab & 5 & 0.038 & 21.189 & 20.793 & 0.557 & 13 & 20.463 & 21.669 & 1.206 & 238.6125 & 0.660 & 0.015 & 6 \\
F3V14 & 0 & 421.69 & 596.34 & 7.61 & 0.89 & 0.479 & RRab & 2 & 0.018 & 21.126 & 20.690 & 0.487 & 14 & 20.587 & 21.520 & 0.933 & 238.6336 & 0.562 & 0.018 & 7 \\
F3V15 & 0 & 432.78 & 450.30 & 2.31 & 0.81 & 0.298 & RRc? & 7 & 0.047 & 21.025 & 20.655 & 0.407 & 13 & 20.859 & 21.187 & 0.328 & 238.7009 & 0.453 & 0.029 & 6 \\
F3V16 & 1 & 357.21 & 283.56 & 43.01 & 1.00 & 0.373 & RRc & 7 & 0.023 & 21.243 & 20.772 & 0.447 & 14 & 20.986 & 21.503 & 0.517 & 238.8264 & 0.489 & 0.016 & 3 \\
F3V17 & 3 & 256.03 & 743.78 & 65.77 & 1.00 & 0.633 & RRab & 5 & 0.040 & 21.371 & 20.977 & 0.553 & 14 & 20.795 & 21.730 & 0.935 & 238.6126 & 0.638 & 0.013 & 6 \\
F3V18 & 0 & 244.85 & 453.79 & 6.44 & 0.91 & 0.383 & RRc & 7 & 0.027 & 21.205 & 20.726 & 0.428 & 13 & 20.983 & 21.425 & 0.442 & 238.5250 & 0.520 & 0.024 & 2 \\
F3V19 & 0 & 369.18 & 340.92 & 4.25 & 1.00 & 0.499$^{*}$ & RRab & 6 & 0.031 & 21.246 & 20.732 & 0.475 & 13 & 20.912 & 21.499 & 0.587 & 238.4831 & 0.532 & 0.073 & 1 \\
F3V20 & 0 & 360.55 & 481.99 & 2.49 & 0.81 & 0.383 & RRc & 7 & 0.019 & 21.242 & 20.796 & 0.451 & 14 & 20.972 & 21.516 & 0.544 & 238.7924 & 0.454 & 0.027 & 3 \\
F3V21 & 0 & 397.75 & 561.52 & 5.87 & 1.00 & 0.526$^{*}$ & RRab & 3 & 0.020 & 21.212 & 20.703 & 0.492 & 14 & 20.843 & 21.419 & 0.576 & 238.4591 & 0.620 & 0.083 & 1 \\
\hline
\end{tabular}
\end{minipage}}}}
\end{table*}

\begin{table*}
\centerline{\resizebox{!}{!}{\rotatebox{90}{
\begin{minipage}{238mm}
\begin{tabular}{@{}lcccccclccccccccccccc}
\hline \hline
Name & Chip & $X_c$ & $Y_c$ & $r$ & $\alpha_c$ & Period & Type & $N_t$ & rms & $\langle V \rangle$ & $\langle I \rangle$ & $\overline{(V-I)}$ & $N_{o}$ & $V_{+}$ & $V_{-}$ & $\Delta V$ & $\degr E_{m}$ & $(V-I)_{m}$ & $\sigma_m$ & $N_{m}$ \\
 & & (pix) & (pix) & $(\arcsec)$ & & (days) & & & & & & & & & & & & & & \\
\hline
F3V22 & 0 & 482.12 & 420.86 & 4.46 & 1.00 & 0.416 & RRc & 7 & 0.026 & 21.140 & 20.694 & 0.481 & 14 & 20.876 & 21.408 & 0.532 & 238.5678 & 0.568 & 0.019 & 6 \\
F3V23 & 0 & 290.78 & 245.09 & 9.56 & 1.00 & 0.410 & RRc & 8 & 0.021 & 21.264 & 20.780 & 0.469 & 14 & 21.068 & 21.480 & 0.412 & 238.5615 & 0.493 & 0.022 & 6 \\
F3V24 & 0 & 594.58 & 524.15 & 10.38 & 1.00 & 0.398 & RRc & 7 & 0.012 & 21.211 & 20.729 & 0.461 & 14 & 20.964 & 21.460 & 0.496 & 238.8566 & 0.588 & 0.017 & 4 \\
F3V25 & 0 & 192.58 & 402.32 & 8.86 & 1.00 & 0.387 & RRc & 7 & 0.035 & 21.284 & 20.867 & 0.471 & 14 & 21.001 & 21.574 & 0.573 & 238.6752 & 0.471 & 0.021 & 6 \\
F3V26 & 0 & 459.70 & 504.99 & 4.72 & 1.00 & 0.678 & RRab & 1 & 0.027 & 21.084 & 20.614 & 0.621 & 14 & 20.399 & 21.493 & 1.094 & 238.5383 & 0.629 & 0.019 & 8 \\
F3V27 & 0 & 659.89 & 516.97 & 13.05 & 1.00 & 0.540 & RRab & 3 & 0.026 & 21.238 & 20.726 & 0.556 & 14 & 20.727 & 21.540 & 0.813 & 238.5400 & 0.594 & 0.021 & 6 \\
F3V28 & 0 & 345.50 & 541.99 & 5.27 & 1.00 & 0.389 & RRc & 7 & 0.015 & 21.095 & 20.637 & 0.461 & 14 & 20.856 & 21.335 & 0.479 & 238.5543 & 0.500 & 0.017 & 4 \\
F3V29 & 0 & 576.21 & 538.17 & 9.92 & 1.00 & 0.559 & RRab & 1 & 0.020 & 21.197 & 20.793 & 0.490 & 14 & 20.476 & 21.634 & 1.158 & 238.5992 & 0.614 & 0.020 & 7 \\
F3V30 & 0 & 320.16 & 155.40 & 12.95 & 1.00 & 0.389 & RRc & 7 & 0.026 & 21.205 & 20.782 & 0.467 & 14 & 20.986 & 21.423 & 0.437 & 238.6022 & 0.510 & 0.020 & 6 \\
F3V31 & 2 & 249.62 & 76.60 & 42.32 & 1.00 & 0.554$^{*}$ & RRab & 5 & 0.025 & 21.363 & 20.779 & 0.546 & 13 & 20.999 & 21.572 & 0.573 & 238.9565 & 0.649 & 0.033 & 1 \\
F3V32 & 1 & 179.45 & 98.61 & 23.14 & 1.00 & 0.360 & RRc & 7 & 0.029 & 21.274 & 20.762 & 0.472 & 14 & 21.090 & 21.455 & 0.365 & 238.5399 & 0.536 & 0.015 & 4 \\
F3V33 & 3 & 210.24 & 410.88 & 39.41 & 1.00 & 0.401$^{*}$ & RRab? & 6 & 0.012 & 21.346 & 20.824 & 0.476 & 14 & 21.010 & 21.601 & 0.591 & 238.8575 & 0.568 & 0.016 & 4 \\
F3V34 & 0 & 399.75 & 324.37 & 4.99 & 1.00 & 0.568 & RRab & 6 & 0.009 & 21.166 & 20.656 & 0.567 & 13 & 20.784 & 21.460 & 0.676 & 238.5457 & 0.645 & 0.027 & 5 \\
F3V35 & 0 & 342.01 & 458.86 & 2.28 & 0.81 & 0.607 & RRab & 2 & 0.048 & 21.168 & 20.617 & 0.477 & 14 & 20.664 & 21.532 & 0.868 & 238.3644 & $-$ & $-$ & 0 \\
F3V36 & 0 & 476.83 & 661.28 & 11.17 & 1.00 & 0.460 & RRc & 7 & 0.021 & 21.146 & 20.672 & 0.491 & 14 & 20.954 & 21.334 & 0.380 & 238.7101 & 0.554 & 0.015 & 4 \\
F3V37 & 0 & 426.12 & 503.72 & 3.72 & 0.80 & 0.495$^{*}$ & RRab & 3 & 0.025 & 21.422 & 20.799 & 0.540 & 14 & 20.965 & 21.686 & 0.721 & 238.8745 & 0.637 & 0.028 & 4 \\
F3V38 & 1 & 658.37 & 621.38 & 86.01 & 1.00 & 0.481$^{*}$ & RRab & 1 & 0.032 & 21.581 & 21.057 & 0.546 & 14 & 21.132 & 21.830 & 0.698 & 238.7698 & 0.630 & 0.016 & 6 \\
F3V39 & 3 & 209.41 & 689.64 & 58.73 & 1.00 & 0.380 & RRc & 7 & 0.032 & 21.318 & 20.827 & 0.495 & 14 & 21.096 & 21.539 & 0.443 & 238.7928 & 0.579 & 0.022 & 2 \\
F3V40 & 0 & 368.42 & 380.44 & 2.51 & 0.81 & 0.441 & RRc & 7 & 0.024 & 21.061 & 20.562 & 0.530 & 13 & 20.815 & 21.309 & 0.494 & 238.6757 & 0.598 & 0.032 & 3 \\
F3V41 & 0 & 96.81 & 294.20 & 14.51 & 1.00 & 0.380 & RRc & 7 & 0.017 & 21.346 & 20.912 & 0.507 & 13 & 21.024 & 21.682 & 0.658 & 238.7354 & 0.616 & 0.023 & 4 \\
F3V42 & 0 & 545.33 & 407.93 & 7.38 & 0.91 & 0.425$^{*}$ & RRab & 6 & 0.022 & 21.124 & 20.575 & 0.509 & 14 & 20.632 & 21.517 & 0.885 & 238.4765 & 0.636 & 0.016 & 4 \\
F3V43 & 0 & 376.87 & 357.41 & 3.46 & 0.81 & 0.412 & RRc & 7 & 0.018 & 21.161 & 20.732 & 0.476 & 13 & 20.931 & 21.390 & 0.459 & 238.6587 & 0.483 & 0.026 & 6 \\
F3V44 & 0 & 740.08 & 381.46 & 16.27 & 0.88 & 0.453 & RRc & 7 & 0.007 & 21.207 & 20.800 & 0.470 & 14 & 20.934 & 21.486 & 0.552 & 238.7992 & 0.515 & 0.023 & 4 \\
F3V45 & 2 & 401.16 & 207.53 & 61.86 & 0.99 & 0.660 & RRab & 3 & 0.029 & 21.385 & 20.779 & 0.517 & 14 & 20.888 & 21.676 & 0.788 & 238.2720 & $-$ & $-$ & 0 \\
F3V46 & 0 & 720.95 & 664.39 & 18.45 & 1.00 & 0.670 & RRab & 1 & 0.024 & 21.173 & 20.566 & 0.545 & 14 & 20.683 & 21.449 & 0.766 & 238.9354 & $-$ & $-$ & 0 \\
F3V47 & 0 & 350.50 & 195.74 & 10.90 & 1.00 & 0.555 & RRab & 2 & 0.024 & 21.032 & 20.458 & 0.563 & 14 & 20.568 & 21.362 & 0.794 & 238.9554 & 0.644 & 0.025 & 1 \\
F3V48 & 0 & 379.41 & 551.08 & 5.38 & 1.00 & 0.410 & RRc & 7 & 0.014 & 21.197 & 20.690 & 0.517 & 14 & 21.006 & 21.384 & 0.378 & 238.7094 & 0.613 & 0.040 & 4 \\
F3V49 & 3 & 451.63 & 361.86 & 59.77 & 1.00 & 0.403$^{*}$ & RRab & 5 & 0.015 & 21.249 & 20.694 & 0.502 & 13 & 20.818 & 21.503 & 0.685 & 238.8996 & 0.628 & 0.020 & 2 \\
F3V50 & 0 & 379.45 & 557.58 & 5.67 & 1.00 & 0.447$^{*}$ & RRab? & 3 & 0.025 & 21.170 & 20.669 & 0.468 & 14 & 20.692 & 21.449 & 0.757 & 238.9094 & 0.516 & 0.019 & 4 \\
F3V51 & 0 & 94.20 & 588.37 & 14.94 & 1.00 & 0.535 & RRc? & 7 & 0.042 & 21.409 & 20.851 & 0.528 & 14 & 21.103 & 21.728 & 0.625 & 238.4568 & 0.547 & 0.058 & 3 \\
F3V52 & 1 & 780.06 & 396.32 & 75.82 & 1.00 & 0.576 & RRc? & 7 & 0.019 & 21.225 & 20.801 & 0.512 & 14 & 20.949 & 21.508 & 0.559 & 238.7226 & 0.548 & 0.013 & 6 \\
F3V53 & 0 & 517.90 & 431.75 & 6.05 & 0.91 & 0.367$^{*}$ & RRab? & 5 & 0.015 & 21.153 & 20.621 & 0.500 & 14 & 20.798 & 21.356 & 0.558 & 238.5377 & 0.603 & 0.015 & 4 \\
F3V54 & 0 & 403.01 & 484.67 & 2.49 & 0.81 & 0.364 & RRc & 7 & 0.024 & 21.212 & 20.715 & 0.488 & 14 & 21.049 & 21.369 & 0.320 & 238.7943 & 0.662 & 0.035 & 2 \\
F3V55 & 0 & 528.34 & 369.69 & 7.13 & 0.91 & 0.466 & RRc & 7 & 0.015 & 21.270 & 20.826 & 0.474 & 11 & 21.060 & 21.477 & 0.417 & 238.4885 & 0.587 & 0.023 & 2 \\
F3V56 & 0 & 297.03 & 330.07 & 6.16 & 0.91 & 0.585 & RRc? & 8 & 0.030 & 21.305 & 20.760 & 0.524 & 14 & 21.151 & 21.472 & 0.321 & 238.5482 & 0.535 & 0.018 & 7 \\
F3V57 & 0 & 477.99 & 779.51 & 16.24 & 1.00 & 0.408 & RRc? & 8 & 0.049 & 21.297 & 20.748 & 0.556 & 14 & 21.041 & 21.588 & 0.547 & 238.5237 & 0.503 & 0.025 & 4 \\
F3V58 & 0 & 99.56 & 359.59 & 13.38 & 1.00 & 0.381 & RRc & 8 & 0.023 & 21.352 & 20.899 & 0.508 & 14 & 21.087 & 21.655 & 0.568 & 238.7387 & 0.562 & 0.017 & 4 \\
F3V59 & 0 & 490.48 & 706.09 & 13.27 & 1.00 & 0.613 & RRab & 6 & 0.031 & 21.024 & 20.433 & 0.561 & 14 & 20.630 & 21.330 & 0.700 & 238.3681 & $-$ & $-$ & 0 \\
F3V60 & 0 & 202.59 & 472.21 & 8.48 & 1.00 & 0.394$^{*}$ & RRab & 2 & 0.019 & 21.163 & 20.607 & 0.550 & 14 & 20.747 & 21.455 & 0.708 & 238.5063 & 0.620 & 0.020 & 4 \\
F3V61 & 0 & 562.75 & 509.53 & 8.80 & 1.00 & 0.590 & RRab & 5 & 0.026 & 21.329 & 20.783 & 0.533 & 13 & 21.048 & 21.485 & 0.437 & 238.4022 & $-$ & $-$ & 0 \\
\hline
\end{tabular}
\end{minipage}}}}
\end{table*}

\begin{table*}
\centerline{\resizebox{!}{!}{\rotatebox{90}{
\begin{minipage}{238mm}
\begin{tabular}{@{}lcccccclccccccccccccc}
\hline \hline
Name & Chip & $X_c$ & $Y_c$ & $r$ & $\alpha_c$ & Period & Type & $N_t$ & rms & $\langle V \rangle$ & $\langle I \rangle$ & $\overline{(V-I)}$ & $N_{o}$ & $V_{+}$ & $V_{-}$ & $\Delta V$ & $\degr E_{m}$ & $(V-I)_{m}$ & $\sigma_m$ & $N_{m}$ \\
 & & (pix) & (pix) & $(\arcsec)$ & & (days) & & & & & & & & & & & & & & \\
\hline
F3V62 & 0 & 338.62 & 432.19 & 2.11 & 0.81 & 0.463 & RRc & 7 & 0.024 & 21.208 & 20.694 & 0.514 & 13 & 20.982 & 21.432 & 0.450 & 238.7159 & 0.637 & 0.057 & 3 \\
F3V63 & 0 & 320.81 & 404.73 & 3.19 & 0.81 & 0.531$^{*}$ & RRab & 4 & 0.024 & 21.122 & 20.544 & 0.564 & 13 & 20.735 & 21.411 & 0.676 & 238.4886 & 0.633 & 0.061 & 4 \\
F3V64 & 0 & 243.38 & 328.13 & 8.01 & 1.00 & 0.385$^{*}$ & RRab? & 6 & 0.023 & 21.108 & 20.584 & 0.555 & 13 & 20.799 & 21.341 & 0.542 & 238.5344 & 0.569 & 0.023 & 2 \\
F3V65 & 0 & 294.80 & 572.98 & 7.57 & 0.91 & 0.416 & RRc & 7 & 0.011 & 21.249 & 20.732 & 0.526 & 14 & 21.067 & 21.425 & 0.358 & 238.5482 & 0.553 & 0.028 & 3 \\
F3V66 & 2 & 75.95 & 275.55 & 44.77 & 1.00 & 0.412$^{*}$ & RRab & 6 & 0.030 & 21.303 & 20.742 & 0.512 & 14 & 20.967 & 21.557 & 0.590 & 238.8661 & 0.585 & 0.015 & 4 \\
F3V67 & 3 & 98.01 & 707.04 & 54.71 & 1.00 & 0.618 & RRab & 5 & 0.027 & 21.433 & 20.831 & 0.595 & 14 & 21.075 & 21.639 & 0.564 & 238.4486 & 0.756 & 0.049 & 4 \\
F3V68 & 0 & 279.88 & 332.29 & 6.62 & 0.91 & 0.510 & RRc & 8 & 0.017 & 21.197 & 20.642 & 0.542 & 13 & 20.987 & 21.428 & 0.441 & 238.9672 & 0.613 & 0.026 & 2 \\
F3V69 & 3 & 103.09 & 149.54 & 23.75 & 0.91 & 0.555 & RRab & 5 & 0.044 & 21.363 & 20.822 & 0.639 & 14 & 20.709 & 21.784 & 1.075 & 238.6336 & 0.689 & 0.016 & 4 \\
F3V70 & 0 & 207.17 & 481.52 & 8.38 & 1.00 & 0.664 & RRab & 5 & 0.025 & 21.308 & 20.625 & 0.539 & 13 & 20.867 & 21.569 & 0.702 & 238.9016 & 0.699 & 0.031 & 1 \\
F3V71 & 2 & 481.32 & 510.67 & 87.41 & 1.00 & 0.701 & RRab & 6 & 0.031 & 21.403 & 20.852 & 0.581 & 14 & 21.122 & 21.612 & 0.490 & 238.4729 & 0.642 & 0.015 & 8 \\
F3V72 & 0 & 332.84 & 570.22 & 6.67 & 0.89 & 0.523$^{*}$ & RRab & 6 & 0.027 & 21.055 & 20.458 & 0.590 & 14 & 20.817 & 21.229 & 0.412 & 238.4555 & $-$ & $-$ & 0 \\
F3V73 & 0 & 400.86 & 340.76 & 4.26 & 1.00 & 0.415$^{*}$ & RRab & 6 & 0.037 & 21.337 & 20.706 & 0.580 & 14 & 20.929 & 21.654 & 0.725 & 238.8752 & 0.713 & 0.036 & 4 \\
F3V74 & 0 & 220.40 & 692.51 & 13.93 & 1.00 & 0.412 & RRc & 7 & 0.028 & 21.266 & 20.705 & 0.564 & 13 & 21.046 & 21.485 & 0.439 & 238.7453 & 0.672 & 0.022 & 4 \\
F3V75 & 0 & 382.96 & 385.97 & 2.14 & 0.81 & 0.391$^{*}$ & RRab? & 6 & 0.024 & 21.142 & 20.575 & 0.542 & 14 & 20.962 & 21.271 & 0.309 & 238.5093 & 0.582 & 0.035 & 4 \\
F3V76 & 2 & 126.87 & 746.74 & 90.31 & 1.00 & 0.528 & RRab & 1 & 0.049 & 21.489 & 20.819 & 0.625 & 14 & 20.697 & 21.980 & 1.283 & 238.3870 & 0.694 & 0.039 & 2 \\
F3V77 & 0 & 478.84 & 443.13 & 4.30 & 1.00 & 0.590 & RRab & 5 & 0.025 & 21.082 & 20.476 & 0.599 & 14 & 20.781 & 21.252 & 0.471 & 238.4935 & 0.632 & 0.048 & 5 \\
F3V78 & 0 & 238.81 & 426.41 & 6.66 & 0.91 & 0.721 & RRab & 2 & 0.020 & 21.126 & 20.549 & 0.557 & 14 & 20.689 & 21.434 & 0.745 & 239.0515 & 0.646 & 0.038 & 4 \\
F3V79 & 3 & 364.46 & 109.74 & 49.93 & 1.00 & 0.655 & RRab & 3 & 0.012 & 21.485 & 20.858 & 0.572 & 14 & 21.117 & 21.691 & 0.574 & 238.9895 & $-$ & $-$ & 0 \\
F3V80 & 1 & 276.06 & 192.60 & 32.34 & 1.00 & 0.621 & RRab & 1 & 0.042 & 21.265 & 20.757 & 0.624 & 14 & 20.602 & 21.657 & 1.055 & 238.5658 & 0.638 & 0.015 & 8 \\
F3V81 & 0 & 53.44 & 643.37 & 17.76 & 0.88 & 0.657 & RRab & 4 & 0.020 & 21.488 & 20.970 & 0.504 & 10 & 21.006 & 21.859 & 0.853 & 238.3446 & $-$ & $-$ & 0 \\
F3V82 & 0 & 443.58 & 442.36 & 2.70 & 0.81 & 0.509 & RRc & 7 & 0.019 & 21.046 & 20.501 & 0.500 & 14 & 20.935 & 21.150 & 0.215 & 238.5657 & 0.518 & 0.024 & 6 \\
F3V83 & 0 & 295.07 & 466.47 & 4.37 & 1.00 & 0.444 & RRc & 7 & 0.020 & 21.197 & 20.676 & 0.545 & 14 & 20.975 & 21.416 & 0.441 & 238.6740 & 0.639 & 0.019 & 4 \\
F3V84 & 0 & 419.13 & 367.98 & 3.34 & 0.80 & 0.312$^{*}$ & RRab? & 1 & 0.018 & 21.345 & 20.716 & 0.596 & 14 & 21.128 & 21.457 & 0.329 & 238.8683 & 0.630 & 0.034 & 2 \\
F3V85 & 0 & 437.33 & 355.86 & 4.24 & 1.00 & 0.509$^{*}$ & RRab & 1 & 0.040 & 21.148 & 20.691 & 0.560 & 12 & 20.567 & 21.483 & 0.916 & 238.6852 & 0.632 & 0.020 & 3 \\
F3V86 & 2 & 539.43 & 80.14 & 69.02 & 1.00 & 0.648 & RRab & 1 & 0.036 & 21.504 & 20.949 & 0.632 & 14 & 20.942 & 21.828 & 0.886 & 238.5342 & 0.686 & 0.023 & 8 \\
F3V87 & 0 & 358.82 & 594.17 & 7.42 & 1.00 & 0.397$^{*}$ & RRab & 6 & 0.017 & 21.386 & 20.753 & 0.587 & 14 & 21.097 & 21.601 & 0.504 & 238.4966 & 0.666 & 0.017 & 4 \\
F3V88 & 1 & 647.16 & 88.87 & 47.12 & 1.00 & 0.524$^{*}$ & RRab & 6 & 0.029 & 20.469 & 19.968 & 0.498 & 14 & 20.128 & 20.728 & 0.600 & 238.8568 & 0.595 & 0.015 & 4 \\
F3V89 & 0 & 282.57 & 458.47 & 4.80 & 1.00 & 0.440$^{*}$ & RRab & 6 & 0.027 & 21.379 & 20.737 & 0.616 & 13 & 21.107 & 21.580 & 0.473 & 238.8521 & 0.669 & 0.021 & 3 \\
F3V90 & 0 & 434.05 & 379.93 & 3.29 & 0.81 & 0.664 & RRab & 1 & 0.026 & 21.214 & 20.755 & 0.597 & 14 & 20.593 & 21.577 & 0.984 & 238.5428 & 0.551 & 0.022 & 8 \\
F3V91 & 0 & 432.48 & 394.74 & 2.78 & 0.80 & 0.453$^{*}$ & RRab & 5 & 0.033 & 21.356 & 20.727 & 0.616 & 14 & 21.115 & 21.488 & 0.373 & 238.8540 & 0.637 & 0.019 & 4 \\
F3V92 & 0 & 343.88 & 428.18 & 1.89 & 0.04 & 0.688 & RRab & 5 & 0.027 & 20.968 & 20.408 & 0.670 & 14 & 20.523 & 21.231 & 0.708 & 238.5377 & 0.705 & 0.023 & 7 \\
F3V93 & 1 & 645.48 & 496.28 & 74.83 & 1.00 & 0.536$^{*}$ & RRab & 4 & 0.026 & 21.476 & 20.851 & 0.653 & 14 & 21.183 & 21.688 & 0.505 & 238.7887 & 0.670 & 0.029 & 4 \\
F3V94 & 0 & 712.80 & 327.71 & 15.62 & 1.00 & 0.362$^{*}$ & RRab? & 6 & 0.011 & 20.650 & 20.061 & 0.587 & 14 & 20.367 & 20.861 & 0.494 & 238.8351 & 0.677 & 0.014 & 3 \\
F3V95 & 0 & 473.68 & 389.27 & 4.50 & 0.88 & 0.394$^{*}$ & RRab & 6 & 0.014 & 21.090 & 20.421 & 0.652 & 14 & 20.907 & 21.221 & 0.314 & 238.9062 & 0.669 & 0.016 & 3 \\
F3V96 & 1 & 219.83 & 284.31 & 41.08 & 1.00 & 0.659 & RRab & 5 & 0.029 & 21.361 & 20.816 & 0.633 & 14 & 20.896 & 21.638 & 0.742 & 238.5711 & 0.629 & 0.014 & 5 \\
F3V97 & 0 & 478.20 & 456.30 & 4.37 & 0.82 & 0.397$^{*}$ & RRab & 6 & 0.024 & 21.361 & 20.652 & 0.661 & 14 & 21.110 & 21.545 & 0.435 & 238.4847 & 0.717 & 0.028 & 4 \\
F3V98 & 0 & 552.32 & 766.78 & 16.89 & 1.00 & 0.451$^{*}$ & RRab & 3 & 0.057 & 21.351 & 20.669 & 0.646 & 14 & 21.062 & 21.508 & 0.446 & 238.4394 & 0.692 & 0.027 & 4 \\
F3V99 & 0 & 502.74 & 389.31 & 5.72 & 0.82 & 0.383$^{*}$ & RRab & 6 & 0.031 & 21.368 & 20.695 & 0.649 & 13 & 21.083 & 21.580 & 0.497 & 238.5170 & 0.722 & 0.019 & 4 \\
F5V01 & 0 & 420.01 & 498.13 & 2.69 & 0.99 & 0.405 & RRc & 8 & 0.041 & 21.007 & 20.676 & 0.345 & 14 & 20.771 & 21.272 & 0.501 & 237.9404 & 0.471 & 0.025 & 6 \\
\hline
\end{tabular}
\end{minipage}}}}
\end{table*}

\begin{table*}
\centerline{\resizebox{!}{!}{\rotatebox{90}{
\begin{minipage}{238mm}
\begin{tabular}{@{}lcccccclccccccccccccc}
\hline \hline
Name & Chip & $X_c$ & $Y_c$ & $r$ & $\alpha_c$ & Period & Type & $N_t$ & rms & $\langle V \rangle$ & $\langle I \rangle$ & $\overline{(V-I)}$ & $N_{o}$ & $V_{+}$ & $V_{-}$ & $\Delta V$ & $\degr E_{m}$ & $(V-I)_{m}$ & $\sigma_m$ & $N_{m}$ \\
 & & (pix) & (pix) & $(\arcsec)$ & & (days) & & & & & & & & & & & & & & \\
\hline
F5V02 & 0 & 91.38 & 247.48 & 16.23 & 0.96 & 0.311 & RRc & 7 & 0.009 & 21.493 & 21.199 & 0.343 & 14 & 21.179 & 21.819 & 0.640 & 237.9882 & 0.433 & 0.032 & 6 \\
F5V03 & 0 & 445.78 & 401.77 & 3.43 & 0.95 & 0.278 & RRc & 7 & 0.009 & 21.326 & 21.049 & 0.320 & 14 & 21.121 & 21.528 & 0.407 & 238.0563 & 0.410 & 0.027 & 5 \\
F5V04 & 3 & 64.07 & 311.29 & 22.53 & 0.91 & 0.315 & RRc & 7 & 0.019 & 21.418 & 21.041 & 0.378 & 14 & 21.284 & 21.545 & 0.261 & 238.1354 & 0.413 & 0.017 & 3 \\
F5V05 & 0 & 347.30 & 687.40 & 10.96 & 1.00 & 0.394 & RRc & 7 & 0.015 & 21.284 & 20.920 & 0.367 & 13 & 21.211 & 21.351 & 0.140 & 238.0430 & 0.384 & 0.017 & 4 \\
F5V06 & 0 & 277.07 & 471.31 & 5.10 & 0.98 & 0.595 & RRab & 2 & 0.011 & 21.215 & 20.631 & 0.536 & 14 & 20.713 & 21.577 & 0.864 & 237.7311 & $-$ & $-$ & 0 \\
F5V07 & 0 & 290.49 & 593.95 & 7.91 & 1.00 & 0.300 & RRc & 8 & 0.017 & 21.306 & 20.904 & 0.377 & 14 & 21.205 & 21.412 & 0.207 & 237.9470 & 0.329 & 0.020 & 4 \\
F5V08 & 0 & 391.59 & 375.00 & 3.37 & 0.99 & 0.419 & RRc & 7 & 0.023 & 21.237 & 20.848 & 0.410 & 14 & 20.911 & 21.578 & 0.667 & 238.0836 & 0.532 & 0.026 & 3 \\
F5V09 & 0 & 349.28 & 424.61 & 2.04 & 0.99 & 0.378 & RRc & 7 & 0.014 & 21.247 & 20.836 & 0.400 & 13 & 21.016 & 21.477 & 0.461 & 238.0961 & 0.545 & 0.025 & 4 \\
F5V10 & 0 & 422.69 & 317.87 & 6.18 & 1.00 & 0.340 & RRc & 7 & 0.020 & 21.325 & 20.915 & 0.401 & 14 & 21.038 & 21.619 & 0.581 & 238.1377 & 0.536 & 0.026 & 2 \\
F5V11 & 0 & 325.60 & 391.00 & 3.84 & 0.99 & 0.364 & RRc & 7 & 0.018 & 21.191 & 20.786 & 0.451 & 14 & 20.919 & 21.468 & 0.549 & 237.9517 & 0.548 & 0.023 & 6 \\
F5V12 & 2 & 323.81 & 545.98 & 81.04 & 1.00 & 0.396 & RRc & 7 & 0.025 & 21.143 & 20.743 & 0.465 & 14 & 20.838 & 21.459 & 0.621 & 238.0049 & 0.551 & 0.021 & 6 \\
F5V13 & 0 & 396.96 & 424.02 & 1.22 & 0.39 & 0.368 & RRc & 7 & 0.026 & 21.273 & 20.846 & 0.439 & 14 & 21.044 & 21.501 & 0.457 & 237.9100 & 0.509 & 0.027 & 4 \\
F5V14 & 0 & 380.64 & 327.69 & 5.53 & 1.00 & 0.372 & RRc & 7 & 0.028 & 21.317 & 20.899 & 0.469 & 14 & 21.055 & 21.583 & 0.528 & 238.0498 & 0.518 & 0.020 & 3 \\
F5V15 & 0 & 465.75 & 515.87 & 4.70 & 1.00 & 0.362 & RRc & 8 & 0.016 & 21.291 & 20.828 & 0.425 & 14 & 21.058 & 21.552 & 0.494 & 237.8789 & 0.568 & 0.017 & 4 \\
F5V16 & 0 & 405.52 & 290.88 & 7.24 & 1.00 & 0.520 & RRab & 3 & 0.014 & 21.332 & 20.792 & 0.436 & 14 & 20.716 & 21.708 & 0.992 & 238.2011 & 0.554 & 0.017 & 4 \\
F5V17 & 0 & 554.79 & 363.72 & 8.56 & 1.00 & 0.353 & RRc & 7 & 0.022 & 21.310 & 20.870 & 0.397 & 14 & 21.055 & 21.566 & 0.511 & 237.8923 & 0.544 & 0.034 & 4 \\
F5V18 & 1 & 385.96 & 81.77 & 25.97 & 1.00 & 0.361 & RRc & 7 & 0.015 & 21.382 & 20.960 & 0.475 & 14 & 21.131 & 21.635 & 0.504 & 237.9931 & 0.558 & 0.019 & 6 \\
F5V19 & 3 & 378.12 & 714.40 & 71.86 & 1.00 & 0.404 & RRc & 7 & 0.023 & 21.47 & 20.995 & 0.486 & 14 & 21.256 & 21.681 & 0.425 & 238.0630 & 0.570 & 0.018 & 4 \\
F5V20 & 0 & 694.33 & 577.77 & 15.10 & 1.00 & 0.422 & RRc & 7 & 0.011 & 21.356 & 20.914 & 0.416 & 14 & 21.068 & 21.652 & 0.584 & 238.2059 & 0.573 & 0.019 & 4 \\
F5V21 & 0 & 417.89 & 512.44 & 3.21 & 0.94 & 0.351 & RRc & 7 & 0.010 & 21.171 & 20.748 & 0.482 & 14 & 20.931 & 21.412 & 0.481 & 237.9694 & 0.552 & 0.025 & 6 \\
F5V22 & 0 & 468.89 & 628.90 & 8.97 & 1.00 & 0.402 & RRc & 7 & 0.016 & 21.367 & 20.905 & 0.482 & 14 & 21.129 & 21.605 & 0.476 & 237.9445 & 0.577 & 0.022 & 6 \\
F5V23 & 1 & 476.49 & 152.30 & 37.03 & 1.00 & 0.379 & RRc & 7 & 0.033 & 21.362 & 20.91 & 0.459 & 14 & 21.126 & 21.599 & 0.473 & 238.1427 & 0.476 & 0.017 & 4 \\
F5V24 & 0 & 437.65 & 424.47 & 2.56 & 0.99 & 0.445 & RRc & 7 & 0.015 & 21.413 & 20.934 & 0.452 & 14 & 21.278 & 21.543 & 0.265 & 237.8350 & 0.456 & 0.034 & 2 \\
F5V25 & 0 & 463.33 & 340.50 & 6.04 & 1.00 & 0.494 & RRab & 2 & 0.020 & 21.18 & 20.685 & 0.521 & 14 & 20.630 & 21.584 & 0.954 & 237.8604 & 0.517 & 0.050 & 4 \\
F5V26 & 0 & 227.21 & 589.26 & 9.66 & 0.98 & 0.571 & RRab & 2 & 0.016 & 21.28 & 20.712 & 0.538 & 14 & 20.680 & 21.726 & 1.046 & 237.7202 & $-$ & $-$ & 0 \\
F5V27 & 0 & 432.85 & 616.59 & 7.89 & 1.00 & 0.382 & RRc & 7 & 0.014 & 21.281 & 20.837 & 0.489 & 14 & 21.055 & 21.506 & 0.451 & 238.0194 & 0.518 & 0.021 & 6 \\
F5V28 & 0 & 115.32 & 737.13 & 17.89 & 0.95 & 0.326 & RRc & 7 & 0.021 & 21.434 & 21.073 & 0.454 & 13 & 21.152 & 21.722 & 0.570 & 238.0419 & 0.471 & 0.020 & 8 \\
F5V29 & 0 & 471.83 & 291.22 & 8.15 & 1.00 & 0.420 & RRc & 8 & 0.019 & 21.367 & 20.874 & 0.501 & 13 & 21.212 & 21.534 & 0.322 & 237.9307 & 0.602 & 0.038 & 4 \\
F5V30 & 3 & 211.47 & 345.73 & 37.21 & 1.00 & 0.433$^{*}$ & RRab & 4 & 0.019 & 21.297 & 20.728 & 0.518 & 14 & 20.730 & 21.746 & 1.016 & 238.2367 & 0.644 & 0.015 & 4 \\
F5V31 & 0 & 437.38 & 321.31 & 6.25 & 1.00 & 0.572 & RRab & 3 & 0.012 & 21.31 & 20.706 & 0.557 & 14 & 20.841 & 21.582 & 0.741 & 237.6900 & 0.682 & 0.029 & 2 \\
F5V32 & 0 & 347.33 & 243.78 & 9.51 & 1.00 & 0.603 & RRab & 3 & 0.032 & 21.296 & 20.794 & 0.534 & 14 & 20.850 & 21.553 & 0.703 & 238.1314 & 0.573 & 0.016 & 5 \\
F5V33 & 0 & 293.43 & 457.39 & 4.28 & 1.00 & 0.428 & RRc? & 7 & 0.011 & 21.234 & 20.685 & 0.538 & 14 & 21.125 & 21.337 & 0.212 & 238.2763 & 0.591 & 0.021 & 2 \\
F5V34 & 0 & 326.89 & 674.77 & 10.60 & 1.00 & 0.720 & RRab & 3 & 0.049 & 21.433 & 20.74 & 0.490 & 13 & 20.747 & 21.864 & 1.117 & 238.2507 & $-$ & $-$ & 0 \\
F5V35 & 0 & 351.30 & 494.77 & 2.64 & 0.94 & 0.415 & RRc? & 7 & 0.014 & 21.066 & 20.547 & 0.536 & 14 & 20.951 & 21.174 & 0.223 & 238.0014 & 0.539 & 0.023 & 6 \\
F5V36 & 3 & 501.81 & 655.42 & 77.55 & 1.00 & 0.540 & RRab & 3 & 0.016 & 21.488 & 20.804 & 0.559 & 14 & 20.971 & 21.792 & 0.821 & 237.6789 & 0.674 & 0.023 & 2 \\
F5V37 & 0 & 378.58 & 336.97 & 5.11 & 0.98 & 0.532$^{*}$ & RRab & 5 & 0.015 & 21.441 & 20.923 & 0.551 & 14 & 21.009 & 21.696 & 0.687 & 238.1578 & 0.640 & 0.015 & 6 \\
F5V38 & 0 & 336.72 & 619.27 & 8.06 & 0.98 & 0.392$^{*}$ & RRab & 5 & 0.015 & 21.524 & 20.86 & 0.627 & 14 & 21.279 & 21.659 & 0.380 & 237.8331 & 0.656 & 0.017 & 4 \\
F5V39 & 0 & 616.54 & 781.25 & 18.24 & 1.00 & 0.513$^{*}$ & RRab & 5 & 0.017 & 21.378 & 20.836 & 0.542 & 14 & 21.168 & 21.493 & 0.325 & 238.1786 & 0.636 & 0.031 & 4 \\
F5V40 & 0 & 418.38 & 479.85 & 2.00 & 0.95 & 0.466$^{*}$ & RRab & 6 & 0.021 & 21.373 & 20.746 & 0.617 & 14 & 21.163 & 21.527 & 0.364 & 238.1904 & 0.627 & 0.017 & 5 \\
\hline
\end{tabular}
\end{minipage}}}}
\end{table*}

\subsection{Remaining candidates}
\label{candidates}

\begin{table*}
\begin{minipage}{133mm}
\caption{Photometric measurements for the candidate RR Lyrae stars.}
\begin{tabular}{@{}lccccccccccc}
\hline \hline
Name & Chip & $X_c$ & $Y_c$ & $r$ & $\alpha_c$ & $\langle V \rangle$ & $\langle I \rangle$ &  $N_{o}$ & $V_{+}$ & $V_{-}$ & $\Delta V$ \\
\hline
F1C01 & 2 & 216.99 & 73.27 & 39.27 & 1.00 & 21.356 & 20.796 & 13 & 21.017 & 21.510 & 0.493 \\
F2C01 & 0 & 270.89 & 750.76 & 14.21 & 0.98 & 21.253 & 20.729 & 14 & 20.995 & 21.514 & 0.519 \\
F2C02 & 0 & 181.01 & 437.51 & 7.69 & 0.97 & 21.395 & 20.877 & 12 & 21.286 & 21.533 & 0.247 \\
F2C03 & 0 & 494.90 & 97.50  & 17.17 & 1.00 & 21.164 & 20.577 & 13 & 20.924 & 21.482 & 0.558 \\
F3C01 & 0 & 292.15 & 545.72 & 6.64 & 1.00 & 21.143 & 20.802 & 14 & 21.057 & 21.266 & 0.209 \\
F3C02 & 0 & 102.44 & 473.52 & 12.95 & 1.00 & 21.318 & 20.825 & 14 & 21.069 & 21.625 & 0.556 \\
F3C03 & 0 & 355.77 & 453.05 & 1.61 & 0.08 & 21.326 & 20.851 & 14 & 20.832 & 21.823 & 0.991 \\
F3C04 & 0 & 425.88 & 793.12 & 16.40 & 0.88 & 21.316 & 20.714 & 9 & 21.097 & 21.484 & 0.387 \\
F3C05 & 3 & 498.62 & 616.54 & 74.58 & 1.00 & 21.496 & 20.878 & 14 & 21.127 & 21.760 & 0.633 \\
F3C06 & 0 & 386.76 & 392.09 & 1.86 & 0.12 & 21.375 & 20.679 & 13 & 20.980 & 21.704 & 0.724 \\
F5C01 & 0 & 471.40 & 480.74 & 4.10 & 0.98 & 21.076 & 20.557 & 14 & 20.639 & 21.444 & 0.805 \\
F5C02 & 0 & 121.56 & 309.44 & 13.62 & 1.00 & 21.519 & 20.984 & 14 & 21.281 & 21.653 & 0.372 \\
F5C03 & 0 & 492.01 & 202.26 & 12.19 & 0.98 & 21.350 & 20.785 & 14 & 21.180 & 21.522 & 0.342 \\
\hline
\label{candphotom}
\end{tabular}
\end{minipage}
\end{table*}

In addition to these 197 objects, a further 13 stars fell within or nearby to the HB instability
strip defined by the RR Lyrae stars. However, suitable template fits could not be obtained for
these\footnote{This was generally because the shape of the variation did not match any of the
templates, or because a template which fit in the $V$-band produced a very poor light curve in
the $I$-band. In many cases, the root cause is probably poor measurements due to crowding or 
other contamination, combined with genuine variability}. Plots of their photometry are presented 
at the end of Fig. \ref{lightcurves}. Most appear to have some form of cyclic variation, and given 
this and their locations on the HB, many are likely RR Lyrae stars of some sort. Their positions and
completeness values are listed in Table \ref{candphotom}, together with the photometric parameters 
$\langle V \rangle$, $\langle I \rangle$, $V_{+}$, $V_{-}$, $\Delta V$, and $N_o$. Unlike for
the RR Lyrae stars, the $V$-band parameters for the present stars were of course calculated
from the measured data. Finder charts are also available for these candidate RR Lyrae stars on-line,
or by request.

\subsection{Comparisons with previous work}
\label{prevwork}
Although we searched carefully, we were unable to find any published quantitative measurements 
of RR Lyrae stars in Fornax globular clusters. There has however been some work on the RR Lyrae
population of the dwarf galaxy itself (e.g., Bersier \& Wood \shortcite{bersier}). Nonetheless, 
several authors have noted the inevitable presence of RR Lyrae stars in the Fornax clusters, 
given their well populated horizontal branches, and on several occasions detections of variability
have been made. 

In one of the earliest CMD studies of the Fornax clusters, Buonanno et al. 
\shortcite{buonold} located several likely HB stars with variable photometry
in their sample, and interpreted these as possible RR Lyrae stars. Buonanno et al. 
\shortcite{buonposter} noted the presence of 17 horizontal branch
stars with variable photometry in Fornax 1, and 40 such stars in Fornax 3. They identified these
as RR Lyrae stars; however there were not enough measurements to obtain significant average 
magnitudes. Smith et al. \shortcite{smith15} detected 21 candidate HB variables in Fornax 1 based
on a comparison of their photometry with that of Demers, Kunkel \& Grondin \shortcite{demers1},
who also found evidence for variability in a couple of HB stars. Similarly, Smith, Rich \& Neill
\shortcite{smith3} found evidence for 7 HB variables in Fornax 3, and one variable above the
Fornax 3 HB. Using the WFPC2 data from the present study, Buonanno et al. \shortcite{buonfora}
identified 8, 39, 66, and 36 candidate RR Lyrae stars in clusters 1, 2, 3, and 5, respectively,
based on the fact that these stars had frame-to-frame variations greater than $3\sigma$. These
numbers are quite similar to ours, as might be expected given that we are using the same 
observations. However, we detect more RR Lyraes in all clusters, demonstrating the sensitivity of
our search strategy.

More recently, Maio et al. \shortcite{maio} have described wide-field observations of a region of
the Fornax dwarf, including cluster 3. They detect 70 candidate variables in a 
$95\arcsec \times 95\arcsec$ box centred on this cluster, most of which lie on the HB. This is 
again fewer variables than we detect in Fornax 3; however, the present WFPC2 observations have 
superior resolution to the terrestrial wide-field observations. Maio et al. \shortcite{maio}
present preliminary uncalibrated light curves for two of their candidate cluster 3 RR Lyraes; 
however neither these nor any of their other variables are identified, and no quantification
of the variability is presented.

We therefore believe that all of the 197 RR Lyrae stars and 13 candidates we describe here have been
newly identified. This is the first time that calibrated light curves and variability data have
been published for any RR Lyrae stars in the globular clusters in the Fornax dwarf galaxy.

\subsection{Effects of under-sampling}
\label{undersampling}
We consider briefly here the effects that our short baselines of observation are expected to have
on the RR Lyrae parameters derived from the fitted light curves. As described in Sections
\ref{templates} and \ref{rrlyraestars}, the primary effect is that $\sim 50$ per cent of the 
calculated RRab periods are lower limits. This has knock-on effects for most of the other 
quantities computed for these stars, which must be accounted for if any use is to be made of
the measurements (e.g., in determining distance moduli). 

For example, suppose an RRab star has a true amplitude $(\Delta V)_t = (V_{-})_t - (V_{+})_t$ and
a true period of pulsation $P_t$. However, because of our limited baseline of observation ($< P_t$),
we have only measured $\Delta V < (\Delta V)_t$ for this star, and hence determined a period
$P < P_t$ (as described in Section \ref{templates}). This will lead to a systematic error in our 
calculation of $\langle V \rangle$ from the fitted light curve. It is possible that we have sampled 
only the brightest portion of the light curve, so that our measurement of $V_{+}$ is accurate, but 
in this case our assumed $V_{-}$ will be brighter than $(V_{-})_t$ and our calculated 
$\langle V \rangle$ will also be brighter than the true intensity-mean $V$ magnitude 
$\langle V \rangle_{t}$. Similarly, it is possible that we have accurately measured $V_{-}$, but 
assumed too faint a value for $V_{+}$. In this case our $\langle V \rangle$ will be fainter than 
$\langle V \rangle_{t}$. For symmetric light curves, such systematic errors should average to zero 
over a large ensemble of stars; however RRab stars have significantly asymmetric pulsations, in 
that they are brighter than $\langle V \rangle_{t}$ for typically only $\sim 30-40$ per cent of a 
cycle. Hence, while determinations of $\langle V \rangle$ for each star individually may be either 
too bright or too faint, we expect on average that our measurements will provide values of 
$\langle V \rangle$ somewhat fainter than $\langle V \rangle_{t}$ (i.e., we measure $V_{-}$ 
accurately more often than we measure $V_{+}$ accurately). The same reasoning applies for 
calculations involving $\langle I \rangle$, which are made directly from the photometry.

For stars with lower limit periods, our measurement of the magnitude-mean colour over the cycle
will also suffer a systematic error. Because we will have preferentially sampled
the RRab light curves around minimum light, on average a measurement of $\overline{(V-I)}$
will be redder than its true value, because minimum light is the reddest part of the pulsation
cycle. We note that individual measurements may, of course, also be too blue -- this is just more
unlikely than a red error. 

It is more complicated to estimate the effect of a lower-limit period on a measurement of 
$(V-I)_m$. Only points with phases in the range $0.5 < \phi < 0.8$ are used in the
calculation of this parameter. Since we have $P < P_t$, the true phase of any given point
will be smaller than the phase calculated using $P$. Hence, it is likely that we have used 
points which have true phases $\phi_t < 0.5$ in our calculations, and that on average our
measured $(V-I)_m$ values will be too blue. We note that this effect will be small,
since the rate of change of colour around the minimum of an RRab pulsation cycle is slow.

There are two ways to account for these systematic errors if any use of the derived data is
made -- for example, in determining the distance modulus to Fornax. The first is simply to
leave the stars flagged as having lower limit periods out of the calculations. Alternatively,
it is possible just to use those stars which have large amplitudes (say $\Delta V > 0.9$),
since these are the most likely to have an accurate determination for $P$. These two methods
allow a consistency check to be made, and the magnitude of any of the systematic errors 
described above to be investigated.

\section{Discussion}
\label{discussion}
Because the RR Lyrae populations of the Fornax globular clusters have not been previously 
investigated in a quantitative manner, it is useful to consider their characteristics. Much
can be learned from the measurements we have made, even bearing in mind the uncertainties in
the period fitting. As discussed in Section \ref{undersampling}, it is possible to investigate
and account for the systematic errors our procedure might introduce. Throughout the following
Sections, Table \ref{clusterprop} keeps a record of the calculations we discuss.

\subsection{Specific frequencies}
\label{freq}
Each of the four Fornax clusters we have examined has a significant population of RR Lyrae
stars. To investigate this quantitatively we calculate the RR Lyrae specific frequency, 
$S_{RR}$, which is the number of RR Lyrae stars in a cluster normalized to a cluster
absolute magnitude of $M_V = -7.5$:
\begin{equation}
S_{RR} = N_{RR}\,\,10^{0.4(7.5 + M_V)}
\end{equation}
Buonanno et al. \shortcite{buonfora} have noted that $S_{RR}$ for Fornax clusters 1, 2, 3, and 5
appears quite high relative to the Milky Way globular clusters. 

We calculate $S_{RR}$ using the integrated luminosities for the four Fornax clusters derived
by us from accurate surface brightness profiles \cite{fornaxpaper}. For the four clusters we
obtained integrated luminosities $\log (L_\infty / L_\odot) = 4.07$, $4.76$, $5.06$, and $4.76$ 
respectively. These correspond to F555W luminosities. Assuming the F555W solar magnitude to be
$V_{555}^{\odot} = +4.85$ \cite{lmcpaper}, we obtain absolute magnitudes of $-5.33$, $-7.05$, 
$-7.80$, and $-7.05$ for the four clusters. These are in good agreement with the values derived
by Buonanno et al. \shortcite{buonold}, and the values listed by Webbink \shortcite{webbink}.

The specific frequencies for the four clusters are then $S_{RR} = 110.7$, $65.1$, $75.1$, and
$60.5$, respectively. We note that these are lower limits, because we have not included information
about detection completeness, or counted the candidate variable stars. We can account for 
completeness effects using the $\alpha_c$ values in Table \ref{rrparams}. To avoid unwarranted
weighting by the few stars with very low completeness fractions, we set any $\alpha_c$ values
of less than $0.25$ equal to $0.25$. If we then assume that all the candidate variable stars
are actually RR Lyrae stars, we obtain specific frequencies of $S_{RR} = 118.2$, $70.5$, $90.2$, and
$68.4$. These are still lower limits, because in no cases did the WFPC2 observations cover an 
entire cluster. Hence, the samples of RR Lyrae stars are certainly incomplete, but in calculating
$S_{RR}$, we have normalized the $N_{RR}$ values to {\em full cluster} luminosities. Accounting
for this effect is not trivial, because we do not know the spatial distribution of RR Lyrae stars
in any cluster, and because the WFPC2 geometry is complicated. Given that we have imaged the centre
of each cluster, we expect the $N_{RR}$ to be $80-90$ per cent complete. Hence, the $S_{RR}$ values
may be $10-20$ per cent greater than quoted above.

These specific frequencies are extremely high compared to those for the galactic globular clusters.
The catalogue of Harris \shortcite{harris} lists only four of the $\sim 150$ galactic clusters
as having $S_{RR} > 60$, and only two of these have $S_{RR} > 100$. The largest value, that for
Palomar 13, is $S_{RR} = 127.5$. Given our sample incompleteness, it is very likely that Fornax 1
has $S_{RR}$ significantly greater than this. It is certainly intriguing that only a tiny fraction
of galactic globular clusters have very high $S_{RR}$, whereas {\em all} of the Fornax clusters
we have investigated do -- this is a fact worthy of further attention.

\subsection{Oosterhoff classification}

\begin{figure*}
\begin{minipage}{175mm}
\begin{center}
\includegraphics[width=80mm]{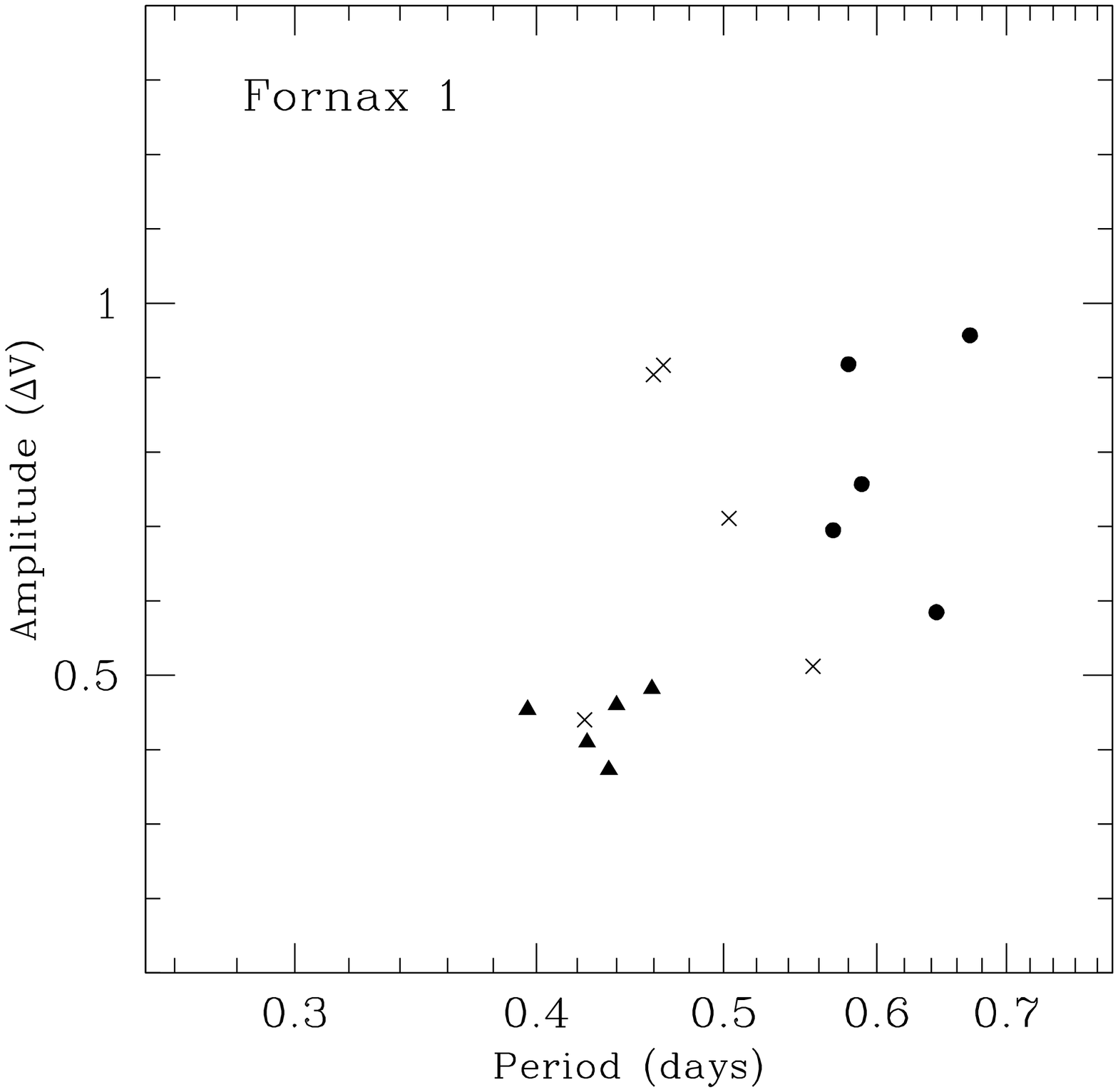}
\hspace{2mm}
\includegraphics[width=80mm]{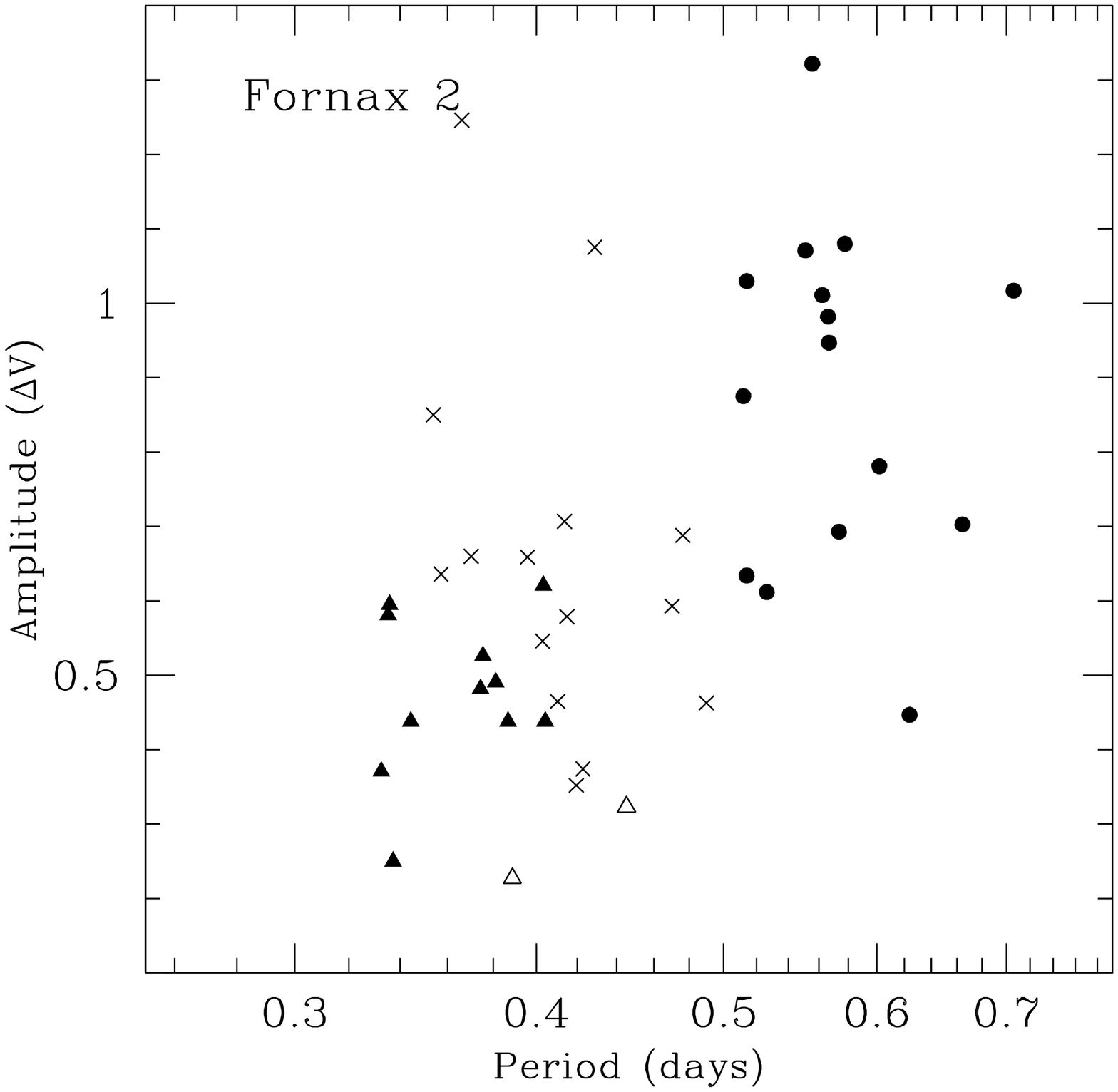} \\
\vspace{1mm}
\includegraphics[width=80mm]{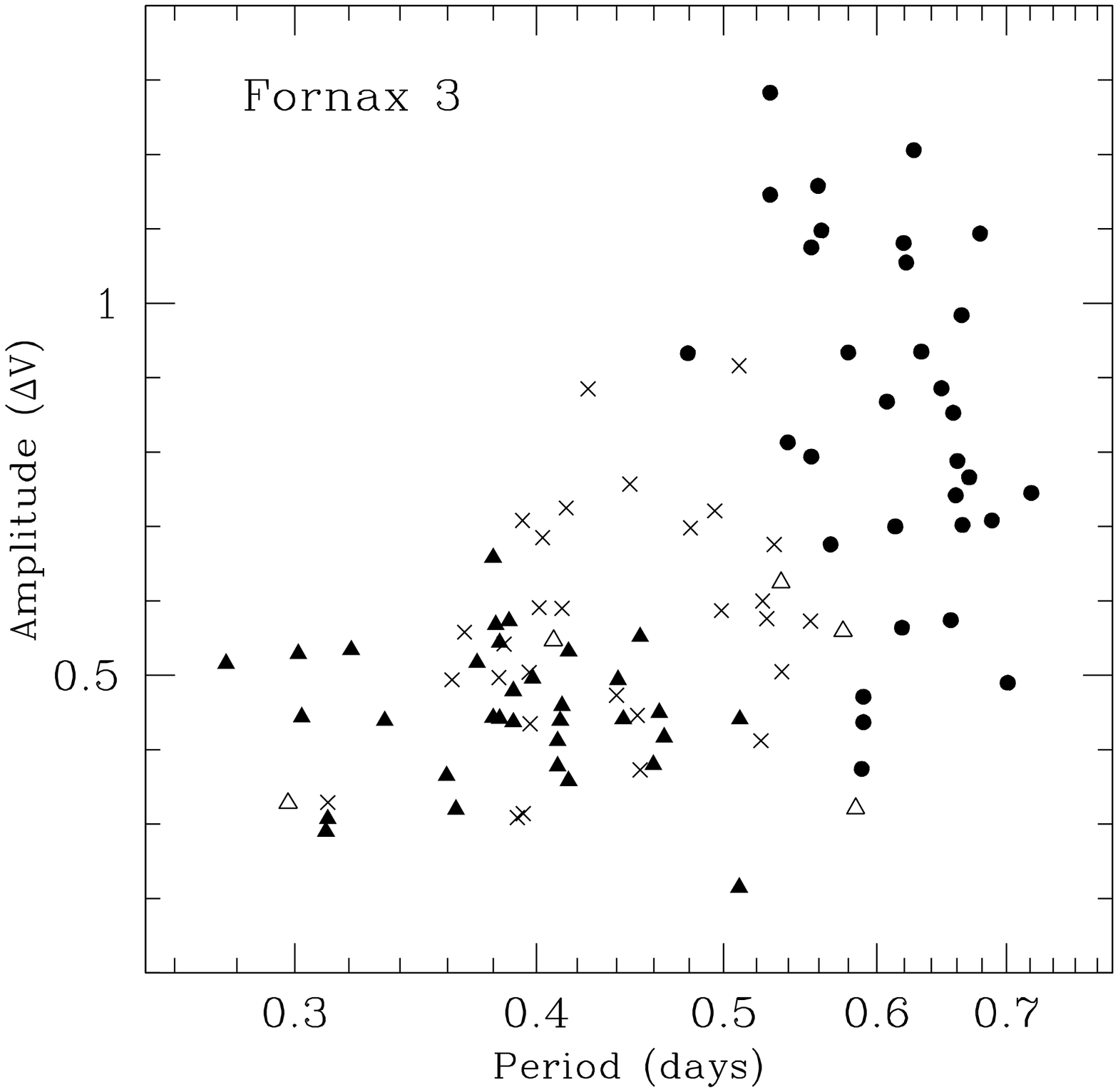}
\hspace{2mm}
\includegraphics[width=80mm]{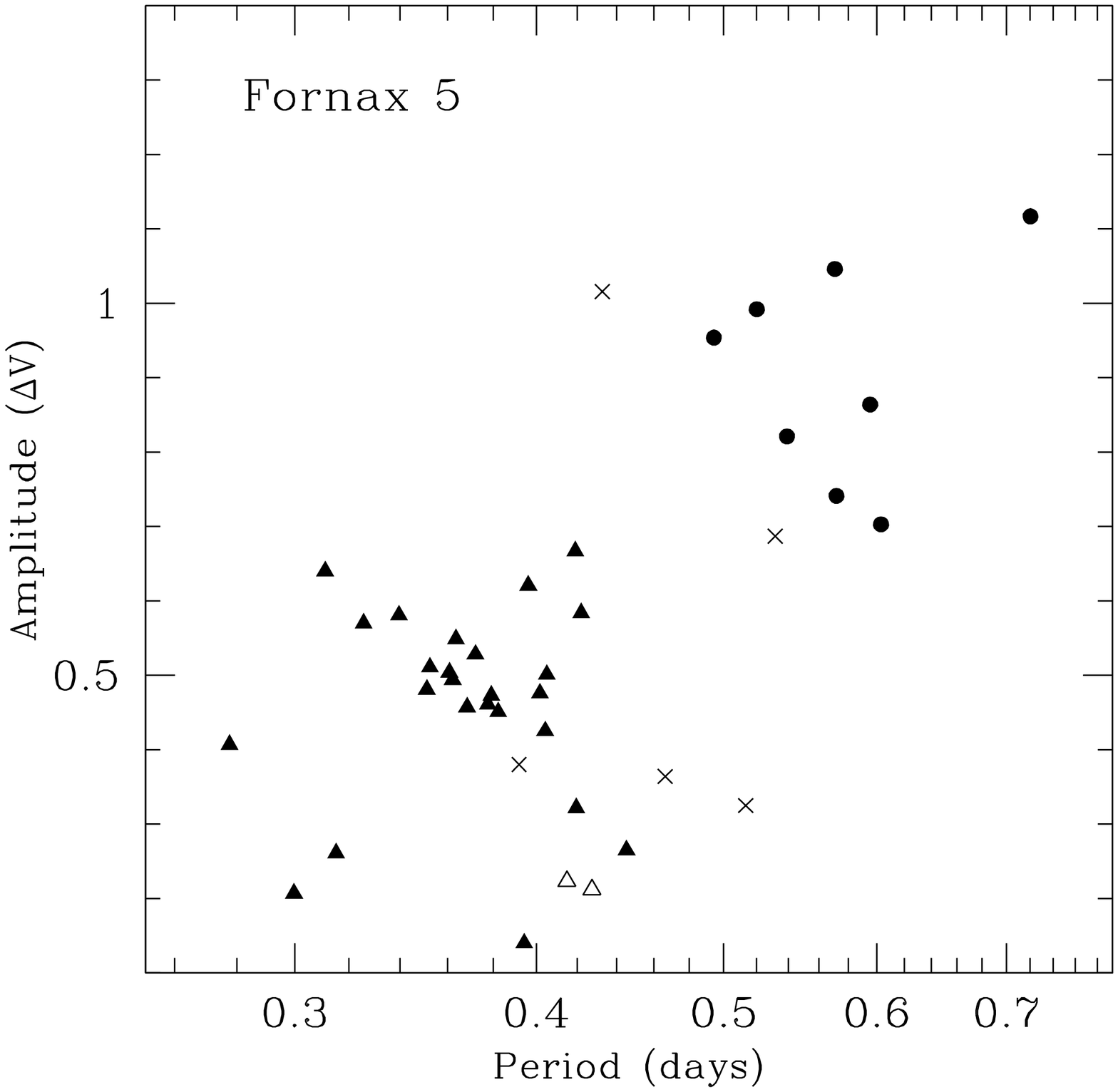}
\caption{Period-amplitude diagrams for each of the four clusters. RRab stars with good periods are plotted as solid circles; those with lower limit periods are crosses. RRc stars with uncertain classification are hollow triangles, while the remainder are solid triangles.}
\label{peramp}
\end{center}
\end{minipage}
\end{figure*}

It is also instructive to investigate the relationship between the derived periods and 
amplitudes for our RR Lyrae populations. This can be done by plotting a period-amplitude
diagram for each cluster. For galactic globular clusters, such diagrams fall naturally into two 
distinct groups -- the Oosterhoff type-I and type-II clusters (OoI and OoII, respectively). 
Each group has characteristic properties, summarized in Smith \shortcite{rrbook}. 
Furthermore, there is in general a close relationship between period and amplitude for the RRab
stars in a cluster -- namely that the shortest period RRab stars have the largest amplitudes,
while those with the smallest amplitudes have the longest periods.

Period-amplitude diagrams for the four Fornax clusters are presented in Fig. \ref{peramp}. RRab
stars with good periods are plotted as filled circles, while those with lower limit periods are
plotted as crosses. RRc stars are plotted as triangles -- those with uncertain classification in
Table \ref{rrparams} are hollow points, while the remainder are filled. From examination of these
diagrams, evidence of the lower limit period problem is immediately clear. The crosses form a 
distinct population between the good RRab stars and the RRc stars in each diagram. Once good
periods are determined for these stars, they will move into the regions occupied by the good 
RRab stars. In the best populated clusters, such as Fornax 3, evidence of the upper part of
the RRab sequence can be seen, although the relationships certainly have significant scatter. 
Improved period measurements should tighten these sequences.

From the period-amplitude diagrams it is not possible to determine conclusively whether the 
Fornax clusters resemble either the OoI or OoII clusters, or (like the old LMC clusters and several
dwarf spheroidal galaxies) have properties which lie somewhere in between the two (e.g., 
Pritzl et al. \shortcite{pritzl}). We can shed some light on this by calculating selected 
parameters for each of our RR Lyrae populations. Smith \shortcite{rrbook} lists the OoI clusters as 
having $\langle P_{ab} \rangle = 0.55$ d, $\langle P_{c} \rangle = 0.32$ d, and 
$N_c / N_{RR} = 0.17$, and the OoII clusters as having 
$\langle P_{ab} \rangle = 0.64$ d, $\langle P_{c} \rangle = 0.37$ d, and 
$N_c / N_{RR} = 0.44$. Pritzl lists these parameters for several old LMC clusters. These have
$\langle P_{ab} \rangle = 0.56 - 0.60$ d, $\langle P_{c} \rangle = 0.33 - 0.38$ d, and 
$N_c / N_{RR} = 0.3 - 0.5$, placing them in general somewhere between the OoI and OoII groups.

We calculated these parameters for the four Fornax clusters -- the results are listed in
Table \ref{clusterprop}. Our $\langle P_{c} \rangle$ values cover the range $0.37 - 0.43$ d,
which are similar to those for OoII clusters. However, excepting Fornax 5, the fraction
$N_c / N_{RR}$ ranges from $0.3 - 0.38$ for the Fornax clusters, which is more similar to the 
intermediate LMC clusters than either of the Oosterhoff groups. Given the presence of the 
many RRab stars with lower limit periods, it does not make sense for us to calculate 
$\langle P_{ab} \rangle$ using all our RRab stars\footnote{although we list this value in Table
\ref{clusterprop} for comparative purposes.}. It is far more useful to calculate this
quantity using only those RRab stars with good periods. If we do this, we obtain values in the
range $0.57 - 0.61$, again most similar to the intermediate LMC clusters. We must bear in mind
however that these averages do not include many stars with long periods, since we have only
measured lower limit periods for these stars. Hence, these mean values will likely 
increase once all stars have good periods. 

One final discriminator we can use is the transition
period $P_{tr}$ of RRab stars -- that is, the shortest period that this class of stars have in a 
given cluster, without being RRc stars. Given the inverse period-amplitude relation for RRab stars, 
those with the shortest periods have the largest amplitudes. Calculating $\langle P_{ab} \rangle$ 
for our RRab stars with $\Delta P > 0.9$ therefore provides a good estimate of $P_{tr}$. Smith 
\shortcite{rrbook} lists $P_{tr} = 0.43$ for OoI clusters and $P_{tr} = 0.55$ for OoII clusters. 
Our values range from $0.54 - 0.58$, which match the OoII clusters well.

Hence we can rule out the possibility that the Fornax globular clusters 1, 2, 3, and 5 resemble
the galactic OoI type clusters. However, with quite a dispersion in their properties, it is not
clear whether they do resemble either the OoII type clusters, or the intermediate LMC type clusters.
Both $N_c / N_{RR}$ and $\langle P_{ab} \rangle$ suggest an intermediate-type classification,
while $\langle P_{c} \rangle$ and $P_{tr}$ suggest a similarity to the OoII clusters. One 
explanation is that the Fornax clusters are, like the LMC clusters, of intermediate classification,
but with somewhat different characteristics. This interpretation is consistent with the
observation that the properties of RR Lyrae stars measured in the {\em fields} of dwarf galaxies
classify these objects also as intermediate, but with a considerable range in properties
(see e.g., Pritzl et al. \shortcite{pritzl}). Bersier \& Wood \shortcite{bersier} have measured the
properties of the field RR Lyrae population in Fornax, and find $\langle P_{ab} \rangle = 0.585$
and a mean RR Lyrae $[$Fe$/$H$] \sim -1.8$\footnote{In fact, they measure 
$[$Fe$/$H$] \sim -1.6$ on the Butler-Blanco metallicity scale, which is $\sim 0.2$ dex more metal
rich than the Zinn \& West \shortcite{zinnwest} scale. We quote the metallicity on the latter scale
for consistency with the measurements in Table \ref{clusterprop}.}. This metallicity is very similar
to those for Fornax clusters 2 and 5 (see Section \ref{distance} and Table \ref{clusterprop}),
which also have $\langle P_{ab} \rangle \sim 0.58$. Clusters 1 and 3 are somewhat more metal poor,
with $[$Fe$/$H$] \sim -2$, and have slightly longer mean RRab periods -- 
$\langle P_{ab} \rangle \sim 0.61$. From the list in Pritzl, these are most similar to the Carina
and Draco dwarf galaxies, which also have $[$Fe$/$H$] \sim -2$, and 
$\langle P_{ab} \rangle \sim 0.62$. In fact, the Carina RR Lyrae stars have recently been studied
in detail by Dall'Ora et al. \shortcite{dallora}, who find that $\langle P_{ab} \rangle$ for these 
stars matches well
the OoII clusters, but that $N_c / N_{RR}$ matches the OoI clusters -- a similar situation to
that for the Fornax clusters. These combinations are the strongest indicators that the
Fornax globular clusters also deserve an intermediate Oosterhoff classification.

\subsection{The strange case of Fornax 5}
It is worth commenting briefly on cluster 5, which has somewhat exceptional properties.
Specifically, it has a very large relative population of RRc stars, so that $N_c / N_{RR} = 0.68$.
This is significantly larger than any of the RR Lyrae populations listed in the compilation
of Pritzl et al. \shortcite{pritzl}, in which the largest belongs to the LMC cluster NGC 2257, with
$N_c / N_{RR} = 0.50$. 

The period-amplitude diagram for Fornax 5 is also unusual, in that
in addition to the well defined RRab and RRc groups, there appears a third group of four stars
in the very lower left of the diagram. This group strongly resembles the putative RRe type
stars (second overtone pulsators) identified by Clement \& Rowe \shortcite{rre} in $\omega$ 
Centauri. Although there has been some debate in the literature over the existence (or not) of
such objects, there have been several instances of observational evidence (see the references in
Clement \& Rowe). The $\sim 20$ RRe candidates identified by Clement \& Rowe have a mean period
of $0.304$ days, and a range $0.28 - 0.35$ days. The four candidates in Fornax 5 (F0503, F0504,
F0505, and F0507) have a range $0.28 - 0.39$ days, and a mean of $0.32$ days. The mean amplitude
of the $\omega$ Cen objects is slightly larger than $0.2$ mag, which also matches the mean
amplitude of the Fornax 5 candidates: $\langle \Delta V \rangle \sim 0.25$ mag. Hence the two
groups of stars occupy the same region on the period-amplitude diagram, and we conclude that
the four Fornax 5 objects are good RRe candidates. We note that these four stars are among the
five bluest RR Lyrae stars in this cluster, adding weight to the argument. There also exist
several similar stars in clusters 2 and 3; however these do not seem as well separated on the
period-amplitude diagrams as the cluster 5 stars, and we cannot claim them as RRe candidates with
certainty. 

Returning to the question of the large number of RRc stars in Fornax 5, even subtracting the
four potential RRe stars from the calculation leaves $N_c / N_{RR} = 0.58$ -- still an
unusually large value. It is interesting to note that Fornax 5 is unique in additional ways --
it is in the very outer regions of the Fornax dwarf, and it is the only post core-collapse
candidate in the system (Mackey \& Gilmore \shortcite{fornaxpaper}). It seems difficult to imagine
some link between the structure of the cluster and its RR Lyrae star population however, especially
given that neither $\omega$ Cen nor NGC 2257 are exceptionally compact objects.

\subsection{Reddening towards the Fornax globular clusters}
\label{reddening}
It is a well known property of ab-type RR Lyrae stars that they have extremely similar intrinsic 
colours at minimum light, independent of metallicity: $[(V-I)_{m}]_0 = 0.58 \pm 0.03$ 
(see e.g., Mateo et al. \shortcite{muskkk} and references therein). This can be used to determine 
the line-of-sight reddening to a group of RRab stars. The validity of this technique has been 
demonstrated for RR Lyrae stars in the Sagittarius dwarf galaxy by both Mateo et al. 
\shortcite{muskkk} and Layden \& Sarajedini \shortcite{laysar}. 

To apply this technique to the four Fornax clusters individually, we calculated the
weighted mean $\langle (V-I)_{m} \rangle$ for each sample of RRab stars, using the $(V-I)_{m}$ and
$\sigma_m$ measurements from Table \ref{rrparams}. We obtained 
$\langle (V-I)_{m} \rangle = 0.68 \pm 0.01$ for Fornax 1, $0.65 \pm 0.01$ for Fornax 2,
$0.63 \pm 0.01$ for Fornax 3, and $0.62 \pm 0.01$ for Fornax 5. To investigate the possibility
of systematic errors due to the lower limit periods, as described in Section \ref{undersampling},
we recalculated these values -- first using only RRab stars not flagged in Table \ref{rrparams} as 
having lower limit periods, and second using only RRab stars with $\Delta V > 0.9$. The results
are listed in Table \ref{clusterprop}. As expected, for the most part the values calculated using
all the RRab stars are slightly bluer than those which exclude the stars with lower limit periods.
However, the offsets are not constant. We note that the recalculations for Fornax 1 and Fornax 5
each involved less than 5 stars (since several objects do not have $(V-I)_{m}$
measurements), so these two values are strongly subject to stochastic effects. Over the combined
ensemble of all stars, the two recalculation values are $8\times10^{-4}$ mag redder and 
$3\times10^{-3}$ mag redder than the original calculation, respectively. We therefore conclude that 
the systematic effect is small (as predicted in Section \ref{undersampling}), and we feel confident 
in using the four original values of $\langle (V-I)_{m} \rangle$ quoted above.

It is then a simple matter to show that for the four clusters $E(V-I) = 0.10$, $0.07$, $0.05$, and
$0.04$. Errors in each of these values are $\pm 0.01$ random and $\pm 0.03$ systematic (from
the value of $[(V-I)_{m}]_0 $ determined by Mateo et al. \shortcite{muskkk}). Adopting 
the reddening laws calculated in Mackey \& Gilmore \shortcite{fornaxpaper}, we used
these colour excesses to calculate $A_V$ and $E(B-V)$ for each cluster. All of these quantities
are listed in Table \ref{clusterprop}. The four $E(V-I)$ values we obtained match previous
measurements well. For example, Buonanno et al. \shortcite{buonfora} determined
$E(V-I) = 0.05$, $0.09$, $0.05$, and $0.08$ (all $\pm 0.06$) for the four clusters, using their
red giant branches.

\subsection{Horizontal Branch morphology}

\begin{figure*}
\begin{minipage}{175mm}
\begin{center}
\includegraphics[width=130mm]{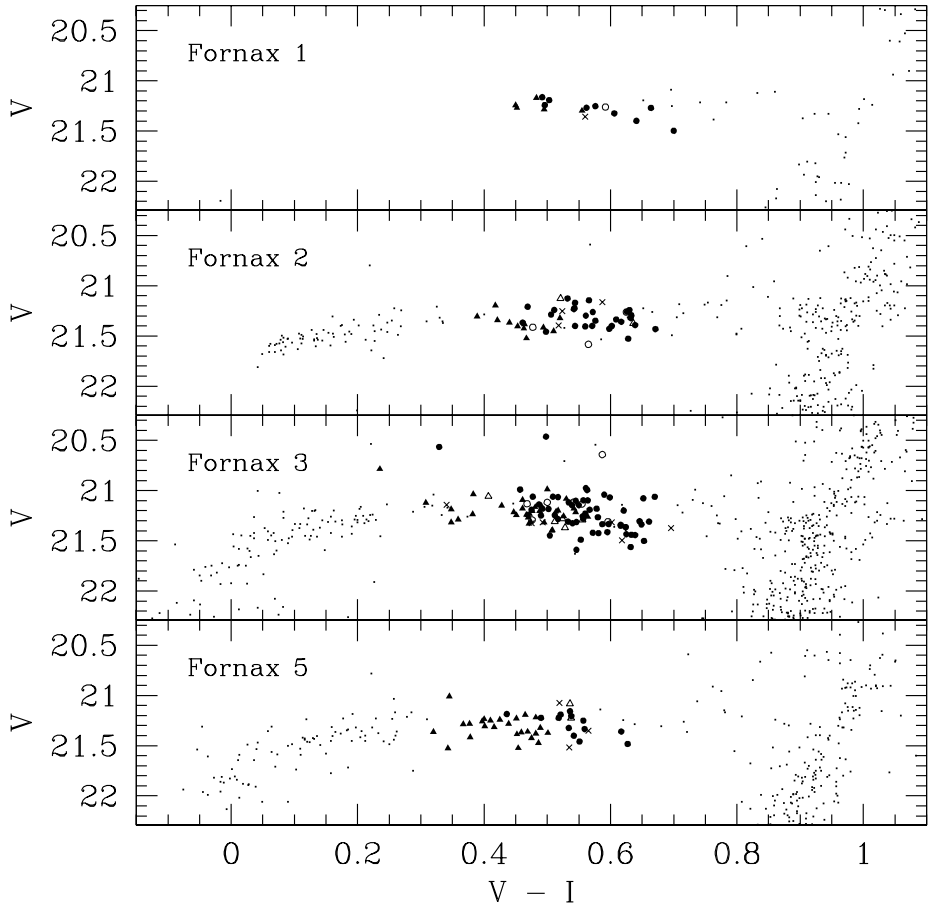}
\caption{Colour-magnitude diagrams for each of the four clusters, showing their horizontal branch regions. The RR Lyrae stars are plotted. RRab stars are solid circles, and RRc stars solid triangles. Those stars with uncertain classification have hollow points. The candidate variable stars are plotted with crosses. Small dots represent non-variable objects.}
\label{rrlyr}
\end{center}
\end{minipage}
\end{figure*}

Quantitative measurements of the numbers and colours of RR Lyrae stars in the Fornax clusters 
allows a new determination of their horizontal branch morphologies and types. Fig. \ref{rrlyr}
shows a CMD of the HB region in each cluster. The RR Lyrae stars and candidate variables are
plotted according to their identification -- RRab stars are circles, RRc stars are triangles, and
candidates are crosses. Stars with uncertain classification have hollow points (circles or 
triangles). For each cluster, the edges of the instability strip appear well defined. RRc stars
fall toward the blue edge and RRab stars toward the red edge, although there appears to be some
mixing of the two. This may be intrinsic, or it might be due to colour and classification
uncertainties (although these effects should be small).

We note that Fornax 3 has four RR Lyrae type objects which are $\sim 0.5$ mag 
brighter than the main HB. The identity of these objects is not clear. It might be that they are 
interlopers from the field of the Fornax dwarf itself -- cluster 3 has the highest level of
background stars of the four clusters considered here (Demers, Irwin \& Kunkel \shortcite{demers2}).
Alternatively, they might be unusual cluster members, such as Population II cepheids or 
anomalous cepheids. Buonanno \shortcite{buonold} also identified one such candidate star in
Fornax 3. We note that these four stars (F3V01, F3V07, F3V88, and F3V94) appear centrally 
concentrated, with three having radial distances $r \le 16\arcsec$. This suggests that at least
some of these stars are cluster members, and worthy of further investigation.

We can measure the intrinsic red and blue edges of the instability strip in each cluster from 
the colours of its RR Lyrae stars, and the colour excesses derived above. The four clusters seem
particularly suitable for this type of measurement, with each possessing a very well populated 
horizontal branch (excepting cluster 1). Examining each edge, we see that in general there is
some small mixing of variable and non-variable stars. In addition, in some cases (e.g., the 
red edge for Fornax 2) one variable star lies quite distinct from the remainder. To account for 
these two effects, we choose an ``edge'' to be the mean colour of the two most extreme variable 
stars. Using this definition, we have calculated  the colours $(V-I)_{RE}$ and $(V-I)_{BE}$ of the 
red and blue edges for each cluster\footnote{We did not include the four ``bright'' variable stars in
the calculation for Fornax 3.}. These quantities are listed in Table \ref{rrparams}, along 
with their de-reddened counterparts. There is good agreement between the de-reddened measurements
of the red edge. We obtain a mean $(V-I)_{RE} = 0.59 \pm 0.02$. The blue edge is more poorly
defined. The HB for Fornax 1 does not extend to non-variable stars, so we can discount the
measurement for this cluster. Similarly, while the Fornax 2 HB has a significant population of 
BHB stars, there is a gap at the blue edge of the RR Lyrae strip. Hence, $(V-I)_{BE}$ for this
cluster is also ill-defined. For the well populated clusters Fornax 3 and 5, we obtain good
agreement in the blue edge colour, with both having $(V-I)_{BE} = 0.28$. Applied to
Fornax 2, this value has excellent consistency, with the reddest non-variable BHB stars having
intrinsic colours $V-I \sim 0.27$. We therefore adopt a mean $(V-I)_{BE} = 0.28 \pm 0.02$.

\begin{figure}
\includegraphics[width=0.5\textwidth]{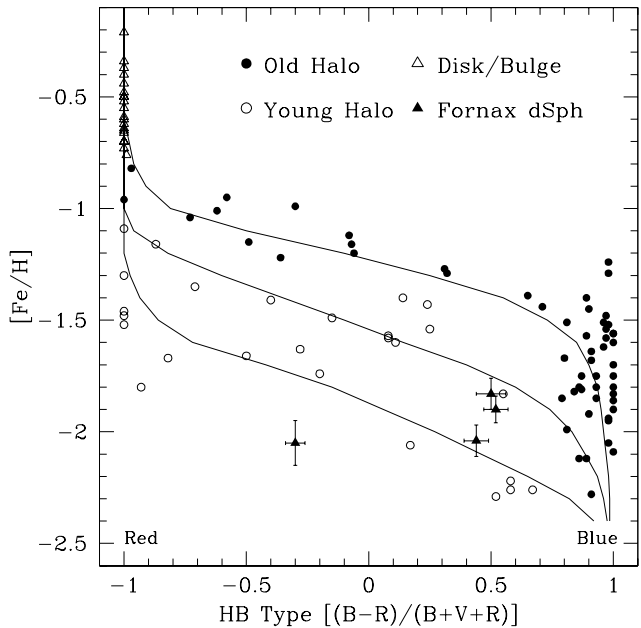}
\caption{HB type versus metallicity diagram for the galactic globular clusters and the Fornax clusters from the present paper. The galactic clusters are split into three subsystems according to the criteria of Zinn \shortcite{zinn}. The Fornax clusters are most similar to the galactic young halo population. The overplotted isochrones are from Rey et al. \shortcite{rey} and constitute the latest versions of the Lee et al. \shortcite{lee} synthetic HB models. The two lower isochrones are respectively 1.1 Gyr and 2.2 Gyr younger than the top isochrone.}
\label{hbmet}
\end{figure}

Finally, we can calculate the index $(B-R)/(B+V+R)$ of Lee, Demarque \& Zinn \shortcite{lee} 
(hereafter referred to as the HB type). In this quantity, $B$ is the number of BHB stars in a 
cluster, $R$ is the number of RHB stars, and $V$ the number of HB variable stars -- in our 
notation these are $N_{BHB}$, $N_{RHB}$, and $N_{VHB}$ respectively. In general, 
$N_{VHB} \neq N_{RR}$, because we include $N_{cand}$ in the count, and have the possibility of
excluding some of the stars counted in $N_{RR}$ -- e.g., the four bright variables in Fornax 3.
If we initially ignore the possibility of field star contamination, we count $N_{RHB}$ as the 
number of stars with $20.8 < V < 21.7$ and $0.59 < (V-I)_0 < 0.77$. The red edge was chosen after
careful consideration of the CMDs from the cluster centres, so as not to include any RGB stars.
The value is in excellent agreement with that adopted for M54 by Layden \& Sarajedini 
\shortcite{laysar}. The BHB stars are easier to count, since this is an isolated area in each
cluster's CMD. We estimated the errors in $N_{RHB}$ and $N_{BHB}$ by counting the number of
stars lying close to the edges of the defined regions. This gives some idea of the number of
stars which might have been scattered in or out of these regions. We estimated the error in $N_{VHB}$
by counting the number of non-variable stars lying in the RR Lyrae strip, and adding this to
$N_{cand}$. This accounts for stars which we may have mis-identified as variable, or
variable stars which were missed by our detection procedure. All star counts are listed in Table
\ref{rrparams}.

Using these quantities, we measured HB types of $-0.30 \pm 0.04$, $0.42 \pm 0.05$, $0.40 \pm 0.05$,
and $0.52 \pm 0.04$ for the four clusters. The errors were derived using the quadrature formula in
Smith et al. \shortcite{smith15}. Our measurements are in reasonable agreement with the values 
derived by Buonanno et al. \shortcite{buonfora} using the same WFPC2 observations, but different
numbers of variable stars (see Section \ref{prevwork}). However, there is the possibility that our
counts as described above have been contaminated by field stars, especially for clusters 2 and 3,
which lie against significant backgrounds (see e.g., Demers et al. \shortcite{demers2}). In 
particular, these two clusters seem to have quite well populated red clumps on the RGB near the RHB,
and it is therefore likely our measured HB types are too red. We have
no nearby field observations, so to try and account for this we recalculated the HB index for each
cluster using only stars measured on the PC, where the cluster centres are imaged. In this
case we derived HB types of $-0.38 \pm 0.06$, $0.50 \pm 0.05$, $0.44 \pm 0.05$, and $0.52 \pm 0.05$, 
respectively. The calculation for cluster 1 involves small numbers of stars, and we prefer the
original value. For the other three clusters, it is clear that Fornax 5 has not suffered any
contamination (as expected -- this cluster lies well separated from the dwarf galaxy), but
Fornax 2 and 3 possess some small field star contamination. For these two clusters we prefer the
HB types as calculated from the PC, rather than the original values. 

Additionally, we note that the CMD for cluster 1 presented by Buonanno et al. \shortcite{buonposter}
apparently shows several HB stars bluer than $V-I \sim 0.3$, whereas these stars are not present
in our data. Hence our derived HB type for this cluster may be too red. It is possible that the 
WFPC2 observations did not include the region of Fornax 1 containing the BHB stars; however we find
this explanation unlikely since the images cover $\sim 75$ per cent of the cluster, including its
core. Even so, adding several BHB stars to the calculation does not produce a large effect. 
Four stars would move the HB type to $\sim -0.1$, still in reasonable agreement with our preferred
value.

These measurements confirm the existence of a ``second parameter'' effect among the Fornax globular
clusters, as highlighted previously by several authors \cite{smith15,buonfora}. For example,
clusters 1 and 3 have very similar metallicities (see Table \ref{rrparams}), but extremely 
disparate HB morphologies. In Fig. \ref{hbmet} we plot a HB type versus metallicity diagram to 
highlight this result. Lee et al. \shortcite{lee} have suggested that age differences can mostly
account for second parameter variations in HB type, in that younger clusters possess redder horizontal
branches at given $[$Fe$/$H$]$. Zinn \shortcite{zinn} split the galactic globular clusters by
metallicity and HB type into three distinct groups. The most metal rich clusters are associated with
the bulge and thick-disk of the galaxy, and the most metal poor with the halo. The halo clusters
are subdivided into ``young'' and ``old'' groups on the basis of their HB type. From Fig. \ref{hbmet}
it is clear that the Fornax clusters are most similar to the galactic young halo clusters.
This is compatible with the assertion of Zinn \shortcite{zinn} that the young halo clusters have
likely been accreted by the galaxy through merger events with satellite dwarf galaxies.

Buonanno et al. \shortcite{buonfora} have measured the ages of Fornax clusters 1, 2, 3, and 5 to be
identical to within $\pm 1$ Gyr, and equally coeval with the metal-poor galactic globular clusters
M68 and M92. The isochrones overplotted on Fig. \ref{hbmet} show that based on HB type, the Fornax
clusters should be more than 1.1 Gyr younger than the galactic old halo clusters, and moreover 
should have an internal age spread greater than $\sim 1.1$ Gyr (in that Fornax 1 should be younger). 
If the Buonanno et al. \shortcite{buonfora} ages are correct, then it is possible that age is not the
sole second parameter operating in the Fornax clusters. Bearing in mind the extreme difference 
highlighted in Section \ref{freq} between the specific frequencies of RR Lyrae stars in the Fornax 
and galactic globular clusters, and the way that the HB index is defined, it seems likely that the 
relative number of RR Lyrae stars in a cluster has a significant bearing on the measured HB type. It 
is certainly interesting to note that the four galactic globular clusters with $S_{RR} > 60$ are all 
members of the galactic young halo population.

\begin{table*}
\begin{minipage}{152mm}
\caption{Globular cluster and RR Lyrae population characteristics.}
\begin{tabular}{@{}llcccc}
\hline \hline
 & & \hspace{6mm}Fornax 1\hspace{6mm} & \hspace{6mm}Fornax 2\hspace{6mm} & \hspace{6mm}Fornax 3\hspace{6mm} & \hspace{6mm}Fornax 5\hspace{6mm} \\
\hline
$N_{RR}$ & & $15$ & $43$ & $99$ & $40$ \\
$N_{cand}$ & & $1$ & $3$ & $6$ & $3$ \\
 & & & & & \\
$M_{V}$ & & $-5.33$ & $-7.05$ & $-7.80$ & $-7.05$ \\
$S_{RR}$ & (Lower limits) & $118.2$ & $70.5$ & $90.2$ & $68.4$ \\
 & & & & & \\
$N_{ab}$ & (All RRab) & $10$ & $30$ & $61$ & $13$ \\
 & (RRab with good $P$) & $5$ & $15$ & $32$ & $8$ \\
 & (RRab with $\Delta V > 0.9$) & $4$ & $10$ & $14$ & $5$ \\
$N_c$ & & $5$ & $13$ & $38$ & $27$ \\
$N_c / N_{RR}$ & & $0.33$ & $0.30$ & $0.38$ & $0.68$ \\
 & & & & & \\
$\langle P_{ab} \rangle$ & (All RRab) & $0.546$ & $0.494$ & $0.532$ & $0.532$ \\
 & (RRab with good $P$) & $0.611$ & $0.574$ & $0.613$ & $0.577$ \\
$P_{tr}$ & (RRab with $\Delta V > 0.9$) & $0.544$ & $0.540$ & $0.582$ & $0.548$ \\
$\langle P_{c} \rangle$ & (All RRc) & $0.431$ & $0.373$ & $0.404$ & $0.374$ \\
 & & & & & \\
$\langle (V-I)_{m} \rangle$ & (All RRab) & $0.68 \pm 0.01$ & $0.65 \pm 0.01$ & $0.63 \pm 0.01$ & $0.62 \pm 0.01$ \\
 & (RRab with good $P$) & $0.70 \pm 0.01$ & $0.66 \pm 0.01$ & $0.63 \pm 0.01$ & $0.60 \pm 0.01$ \\
 & (RRab with $\Delta V > 0.9$) & $0.69 \pm 0.01$ & $0.66 \pm 0.01$ & $0.63 \pm 0.01$ & $0.60 \pm 0.01$ \\
$E(V-I)$ &  & $0.10 \pm 0.01$ & $0.07 \pm 0.01$ & $0.05 \pm 0.01$ & $0.04 \pm 0.01$ \\
$E(B-V)$ &  & $0.07 \pm 0.01$ & $0.05 \pm 0.01$ & $0.04 \pm 0.01$ & $0.03 \pm 0.01$ \\
$A_V$ &  & $0.23 \pm 0.02$ & $0.16 \pm 0.02$ & $0.12 \pm 0.02$ & $0.10 \pm 0.02$ \\
 & & & & & \\
$(V-I)_{RE}$ & Measured & $0.68 \pm 0.02$ & $0.66 \pm 0.02$ & $0.66 \pm 0.02$ & $0.62 \pm 0.02$ \\
 & De-reddened & $0.58 \pm 0.03$ & $0.59 \pm 0.03$ & $0.61 \pm 0.03$ & $0.58 \pm 0.03$ \\
$(V-I)_{BE}$ & Measured & $0.45 \pm 0.01$ & $0.39 \pm 0.02$ & $0.33 \pm 0.02$ & $0.32 \pm 0.01$ \\
 & De-reddened & $0.35 \pm 0.02$ & $0.32 \pm 0.03$ & $0.28 \pm 0.03$ & $0.28 \pm 0.02$ \\
$N_{BHB}$ & $(\equiv B)\,\,\,$ (All chips) & $0 \pm 0$ & $68 \pm 4$ & $114 \pm 15$ & $83 \pm 6$ \\
$N_{RHB}$ & $(\equiv R)\,\,\,$ (All chips) & $7 \pm 1$ & $14 \pm 3$ & $20 \pm 4$ & $12 \pm 3$ \\
$N_{VHB}$ & $(\equiv V)\,\,\,$ (All chips) & $16 \pm 2$ & $46 \pm 9$ & $101 \pm 15$ & $43 \pm 5$ \vspace{1mm} \\ 
HB index & $\frac{(B-R)}{(B+V+R)}\,\,\,$ (All chips)& $-0.30 \pm 0.04$ & $0.42 \pm 0.05$ & $0.40 \pm 0.05$ & $0.52 \pm 0.04$ \vspace{1mm} \\
 & (PC only) & $-0.38 \pm 0.06$ & $0.50 \pm 0.06$ & $0.44 \pm 0.05$ & $0.52 \pm 0.05$ \\
 & & & & & \\
$[$Fe$/$H$]$ & & $-2.05 \pm 0.10$ & $-1.83 \pm 0.07$ & $-2.04 \pm 0.07$ & $-1.90 \pm 0.06$ \\
$M_V (\rmn{RR})$ & & $0.46 \pm 0.02$ & $0.51 \pm 0.02$ & $0.46 \pm 0.02$ & $0.49 \pm 0.02$ \\
 & & & & & \\
$\langle V(\rmn{RR}) \rangle$ & (All RR) & $21.27 \pm 0.01$ & $21.34 \pm 0.01$ & $21.24 \pm 0.01$ & $21.33 \pm 0.01$ \\
 & (All RRab) & $21.29 \pm 0.01$ & $21.32 \pm 0.01$ & $21.25 \pm 0.01$ & $21.35 \pm 0.01$ \\
 & (RRab with good $P$) & $21.29 \pm 0.01$ & $21.31 \pm 0.01$ & $21.21 \pm 0.01$ & $21.32 \pm 0.01$ \\
 & (RRab with $\Delta V > 0.9$) & $21.29 \pm 0.01$ & $21.29 \pm 0.01$ & $21.20 \pm 0.01$ & $21.30 \pm 0.01$ \\
 & (All RRc) & $21.26 \pm 0.01$ & $21.36 \pm 0.01$ & $21.21 \pm 0.01$ & $21.30 \pm 0.01$ \\
 & & & & & \\
$(m-M)_{0}$ & & $20.58 \pm 0.05$ & $20.67 \pm 0.05$ & $20.66 \pm 0.05$ & $20.74 \pm 0.05$ \\
Distance & (kpc) & $130.6 \pm 3.0$ & $136.1 \pm 3.1$ & $135.5 \pm 3.1$ & $140.6 \pm 3.2$ \\
\hline
\label{clusterprop}
\end{tabular}
\end{minipage}
\end{table*}

\subsection{Distance and structure of the Fornax dwarf}
\label{distance}
Using the mean $V$ magnitudes we measured for the RR Lyrae stars, it is also possible to determine
a distance modulus to each cluster. To do this, we need to know the absolute magnitude $M_V(RR)$
of the RR Lyrae stars. This quantity is a function of metallicity, so that more metal poor stars
are intrinsically brighter. We adopt the relation from the review of Chaboyer \shortcite{chaboyer}:
\begin{equation}
M_V(RR) = 0.23([\rmn{Fe} / \rmn{H}] + 1.6) + 0.56
\end{equation}
which includes {\em Hipparcos} results. This is the same relation used by Layden \& Sarajedini
\cite{laysar} to determine the distance to M54 and the Sagittarius dwarf galaxy. We note that the
relation has an uncertainty of $\pm 0.12$ mag in the zero-point term, reflecting the discrepancies
which exist between the different techniques to determine $M_V(RR)$. 

To obtain metallicities for each cluster, we used the recent measurements and literature collation 
of Strader et al. \shortcite{strader}. From their Table 5, we calculated the mean measured
metallicity for each cluster. This combines the results of several different measurement
techniques, ranging from integrated spectroscopy to the slope of the RGB. For the four clusters,
we obtain $[$Fe$/$H$] = -2.05 \pm 0.10$, $-1.83 \pm 0.07$, $-2.04 \pm 0.07$, and $-1.90 \pm 0.06$.
This results in RR Lyrae absolute magnitudes of $M_V(RR) = 0.46 \pm 0.02$, $0.51 \pm 0.02$,
$0.46 \pm 0.02$, and $0.49 \pm 0.02$, respectively. 

The next step is to obtain a mean $V$ RR Lyrae magnitude, $\langle V(RR) \rangle$  for each cluster.
We determined these from the $\langle V \rangle$ measurements in Table \ref{rrparams}. Stars 
with $\langle V \rangle$ deviant by more that $2\sigma$ were not included in the final average. 
The mean $V$ magnitudes so obtained are listed in Table \ref{clusterprop}. Using the extinction 
values calculated above, together with the RR Lyrae absolute magnitudes, we then determined a 
distance modulus for each cluster. These results are also listed in Table \ref{clusterprop}. The 
mean distance modulus is $20.66 \pm 0.03$ mag, which is in good agreement with previously determined 
values. In particular, Buonanno et al. \shortcite{buonforb} obtain an estimate of $20.68 \pm 0.20$ 
to the centre of the Fornax dwarf (i.e., cluster 4) after a thorough discussion of several 
techniques and literature measurements. We note that the uncertainties quoted for our results are 
due to random errors only, and that there are also systematic errors totalling approximately 
$\pm 0.15$ mag. 

In addition we briefly investigated the likelihood that our $\langle V(RR) \rangle$ values are 
systematically too faint, based on the discussion in Section \ref{undersampling}. We recalculated 
$\langle V(RR) \rangle$ using only RRab stars, then only RRab stars with good periods, and then
only RRab stars with $\Delta V > 0.9$. These values are also listed in Table \ref{clusterprop}. 
On average, the $\langle V(RR) \rangle$ calculated using all the RRab stars are $\sim 0.03$ mag
fainter than those calculated using the RRab stars with good periods, and $\sim 0.04$ mag
fainter than those calculated using the RRab stars with $\Delta V > 0.9$. This demonstrates 
exactly what we argued in Section \ref{undersampling} -- that we expect a systematic error in
$\langle V(RR) \rangle$ due to our short observation baselines. If we include the RRc stars
in our calculations, the offsets are moderated somewhat. The two worst affected clusters are
Fornax 3 and 5, which may have distance moduli $0.03$ mag closer than those we have quoted.

Our distance moduli correspond to distances of $130.6 \pm 3.0$ kpc, $136.1 \pm 3.1$ kpc, 
$135.5 \pm 3.1$ kpc, and $140.6 \pm 3.2$ kpc for the four clusters. Again, the quoted uncertainties
reflect random errors only. These measurements are consistent with, but do not require, a 
significant line of sight depth to the Fornax system, with a value of $\sim 10$ kpc. The outlying 
clusters are Fornax 1 and 5, which are also the clusters with the greatest angular separation from 
the centre of the Fornax dwarf (see Mackey \& Gilmore \shortcite{fornaxpaper}). We note that if
Fornax 5 is indeed $0.03$ mag closer than we have listed above, the line of sight depth shrinks
slightly to $\sim 8$ kpc. Demers et al. \shortcite{demers2} obtained a tidal radius for Fornax
of $r_t = 77 \pm 10$ arcmin along the minor axis, and an ellipticity of $0.27$ at this radius.
Hence the tidal radius along the major axis must be $105 \pm 14$ arcmin. These radii correspond
to linear distances of $7.0 \pm 0.9$ kpc and $9.6 \pm 1.3$ kpc at our mean distance modulus of
$20.66$. These are of a similar magnitude to the $\sim 4.5$ kpc radius obtained from our line of 
sight depth, suggesting that Fornax is close to spherical. This lack of extended line of sight
depth argues against models of dSph galaxies which invoke very large line of sight depth and/or
tides to explain the observed high stellar velocity dispersions. Other recent studies
have reached a similar conclusion (e.g., Klessen, Grebel \& Harbeck \shortcite{klessen}, who
studied the thickness of the Draco dSph HB). It seems most likely that dSph galaxies with
high stellar velocity dispersions constitute dark matter dominated systems. 

\section{Summary and Conclusions}
We have identified and measured 197 RR Lyrae stars in four of the globular clusters in the Fornax 
dwarf galaxy, using archival WFPC2 observations. We also located 13 variable HB objects, which
are likely RR Lyrae stars. A typical star in our sample has $14$ paired F555W and F814W 
observations spread over $\sim 0.36$ days. Such a short baseline has compromised our ability
to measure accurate periods for $\sim 50$ per cent of the RRab type stars we identified -- for
these objects we instead calculated lower limits for the oscillation periods and amplitudes.
Nonetheless, we have successfully determined periods for all the RRc and $50$ per cent of the
RRab stars in the ensemble. We also present measurements of mean magnitudes and colours for all
the objects in our sample, and discuss the introduction of systematic errors into these
calculations due to our short observational baseline. This is the first time that RR Lyrae 
stars have been identified in these clusters, and quantitative measurements of their variability 
presented. Naturally, additional observations at more than one epoch will be extremely useful
in refining our classifications and measurements, and in particular in allowing good periods to
be determined for all stars.

Even given the difficulties associated with measuring periods for some of the sample,
we have been able to obtain a large amount of useful information about the Fornax globular clusters
from the RR Lyrae stars. The specific frequencies of the four clusters are exceptionally large
in comparison to those for the galactic globular clusters. All of the four Fornax clusters have
$S_{RR} > 80$, while only four of the $\sim 150$ galactic clusters have such large values.
It is likely that Fornax 1 has the largest specific frequency ever measured in a globular cluster.
The Fornax clusters are also unusual in that their RR Lyrae populations possess mean properties
which lie intermediate between the two Oosterhoff groups defined by the galactic globular clusters.
In this respect the RR Lyrae stars in the Fornax clusters are similar to those in the old 
LMC clusters, but most resemble the field populations of several dwarf galaxies. Fornax cluster 5
stands out in having an extremely high fraction of RRc stars ($N_{c} / N_{RR} \sim 0.6$),
and possessing four strong RRe (second overtone pulsator) candidates.

Using the mean RR Lyrae colours at minimum light, we have measured colour excesses and line of
sight extinction values towards the four clusters. With a significant catalogue of HB variable
stars, we have also been able to provide new estimates of the HB morphology in the these clusters.
We have confirmed the presence of the second parameter effect, in particular between clusters 1
and 3, which have identical metallicities and ages, but extremely different HB morphology. The
HB morphologies of the Fornax clusters are in general redder than many galactic clusters of 
similar metallicity -- in this respect they most resemble the ``young'' galactic halo population.
We also find good agreement between measurements of the intrinsic red and blue edges to the 
instability strip at the HB. We determine $(V-I)_{RE} = 0.59 \pm 0.02$ and 
$(V-I)_{BE} = 0.28 \pm 0.02$.

Finally, we have used the mean RR Lyrae brightnesses to determine the distance to the 
Fornax dwarf galaxy. We measure a mean distance modulus of $(m-M)_0 = 20.66 \pm 0.03$ mag, which 
is in excellent agreement with previous measurements. Our calculations are consistent with
a line of sight depth of $\sim 8-10$ kpc for the Fornax dwarf galaxy. This value is in good 
accordance with the dimensions of this galaxy as measured in the plane of the sky, and argues
against the tidal-remnant models of dSph galaxies which are invoked to explain their observed 
internal stellar dynamics.

\section*{Acknowledgments}
ADM would like to acknowledge the support of a Trinity College ERS 
grant and a British government ORS award. We would also like to thank
Andrew Layden for making his template-fitting programs publicly available.
This paper is based on observations made with the NASA/ESA 
{\em Hubble Space Telescope}, obtained from the data archive at the 
Space Telescope Institute. STScI is operated by the association of 
Universities for Research in Astronomy, Inc. under the NASA contract 
NAS 5-26555.



\bsp 

\label{lastpage}

\end{document}